\documentclass[aps,prd,onecolumn,fleqn,superscriptaddress,groupedaddress,nofootinbib,preprintnumbers]{revtex4}
\pdfoutput=1
\usepackage{amsmath,amssymb,pstricks,graphicx,todonotes,xspace}
\usepackage{pgfplots}
\presetkeys{todonotes}{inline}{}

\newcommand{\nnlojet}{NNLOJET\xspace}

\begin{document}

\hfill FERMILAB-PUB-19-100-T

\hfill CERN-TH-2019-038

\hfill SLAC-PUB-17411

\hfill UWTHPH-2019-9

\hfill LAPTH-019/19

\hfill CFTP/19-006

\hfill ZU-TH 16/19

\hfill LU-TP-19-10

\hfill MCnet-19-06

\hfill IPPP/19/22

\title{Jet cross sections at the LHC and the quest for higher precision}

\author{Johannes Bellm}
\affiliation{Department of Astronomy and Theoretical Physics,
 Lund University, S-223 62 Lund, Sweden}
\author{Andy Buckley}
\affiliation{School of Physics and Astronomy, 
 University of Glasgow, Glasgow, G12 8QQ, UK}
\author{Xuan Chen}
\affiliation{Institut f{\"u}r Theoretische Physik, 
 Universit{\"a}t Z{\"u}rich, CH-8057 Z{\"u}rich, Switzerland}
\author{Aude Gehrmann-De Ridder}
\affiliation{Institute for Theoretical Physics,
 ETH, CH-8093 Z{\"u}rich, Switzerland}
\affiliation{Institut f{\"u}r Theoretische Physik, 
 Universit{\"a}t Z{\"u}rich, CH-8057 Z{\"u}rich, Switzerland}
\author{Thomas Gehrmann}
\affiliation{Institut f{\"u}r Theoretische Physik, 
 Universit{\"a}t Z{\"u}rich, CH-8057 Z{\"u}rich, Switzerland}
\author{Nigel Glover}
\affiliation{Institute for Particle Physics Phenomenology,
 Durham University, Durham DH1 3LE, UK}
\author{Alexander Huss}
\affiliation{Theoretical Physics Department,
 CERN, 1211 Geneva 23, Switzerland}
\author{Joao Pires}
\affiliation{CFTP, Instituto Superior T\'ecnico,
 Universidade de Lisboa, P-1049-001 Lisboa, Portugal}
\author{Stefan H{\"o}che}
\affiliation{SLAC National Accelerator Laboratory,
 Menlo Park, CA, 94025, USA}
\affiliation{Fermi National Accelerator Laboratory,
 Batavia, IL, 60510-0500, USA}
\author{Joey Huston}
\affiliation{Department of Physics and Astronomy,
 Michigan State University, East Lansing, MI 48824, USA}
\author{Silvan Kuttimalai}
\affiliation{SLAC National Accelerator Laboratory,
 Menlo Park, CA, 94025, USA}
\author{Simon Pl{\"a}tzer}
\affiliation{Fakult{\"a}t Physik,
 University of Vienna, 1010 Vienna, Austria}
\author{Emanuele Re}
\affiliation{Theoretical Physics Department,
 CERN, 1211 Geneva 23, Switzerland}
\affiliation{Laboratoire d'Annecy-le-Vieux de Physique Th\'eorique,
 Universit\'e Grenoble Alpes, Universit\'e Savoie Mont Blanc, CNRS, 74940 Annecy, France}

\begin{abstract} We perform a phenomenological study of $Z$ plus jet, Higgs
plus jet and di-jet production at the Large Hadron Collider. We investigate in
particular the dependence of the leading jet cross section on the jet radius as
a function of the jet transverse momentum. Theoretical predictions are obtained
using perturbative QCD calculations at the next-to and next-to-next-to-leading
order, using a range of renormalization and factorization scales. The fixed
order predictions are compared to results obtained from matching
next-to-leading order calculations to parton showers. A study of the scale
dependence as a function of the jet radius is used to provide a better estimate
of the scale uncertainty for small jet sizes. The non-perturbative corrections
as a function of jet radius are estimated from different generators.  
\end{abstract}

\maketitle

\section{Introduction}

The production of a single object like a $Z$ or Higgs boson, or a jet, at high
transverse momentum has been studied intensely in hadron collider environments,
both theoretically and experimentally. These processes are used for measuring
standard-model parameters, to constrain parton distribution functions (PDFs),
and to understand backgrounds to new physics searches. They probe the structure
of the QCD interactions in great detail. On the one hand, the large scales
associated with the production of a high-$p_T$ object make QCD perturbation
theory a prime analysis tool. For $H/Z+\ge1$ jet production, the large boson
mass also provides a large scale to further stabilize the QCD prediction. On
the other hand, the exclusive nature of the reactions may induce
logarithmically enhanced higher-order corrections, which must be resummed to
all orders. Both aspects must be incorporated into simulations used for
experimental and phenomenological analyses to provide accurate predictions.

In the past few years, the state of the art in QCD perturbation theory has
advanced considerably. Next-to-next-to-leading order QCD predictions are now
available for $Z$-boson plus jet~\cite{Ridder:2015dxa,
Boughezal:2015ded,Gehrmann-DeRidder:2016jns,Ridder:2016nkl,
Gehrmann-DeRidder:2017mvr}, Higgs-boson plus
jet~\cite{Boughezal:2013uia,Chen:2014gva,Boughezal:2015dra,Chen:2016zka,Chen:2016vqn},
and for inclusive jet and di-jet production~\cite{Ridder:2013mf,Currie:2013dwa,
Currie:2016bfm,Currie:2017eqf,Currie:2018xkj}. Next-to-leading order accurate
results have been available for some
time~\cite{Ellis:1981hk,Altarelli:1984pt,Ellis:1985er}. They can now be
computed in an automatic fashion using general-purpose event
generators~\cite{Gleisberg:2007md,Frederix:2008hu,Frederix:2009yq,
Cascioli:2011va,Hirschi:2011pa,Cullen:2011ac,Platzer:2011bc,
Cullen:2014yla,Alwall:2014hca} and the matching to parton showers can be
carried out with a number of different
approaches~\cite{Frixione:2002ik,Nason:2004rx}. Analytic results for jet
radius~\cite{Dasgupta:2016bnd} and combined jet radius and threshold
resummation are available as well~\cite{Liu:2017pbb,Liu:2018ktv}.

A contribution to the Les Houches 2017 workshop compared predictions for H+j
production at LO, NLO and NNLO to those from parton-shower matched NLO
calculations using different event generators, for a variety of jet
radii~\cite{Bendavid:2018nar}. The goal for this comparison was multi-fold:
using identical boundary conditions, to check the consistency of the matched
predictions, and to demonstrate that the matched results revert to their
underlying fixed order predictions in kinematic regions where Sudakov
resummation can be neglected. Good agreement was observed between different
parton showers, and between the matched predictions and the fixed order
results. The best agreement of the jet shape was obtained in the comparison
between NLO+PS matched and fixed-order NNLO predictions, which is expected
given the more complete description of the jet shape upon including double-real
radiative corrections at NNLO.

In this study, we follow up with further comprehensive comparisons for
Higgs+jet, and for two additional processes, $Z$+jet and dijet, concentrating
on inclusive observables, such as the lead jet transverse momentum
distribution. Contrary to popular opinion, the agreement among the various
NLO+PS matched predictions (including POWHEG for dijet production) for these
observables is good, as is the underlying agreement with the relevant fixed
order calculation, especially at NNLO, if each prediction is used properly, and
with identical parameters. One of the powerful and indeed unique aspects of
this study is the comparison of jet cross sections for a wide variety of jet
radii (beyond what is commonly used by the ATLAS and CMS experiments). This
allows for a better fundamental understanding of the underlying physics, both
perturbative and non-perturbative.

In addition, we examine the scale uncertainties of the three processes, at LO,
NLO and NNLO, as a function of jet radius, and comment on the implication of
our results on the determination of {\it reasonable} scale uncertainties.

The paper is organized as follows. In Sec.~\ref{sec:setup} we detail the setup
of the generators used in this study. In Sec.~\ref{sec:shapes} we discuss the
the shape of jets for the three processes. Sec.~\ref{sec:kfactors} focuses on
the transverse momentum dependence of the cross-sections and the influence of
higher order corrections at fixed order. In Sec.~\ref{sec:results} we describe
and compare results from the fixed order expansion to parton level Monte Carlo
simulations. In Sec.~\ref{sec:uncertainties} we consider cross section
uncertainties that arise due to the jet definition. Before we give concluding
remarks and an outlook in Sec.~\ref{sec:outlook} we examine the possible
influence of hadronization and multiple parton interactions on the measurement
of cross-sections in Sec.~\ref{sec:hadcorr}. 

\section{Setup} \label{sec:setup} 

We investigate Higgs+jet, $Z$+jet and
inclusive jet production, taking advantage of the NNLO calculations available
for all three processes. The latter two reactions are important for global PDF
fits, where only fixed order predictions (along with the relevant
non-perturbative corrections) have been used so far, and thus it is important
to understand the possible impact of resummation effects. 

The analyses use the anti-$k_T$ jet algorithm~\cite{Cacciari:2008gp}, with
varying jet size as described below, with a jet transverse momentum threshold
of 30~GeV, along with a cut on the jet rapidity of 4.5. To avoid generation cut
effects, the comparisons are performed above a jet transverse momentum of
50~GeV. Further, any cross-section that is sensitive to the colourless system
is to be taken above a transverse momentum of 90~GeV to ensure the possibility
of having three well-separated partons (at NNLO) recoiling and resulting in
generation cut migration.

As a further test of the impact of parton showers versus fixed order, the jet
size was varied across the values 0.3, 0.4, 0.5, 0.6, 0.7, and 1.0, using the
anti-$k_T$ jet algorithm. Indirectly, this tests how well the one (two) extra
parton(s) at NLO (NNLO) reproduce resummation effects. This is of particular
interest as the Higgs ($Z$) boson + jets measurements that have been performed
at the LHC in Run 2 have typically used a jet size of 0.4, which is only
slightly above the region where small $R$ effects become
important~\footnote{Inclusive jet production has been measured at two different
jet radii, for example $R$= 0.4 and 0.6 for ATLAS. The global PDF fiting groups
almost always use the larger jet size, where the jet shape is not as
critical.}. Taking the small $R$ effects into proper account would require
resummation, as discussed in~\cite{Dasgupta:2016bnd,Liu:2017pbb,Liu:2018ktv}.
The NLO+PS predictions provide this resummation by means of the parton showers. 

In this study, predictions from NLO+PS programs were carried out at the parton
shower level to make them comparable to the fixed order
calculations~\footnote{As a reminder, the non-perturbative corrections used for
fixed order predictions are determined from a comparison of the parton shower
predictions with and without the non-perturbative effects, as a function of jet
radius. This implicitly requires the integrated jet shape determined by fixed
order predictions to agree with those determined by parton showers.}.
To the degree to which it was possible, the fiducial setups have been
constrained to be the same for all calculations. We used the PDF4LHCNNLO\_30
PDFs~\cite{Butterworth:2015oua}, with its central value of $\alpha_s(m_Z)$ of
0.118. We do not address PDF uncertainties. The renormalization and
factorization scales used to compute the fixed-order perturbative results have
been chosen as similar as possible, providing a greater level of control than
was available in a similar study during the 2015 Les Houches
workshop~\cite{Badger:2016bpw}. More details will be provided in the
sub-sections. We use the Rivet framework~\cite{Buckley:2010ar} to analyze
events. A CMS routine from the 13~TeV inclusive jet
analysis~\cite{Khachatryan:2016wdh} was modified to add the different $R$
values, as well as additional observables.

\subsection{NNLOJET}

The NNLO corrections to $pp \to X+j$ receive contributions from three types of
parton-level subprocesses: the $X$+5~parton tree-level (double-real
correction), the $X$+4~parton process at one-loop (real--virtual correction),
and the two-loop $X$+3~parton result (double-virtual correction). The
double-real, real-virtual and double virtual corrections to $pp \to H+j$
production were computed in~\cite{DelDuca:2004wt,Dixon:2004za,Badger:2004ty},
\cite{Badger:2009hw,Badger:2009vh,Dixon:2009uk} and \cite{Gehrmann:2011aa},
respectively. The double-real, real-virtual and double virtual corrections to
$pp \to Z+j$ production were computed
in~\cite{Hagiwara:1988pp,Berends:1988yn,Falck:1989uz,Nagy:1998bb},
\cite{Glover:1996eh,Bern:1996ka,Campbell:1997tv,Bern:1997sc} and
\cite{Moch:2002hm,Garland:2001tf,Garland:2002ak,Gehrmann:2011ab}, respectively.
The double-real, real-virtual and double virtual corrections to inclusive jet
and di-jet production were computed in~\cite{Mangano:1990by},
\cite{Bern:1993mq,Bern:1994fz,Kunszt:1994nq} and
\cite{Glover:2001af,Glover:2001rd,Bern:2002tk,Anastasiou:2001sv,
Anastasiou:2000mv,Glover:2003cm,Bern:2003ck}, respectively.

Each of the above components of the NNLO calculations is separately infrared
(IR) divergent, and the divergences cancel upon integration over the unresolved
phase space by virtue of the Kinoshita--Lee--Nauenberg theorem. In order to
compute a fully differential prediction using Monte-Carlo integration
techniques, a procedure for the subtraction of IR singularities is required to
make this cancellation manifest, and to construct a locally finite integrand.
To this end, we employ the antenna subtraction
formalism~\cite{GehrmannDeRidder:2005cm,
GehrmannDeRidder:2005aw,GehrmannDeRidder:2005hi,Daleo:2006xa,Daleo:2009yj,Gehrmann:2011wi,
Boughezal:2010mc,GehrmannDeRidder:2012ja,Currie:2013vh}, which is implemented
in the NNLOJET framework.

For the central predictions of our current study, we use the following
dynamical scale for $H$+jet and $Z$+jet production processes, 

\begin{equation}
\mu_0 = \frac{H_{T,j}}{2} = \frac{1}{2}\bigg(\sqrt{m_X^2+p_{T,X}^2} +\sum_{\rm
jets} p_{T,j}\bigg).\label{eq:SM_Higgs_jet_R:htprimescale} 
\end{equation} 

The
LO and NLO differential cross sections using this dynamical scale choice were
validated against Sherpa and Herwig7. The renormalisation ($\mu_R$) and
factorisation ($\mu_F$) scales are varied independently around $\mu_0$ by
factors of $\tfrac{1}{2}$ and $2$ to estimate the size of missing higher-order
contributions. Here, the two extreme variations are excluded such that we
arrive at the custom 7-point scale variation: 

\begin{equation} (\mu_R,\mu_F) =
\bigl\{ (1,1), \; (2,2), \; (\tfrac{1}{2},\tfrac{1}{2}), \; (\tfrac{1}{2},1),
\; (1,\tfrac{1}{2}), \; (2,1), \; (1,2) \bigr\} \times \mu_0 . 
\end{equation}

The inclusive jet production process has been studied at NNLO in
Ref.~\cite{Currie:2017ctp}, using a standard scale choice of
$\mu_0=p_T^\mathrm{jet}$ where this quantity refers to the transverse momentum
of each individual jet. Thus, for each jet in a NNLO event, there is a
corresponding entry in the plot with the matrix element weight evaluated at the
jet $p_T$ as the scale. This is the very definition of an inclusive cross
section. An alternative choice is to use as a scale the transverse momentum of
the highest $p_{T}$ jet in the event ($p_{T}^{\rm
jetmax}$)~\cite{Currie:2016bfm}. The use of these two scales creates a sizeable
difference at NNLO at low transverse momentum~\cite{Currie:2018xkj}, which is
larger than the nominal scale uncertainty around either scale. This effect was
examined in detail in Ref.~\cite{Currie:2018xkj}, leading to the observation
that large-scale cancellations between different kinematical configurations in
the second jet contribution are aggravated for certain scales. An event based
scale ($H_{T}$) built from the scalar sum of transverse momenta of all partons
in the event was found to be stable, leading to an improved perturbative
convergence on the transverse momentum distributions, with overlapping scale
uncertainty bands between NLO and NNLO. In this study this is the default scale
choice to perform a standard comparison with the ME+PS predictions, unless
otherwise stated in the text.

In order to obtain the results for the various jet sizes in Higgs+jet and
$Z$+jet ($R=0.3$, $0.4$, $0.5$, $0.6$, $0.7$, and $1.0$) required in this
study, we have exploited the fact that the Born-level kinematics for all
processes is insensitive to $R$. As a result, the difference between two cone
sizes can be obtained from a calculation of the $H$+2 jet and $Z$+2 jet process
at NLO accuracy: 

\begin{equation} 
\sigma^\mathrm{NNLO}_{H(Z)+j}(R) -\sigma^\mathrm{NNLO}_{H(Z)+j}(R') = \int \bigl[\mathrm{d}\sigma^\mathrm{NLO}_{H(Z)+2j}(R) -\mathrm{d}\sigma^\mathrm{NLO}_{H(Z)+2j}(R') \bigr]_{N_\mathrm{jets} \geq 1} .
\label{eq:SM_Higgs_jet_R:multi-run} 
\end{equation} 

Note that the difference has
to be taken at the level of the integrand, since one term acts as a local
counter-term of the other in all IR-divergent limits where a jet becomes
unresolved and the $H$+2 jet / $Z$+2 jet configuration degenerates to $H$+jet /
$Z$+jet. Using Eq.~\eqref{eq:SM_Higgs_jet_R:multi-run}, predictions for
different $R$ values can be obtained from a single NNLO computation.

\subsection{Setup for Sherpa} We use a pre-release version of the Sherpa Monte
Carlo event generator~\cite{Gleisberg:2003xi,Gleisberg:2008ta}, based on
version Sherpa-2.2.4. The NLO matching is performed in the S-MC@NLO
approach~\cite{Hoeche:2011fd,Hoeche:2012ft}. We use a modified version of a
parton shower algorithm~\cite{Schumann:2007mg}, which is based on
Catani-Seymour dipole subtraction~\cite{Catani:1996vz,Catani:2002hc}. We use a
running coupling consistent with the PDF, and employ the CMW scheme to include
the two-loop cusp anomalous dimension in the parton-shower
simulation~\cite{Catani:1990rr}. To make the result comparable to the
fixed-order predictions we set the renormalization and factorization scales
to\footnote{Note the slight difference compared to
Eq.~\eqref{eq:SM_Higgs_jet_R:htprimescale}, where the sum runs over partonic
jets, as opposed to all partons. Using the same scale would lead to slightly
better agreement between the two predictions.} 

\begin{eqnarray} 
\mu_0 &=&\frac{H_T}{2} = \frac{1}{2}\bigg(\sqrt{m_X^2+p_{T,X}^2} +\sum_{\rm partons} p_T\bigg)\;\qquad\textrm{for Higgs/Z+jet production},\label{eq:parthtprimescale}\\ 
\mu_0 &=& {H}_T = \sum_{\rm partons} p_T\;\qquad\textrm{for inclusive jet production}\label{eq:parthtprimescale_jet}, 
\end{eqnarray} 

and we set the
resummation scale to $(p_{T,H/Z}+\sum_{\rm partons} p_{T})/2$ for Higgs/Z+jet
production and to $\sum_{\rm partons} p_{T}/2$ for inclusive jet production.
There will be some small differences with respect to NNLO for Higgs/Z+jet
production due to the summing over partons rather than jets. 

\subsection{Setup for Herwig} We used
Herwig 7~\cite{Bahr:2008pv,Bellm:2015jjp,Platzer:2011bc,Bellm:2017bvx} based on version 7.1.4
and ThePEG version 2.1.4 with minor changes to standard Herwig 7 scale settings
to match Eq.~\eqref{eq:parthtprimescale}. The NLO matching was performed with
matrix elements from OpenLoops~\cite{Cascioli:2011va} and
MadGraph~\cite{Alwall:2011uj} interfaced using the BLHA2
standard~\cite{Alioli:2013nda}. For parton distributions the PDF interface from
LHAPDF6~\cite{Buckley:2014ana} was used. In the results we show matched
$\mathrm{NLO} \oplus \mathrm{PS}$ predictions with the $\tilde{Q}$-shower~\cite{Gieseke:2003rz};
using lower statistics it was confirmed that merging according
to~\cite{Platzer:2012bs,Bellm:2017ktr} and matching to the Herwig 7 dipole
shower~\cite{Platzer:2009jq} display similar behaviour. For parton level
comparisions, hadronization and MPI models were switched off and the $\alpha_S$
of the hard process was synchronized with the PDF set. We include the effects
of the CMW scheme~\cite{Catani:1990rr} by an enhanced shower $\alpha_S=0.124$.
The scale used for the core process in the matching is defined as in
Eq.~\eqref{eq:parthtprimescale}, and the resummation scale was set to the
transverse momentum of the hardest jet.

\subsection{Setup for POWHEG BOX} Inclusive jet production was simulated using
POWHEG BOX (v2)~\cite{Nason:2004rx,Frixione:2007vw,Alioli:2010xd}, using the
implementation described in Ref.~\cite{Alioli:2010xa}. The hard matrix elements
entering the $\bar{B}$ function have been evaluated using the scale choice in
Eq.~\eqref{eq:parthtprimescale_jet}. In order to match the setup of Sherpa and
Herwig7, we used the options {\tt btlscalereal 1} and {\tt btlscalect 1},
thereby computing the real matrix elements using values of $\mu_R$ and $\mu_F$
obtained using the corresponding phase space kinematics, rather than the
underlying Born one. The running of the strong coupling is consistent with the
PDF choice, and the CMW scheme is used in the POWHEG Sudakov form factor. The
partonic events were then showered using Pythia8 (version
8.230)~\cite{Sjostrand:2014zea}, using the default tune and hence the default
PDF choice for the showering stage. For this study, vetoing in parton showering
has been achieved using the settings {\tt "SpaceShower:pTmaxMatch = 1"} and
{\tt "TimeShower:pTmaxMatch = 1"}.

\section{Jet shapes} \label{sec:shapes} 
\begin{figure}
\includegraphics[width=\textwidth]{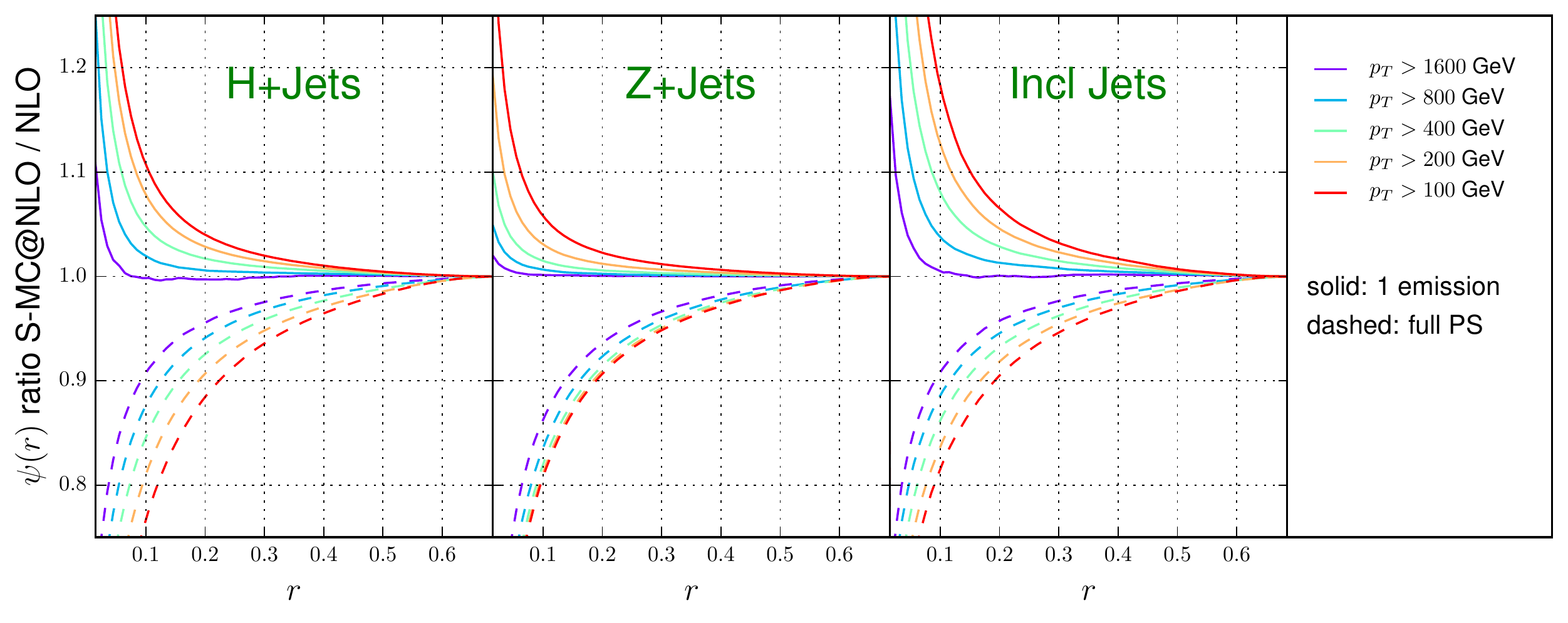} 
\caption{Jet shapes for Higgs and $Z$ plus jets and inclusive jets}
\label{fig:jet_shapes_Higgs_Z_jets} 
\end{figure} 
In Fig.~\ref{fig:jet_shapes_Higgs_Z_jets}, we investigate the difference between
the fixed-order NLO and NLO matched predictions for integrated jet shapes for
Higgs+jets, $Z$+jets and dijet production. The integrated jet shape is defined
as 
\begin{equation} 
\Psi(r) = \frac{1}{N^\mathrm{jet}}\sum\limits_{\mathrm{jets}} \frac{p_\mathrm{T}(0,r)}{p_\mathrm{T}(0,R)}, 
\label{eq:jetshape} 
\end{equation}
with $r$ being the radius of a cone which is concentric to the jet axis and
$p_\mathrm{T}(r_1,r_2)$ being the magnitude of the scalar sum of transverse
momenta in the annulus between radius $r_1$ and $r_2$. 
We also compare to a parton-shower matched prediction, where the number of
final-state partons generated in the simulation is limited to at most two. This
simulation presents the closest possible approximation to the fixed-order NLO
result that we are able to generate using the matching algorithms. It reflects
the kinematical restrictions of the NLO calculation (i.e.\ that only up to one
additional final-state parton can be present), but it also includes additional
approximate higher-order virtual corrections by means of Sudakov factors.
Nevertheless, we observe that the full NLO result and the truncated matched
result approach each other well within the jet cone, and the convergence is
naturally faster for larger jet transverse momenta. Note that the truncated
matched result approaches the full NLO result from above, which indicates that
the NLO calculation predicts less radiation close to the center of the jet.
This is explained by the following effect: The radiation probability in the
fixed-order calculation diverges as $r\to 0$, while it smoothly approaches zero
in the parton-shower matched result, due to Sudakov suppression. If the
parton-shower approximation to the real-emission cross section is good, this
implies that at any given value of $r$, the fixed-order prediction for the
differential jet shape will be larger than the parton-shower result, and
conversely, that the fixed-order prediction for the integrated jet shape will
be smaller than the parton-shower result. This effect is somewhat reduced by
the different scale choice in the two calculations, but it can still be
observed in Fig.~\ref{fig:jet_shapes_Higgs_Z_jets}.

The good agreement between the truncated matched prediction and the fixed-order
calculation strongly suggests that the differences between the fixed-order
predictions and matched results in
Fig.~\ref{fig:SM_Higgs_jet_R:ps_vs_fo_rpt_higgs} below are due to
higher-multiplicity final states. The discrepancies at small and large $R$
should therefore be reduced for higher-order perturbative calculations,
especially at NNLO. The jet shape for the $Z$+jets process is noticeably
narrower. We attribute this to the lead jet accompanying the $Z$-boson being
predominantly a quark jet, which is more collimated than a gluon jet due to its
reduced color charge.

\section{K-Factors and R-dependence at fixed order} 
\label{sec:kfactors}

\begin{figure} 
\includegraphics[width=\textwidth]{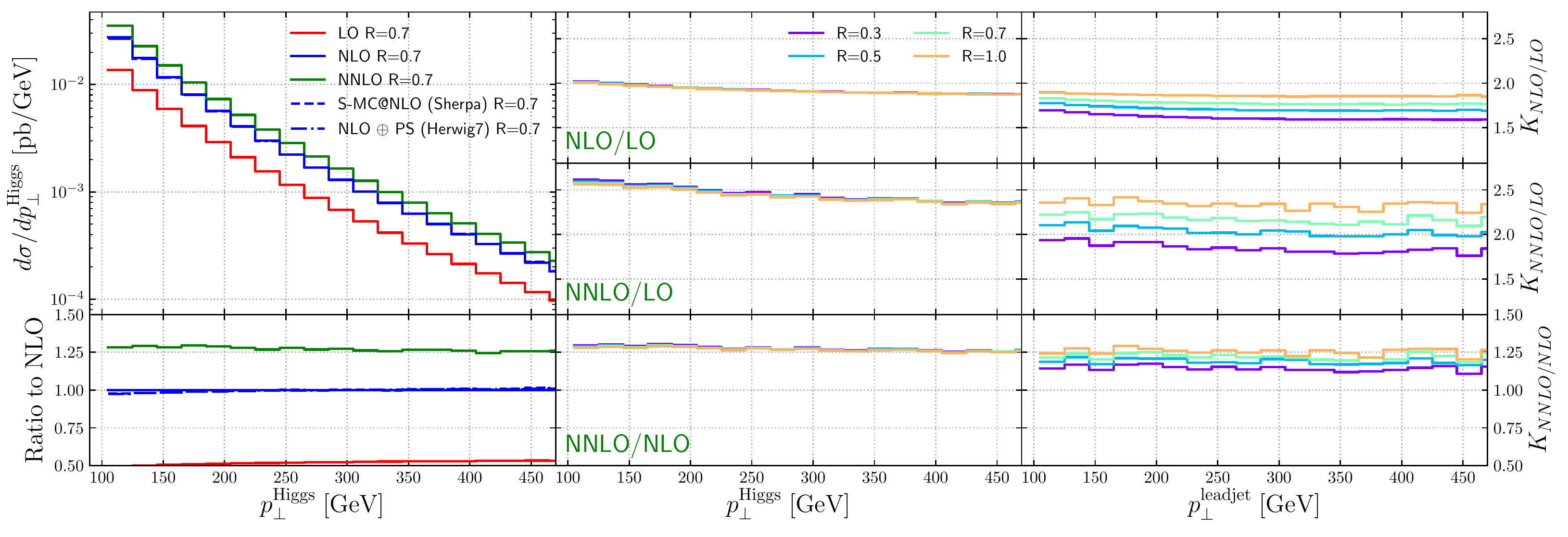}\\
\includegraphics[width=\textwidth]{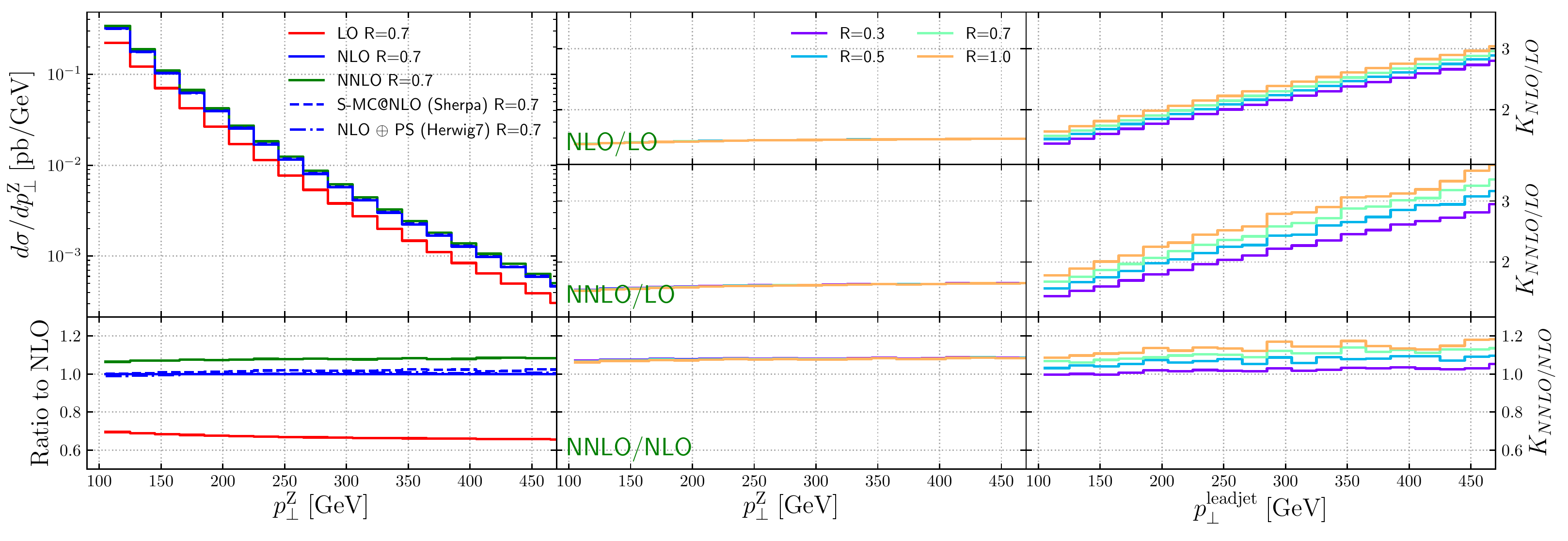}\\
\includegraphics[width=\textwidth]{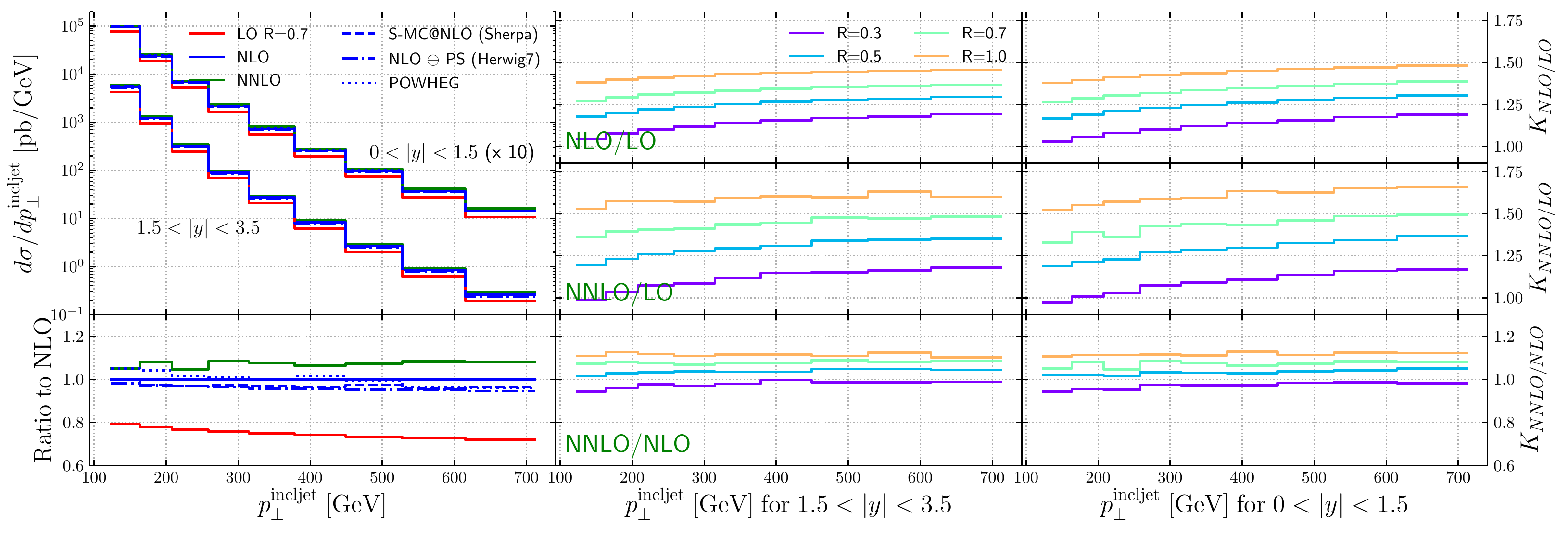} 
\caption{$K$-factors for Higgs plus jets (top), $Z$ plus jets (middle) 
and inclusive jet production (bottom).} 
\label{fig:K-factors} 
\end{figure}

Fig.~\ref{fig:K-factors} top (middle) shows (left) the transverse momentum
spectrum of the Higgs boson ($Z$-boson) as predicted by the fixed-order LO, NLO
and NNLO calculation, as well as the results from an NLO matched computation
using the Sherpa event generator and the NLO-matched 
Herwig result. The NLO, Sherpa and Herwig results are all in very good agreement 
with each other over the entire range of the plot ($\ge$ 100 GeV). 
The NNLO normalizations are larger due to the higher order effects included in 
these calculations. 

Fig.~\ref{fig:K-factors} top (middle) shows (center) the $K$-factors (NLO/LO,
NNLO/LO, NNLO/NLO), from NNLOJET as a function of the Higgs boson ($Z$-boson) $p_T$, 
for different jet radii; as expected 
there is no jet size dependence for this variable in the plotted region. 
Also shown (right) are the
local $K$-factors as a function of the lead jet transverse momentum for the two
processes, for various jet sizes. The $K$-factors for $H+\ge1$ production are
relatively flat as a function of jet $p_T$. The $K$-factors (NLO/LO and
NNLO/LO) for $Z+\ge1$ production grow rapidly with jet $p_T$, due to the
increasing dominance of dijet production, followed by a $Z$-boson emission.
The $K$-factors (NNLO/NLO) are relatively flat, indicating that there are no
substantial new subprocesses being added at NNLO. 
\newpage
Fig.~\ref{fig:K-factors} (bottom left) shows the inclusive jet transverse
momentum spectrum as predicted by the fixed-order LO, NLO and NNLO
calculations, for an $R$-value of 0.7, as well as the results from an NLO
matched computation using the 
Sherpa event generator and the NLO-matched 
Herwig result.
In addition, a prediction from 
Powheg is included as well. The NLO, 
Sherpa, Herwig and Powheg results are all in very good agreement with each
other over the range of the plot ($\ge$ 100 GeV), i.e. there is no significant
$\it parton$ $\it shower$ $\it systematic$ and the predictions with parton
showers reflect the underlying fixed-order NLO results. The NNLO normalizations
are larger due to the higher order effects included in these calculations.
$K$-factors (NLO/LO, NNLO/LO, NNLO/NLO, from 
NNLOJET are shown as a function of jet size, and as a function of the inclusive
jet $p_T$, for two different rapidity intervals. Again, the $K$-factors grow
with increasing jet size, and also have a slight slope (NLO/LO, NNLO/LO) as a
function of the jet transverse momentum. 

\begin{figure} 
\includegraphics[width=\textwidth]{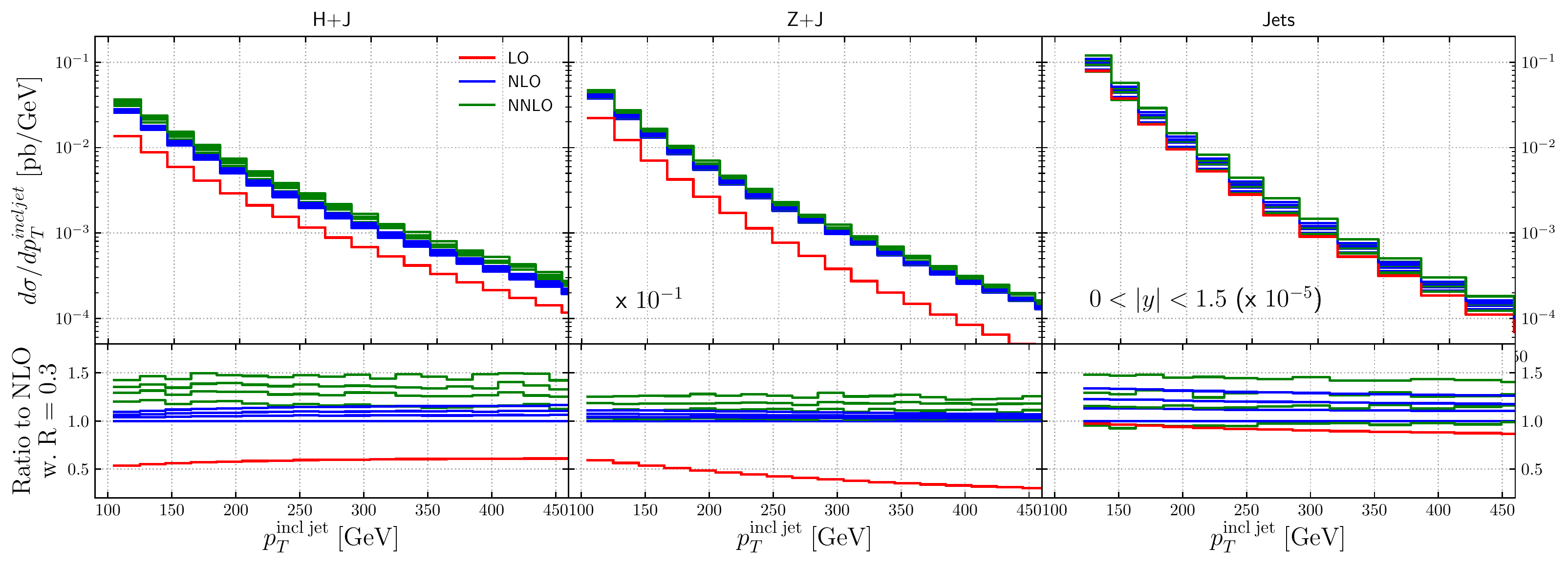}
\caption{The cross sections for $H +\ge1$ jet, $Z +\ge1$ jet, and dijet
production from 
NNLOJET, as a function of the inclusive jet $p_T$ at LO, NLO and NNLO. To
illustrate the spread induced on the cross section, representative values of
$R\in[0.3,0.5,0.7,1.0]$ are shown. } 
\label{fig:Comparison_plot} 
\end{figure}

The cross sections for $H +\ge1$ jet, $Z +\ge1$ jet, and dijet production from 
NNLOJET are shown in Fig.~\ref{fig:Comparison_plot}, as a function of the
inclusive jet $p_T$ at LO, NLO and NNLO. The figure shows representative values
for $R\in[0.3,0.5,0.7,1.0]$ to illustrate the spread induced on the cross
section. 
It is interesting to note that for $H +\ge1$ jet production, the $R$-dependence
is larger at NNLO than at NLO. The $R$-dependence for $Z +\ge1$ jet production
is relatively small both at NLO and NNLO. For dijet production, the
$R$-dependence is relatively large at both NLO and NNLO. The larger
$R$-dependence for $H +\ge1$ jet production at NNLO than at NLO can be traced
back to the large radiative corrections to the signal at NLO. A significant
part of the large higher-order corrections in inclusive $Z$- and Higgs-boson
production originates in the well-known ratio of the Sudakov form factor
between the timelike and the spacelike region~\cite{Magnea:1990zb}. This ratio,
given by
$\left|\Gamma_a(Q^2)/\Gamma_a(-Q^2)\right|^2=1+\alpha_s(Q^2)/(2\pi)\,C_a\pi^2+\mathcal{O}(\alpha_s^2)$
is enhanced by the color charges of the partons annihilating into the
electroweak boson and is therefore larger in Higgs- than in $Z$-boson
production. While the analysis is more complicated in processes with an
additional final-state jet, universal terms of the same form are present there,
which might explain the larger NNLO / NLO $K$-factors in the case of
Higgs-boson plus jet production~\cite{Ahrens:2008qu}.

\section{Results: fixed order vs. parton level Monte Carlo}
\label{sec:results} 

\begin{figure}[t]
\centerline{\includegraphics[width=\textwidth]{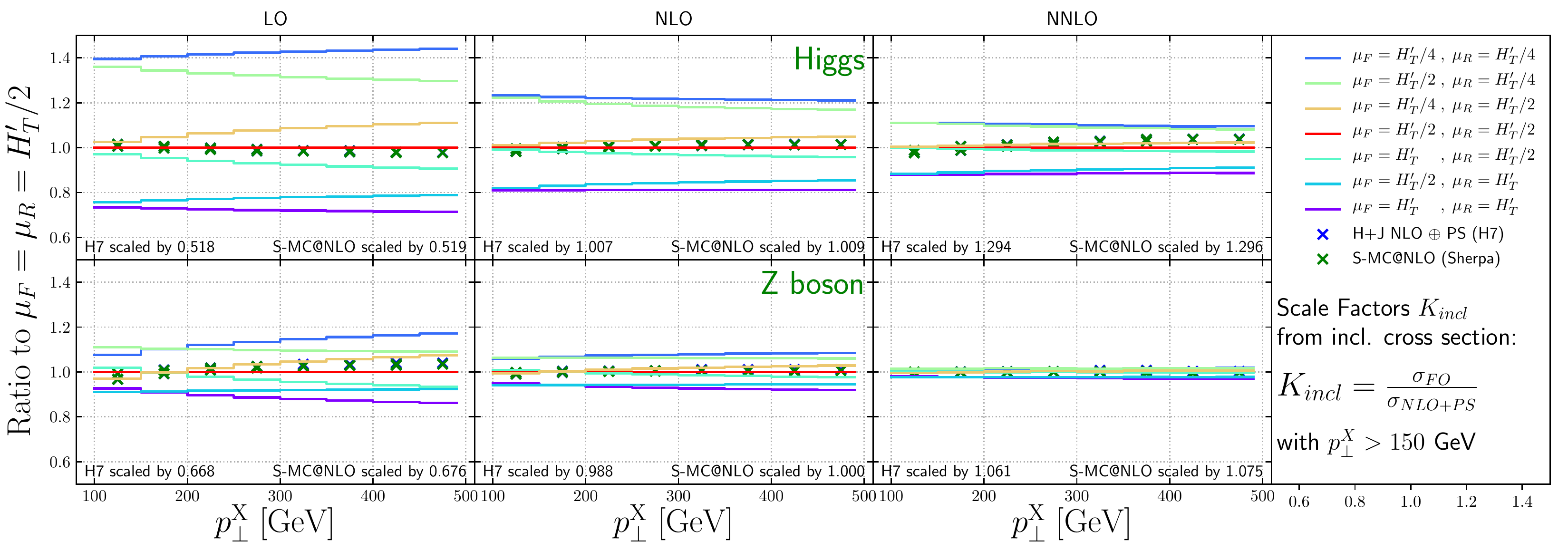}}
\caption{The scale variations at LO, NLO, and NNLO from 
NNLOJET for Higgs and $Z$ boson production, as a function of the boson
transverse momentum, are shown. For comparison, the nominal NLO NLO+PS
predictions are also shown. The generator predictions are scaled with the
inclusive $K_{incl}$ factor with Higgs(Z) $p_\perp > 150$~GeV, see
Fig.~\ref{fig:K-factors}. } 
\label{fig:SM_Higgs_jet_R:ScaleVariationsFOHZ}
\end{figure} 
Fig.~\ref{fig:SM_Higgs_jet_R:ScaleVariationsFOHZ} shows the scale
dependence of the differential cross section as a function of the Higgs or
$Z$-boson transverse momentum. The NLO-matched parton shower predictions have
been scaled by $K$-factors derived from the constraint that the inclusive cross
section for heavy boson transverse momenta above $p_\perp > 150$~GeV match the
fixed-order result. The reduction of the scale dependence in the transition
from LO to NNLO is striking. It is also encouraging that the scaled NLO-matched
parton shower calculations agree very well with the fixed-order results over
the entire range in transverse momentum. This implies that the Monte-Carlo
generators can be utilized to reliably predict the heavy boson transverse
momentum spectra in boosted Higgs and $Z$-boson analyses.

If the renormalization and factorization scales are defined using partonic
variables, we expect a very mild dependence of the Higgs/$Z$ transverse
momentum spectrum on both the jet $p_T$ cut and the jet radius in the plotted
region. At leading order QCD, this transverse momentum is compensated by a
single hard jet. The collinear evolution of the jet is governed by the DGLAP
equations, which prefer highly asymmetric branchings of the jet into softer
sub-jets. Only if this evolution reaches the extremely unlikely final-state
configuration with all jets below the $p_T$ threshold or outside the rapidity
region covered by the detector, the event can be lost and the cross section can
be changed. Due to the restricted final-state multiplicity, the probability for
this is even more reduced at fixed order. In fact, without any jet rapidity
cuts, the cross section could not be modified up to N${}^5$LO, where the
opportunity arises for the first time to have all partons forming individual
jets at $p_T=150\;{\rm GeV}/6<30\;{\rm GeV}$, albeit in a very small phase
space.

\begin{figure}[t]
\centerline{\includegraphics[width=\textwidth]{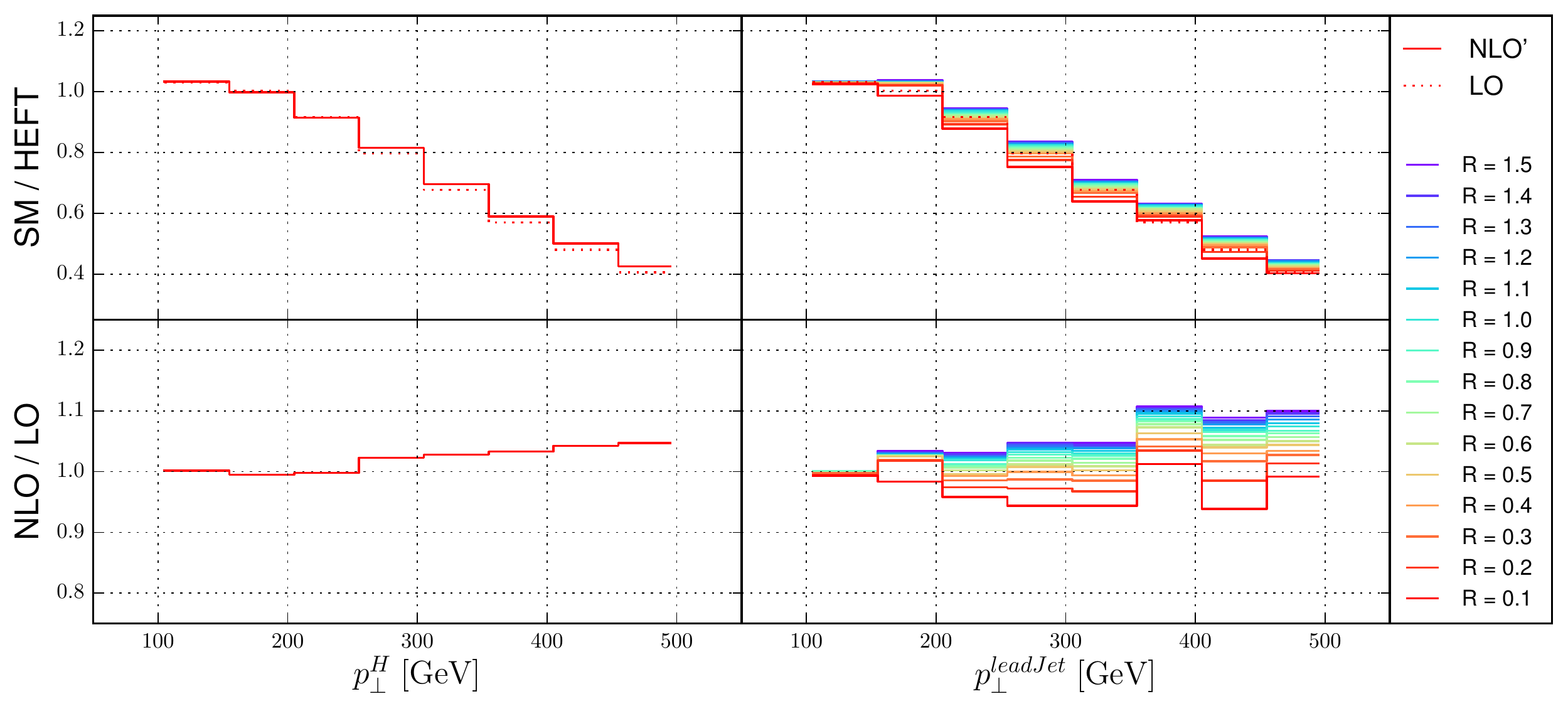}}
\caption{The ratio between results computed in HEFT and the full Standard Model
for the transverse momentum spectrum of the Higgs boson (left) and the leading
jet (right) in Higgs plus jet events. Results labeled NLO' are derived using
the approximation of~\cite{Buschmann:2014sia}.} 
\label{fig:SMvsHEFT}
\end{figure}

In the context of Higgs boson measurements at high transverse momentum, the
difference between predictions computed in the Higgs effective theory (HEFT)
and the full Standard Model, (including dependence on the top-quark mass),
becomes important. The full Standard Model features a significantly steeper
transverse momentum spectrum as well as different scale uncertainties. A
complete calculation at NNLO precision in the full Standard Model is currenly
out of reach. One can, however, assume that top-quark mass effects factorize
from higher-order QCD corrections, such that they can be treated independently.
In Fig.~\ref{fig:SMvsHEFT} we test this hypothesis, both for the Higgs boson
and leading jet transverse momentum spectrum. The results labeled NLO' are
derived using an approximate virtual correction~\cite{Buschmann:2014sia}. This
approximation is motivated by the good agreement between such approximations
and the full NLO result observed in~\cite{Jones:2018hbb}. We note that the
ratio between the full SM and the HEFT result behaves very similar at LO and at
NLO', both as a function of the Higgs and the leading jet transverse momentum.
Note that it has a jet radius dependence as a function of the leading jet
transverse momentum. This originates in the different $p_T$ dependence of the
cross section in the real-emission and Born kinematics at NLO.

\begin{figure}[t]
\centerline{\includegraphics[width=\textwidth]{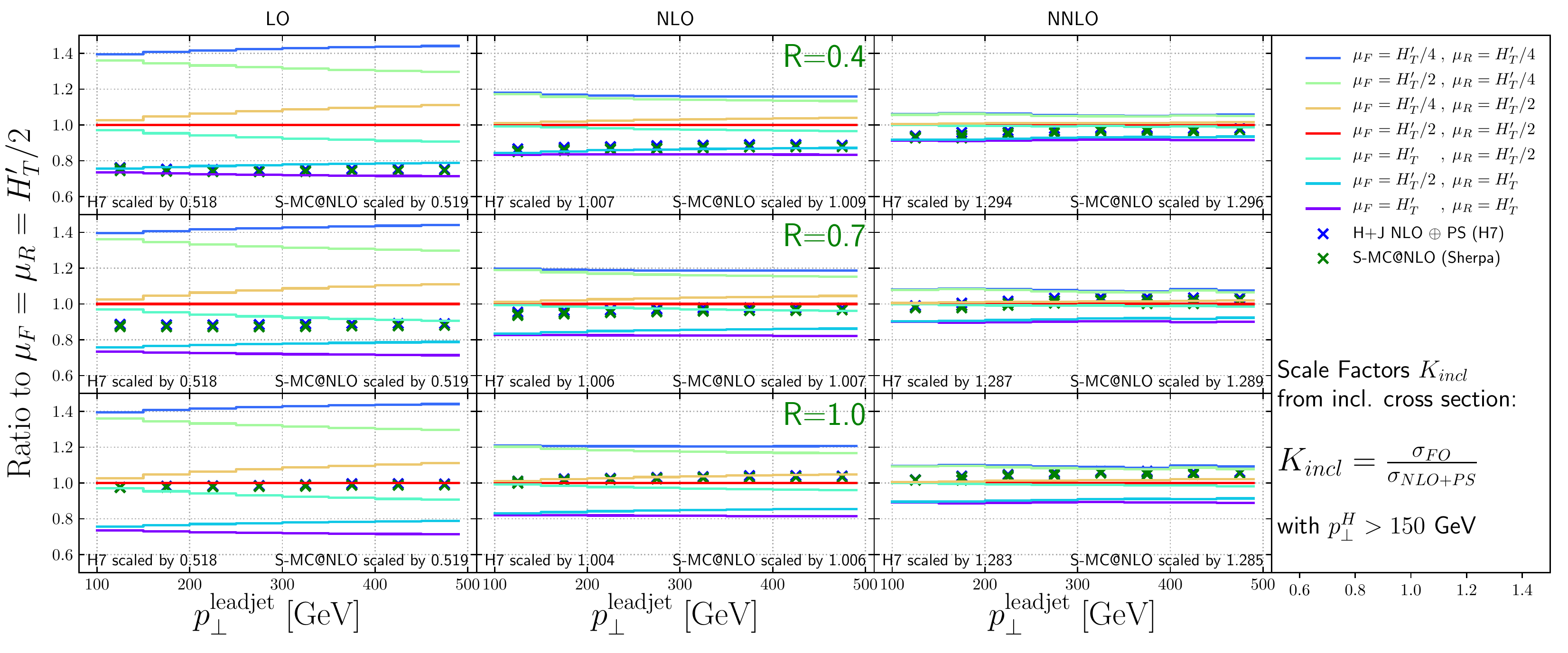}}
\centerline{\includegraphics[width=\textwidth]{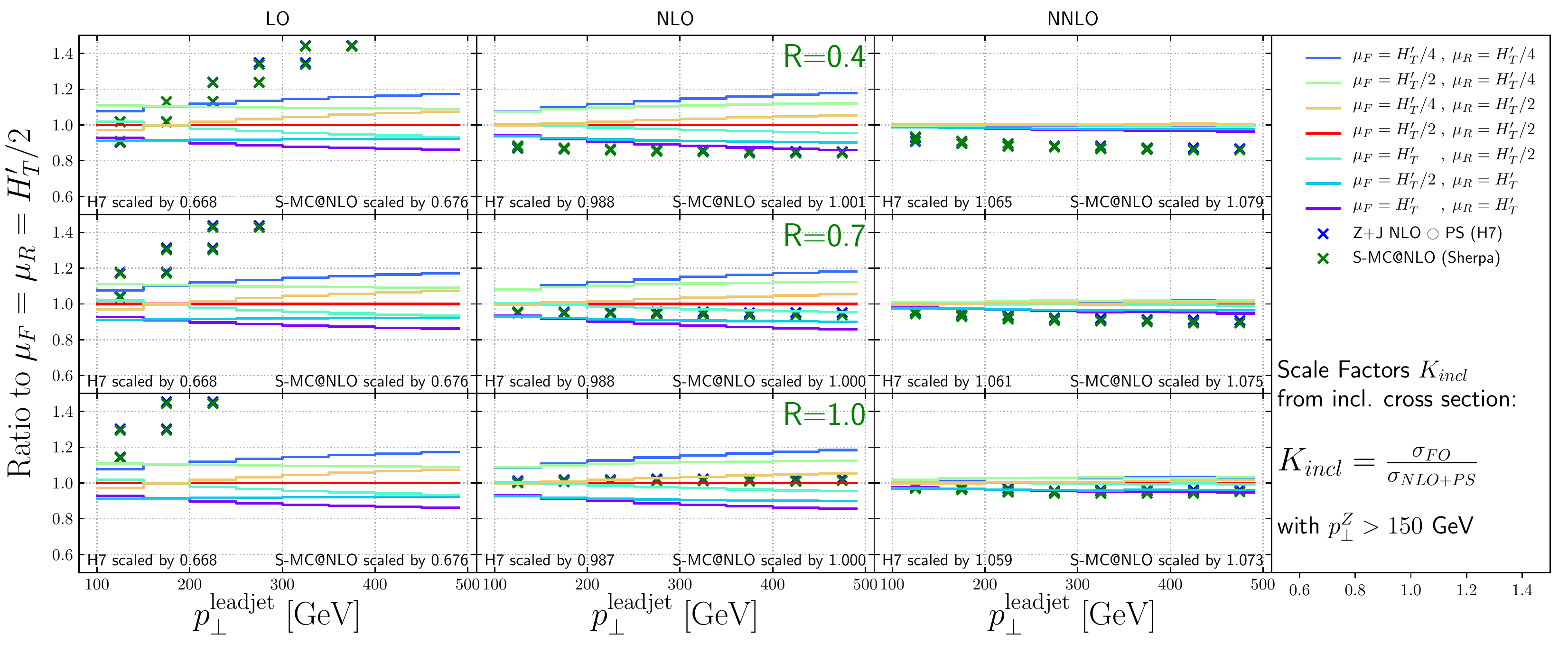}}
\caption{The scale variations at LO, NLO, and NNLO from 
NNLOJET for 3 jet sizes, as a function of leading jet transverse momentum, are
shown. For comparison, the nominal NLO NLO+PS predictions are also shown. The
generator predictions are scaled with the inclusive $K_{incl}$ factor with
Higgs(Z) $p_\perp > 150$~GeV, see Fig.~\ref{fig:K-factors}. }
\label{fig:SM_Higgs_jet_R:ScaleVariationsFO} 
\end{figure}

Fig.~\ref{fig:SM_Higgs_jet_R:ScaleVariationsFO} top (bottom) panel shows the
cross section scale variations at LO, NLO and NNLO for $H(Z) +\ge1$ jet
production, as a function of the leading jet transverse momentum, for various
jet sizes. The uncertainty bands are given by the highest and lowest cross
section predictions at each order. As expected, the uncertainties on the cross
sections decrease from LO to NLO to NNLO. They also decrease (slightly) as the
jet size decreases, perhaps not surprising given that larger jet radii lead to
inclusion of more real radiative corrections. 

The scale uncertainties for $H +\ge1$ jet production are relatively constant as
a function of the lead jet transverse momentum. For $Z +\ge1$ jet production,
the NLO scale uncertainties increase as the lead jet transverse momentum
increases. This can be understood as the effect of new kinematical channels,
which correspond to the radiation of a soft $Z$ boson off a hard di-jet event.
Such configurations arise at leading order in the NLO result, and they become
more important as the lead jet transverse momentum increases. 

\begin{figure}[t]
\centerline{\includegraphics[width=\textwidth]{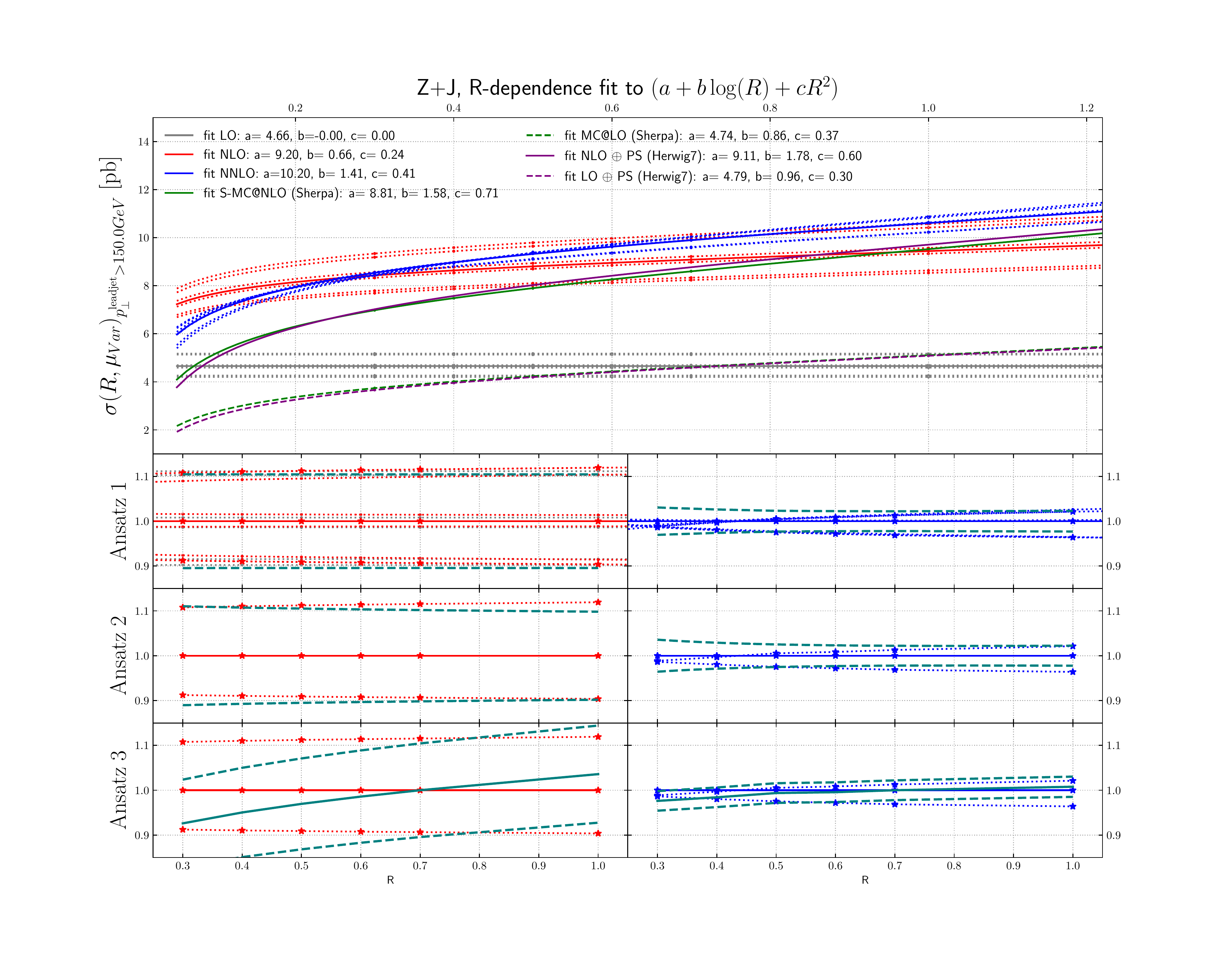}}
\caption{The $R$-dependence of the cross sections at NLO, NNLO and NLO+PS are
shown, for particular scale values, as a function of the jet radius,for $Z
+\ge1$ jet production, for leading jet transverse momenta above 150~GeV.
\label{fig:AccidentalScaleComp_Z}} 
\end{figure}

Also shown for comparison are the predictions from the two NLO+PS calculations.
For the sake of shape comparison, we scale these predictions with inclusive
$K$-factors derived from the Higgs(Z) $p_T$ distribution above 150~GeV. This
has two reasons: First, as described earlier the generation cut on the jet
requires high enough boson $p_T$ cuts to mitigate simulation setup effects.
Second, in the parton shower simulation, we use the jet transverse momentum to
define the shower starting condition. If the Higgs(Z) mass is of similar size
this choice of scale would have an increased ambiguity, which could, however,
be eliminated by performing a multi-jet merged computation.
For $H +\ge1$ jet production, the two NLO+PS predictions agree very well with
each other and tend to be at the lower end of the scale uncertainty bands for
$R$= 0.4, at center of the scale uncertainty bands for $R$= 0.7 and slightly
above the center of the scale uncertainty bands for $R$= 1.0. For $Z +\ge1$ jet
production, the NLO+PS predictions again are in agreement with each other, but
rapidly increase over the LO results as the lead jet $p_T$ increases (again due
to the impact of the dijet contribution, arising only at NLO or above). At NLO,
a similar behavior with respect to the NLO scale uncertainty band is observed
as was seen for $H +\ge1$ jet. At NNLO, the NLO+PS predictions are close to the
scale uncertainty bands, which are extremely small, especially for $R$= 0.4.
This will be discussed further in the context of
Fig.~\ref{fig:AccidentalScaleComp_Z} and in Sec.~\ref{sec:uncertainties}.

\begin{figure}[t]
\centerline{\includegraphics[width=\textwidth]{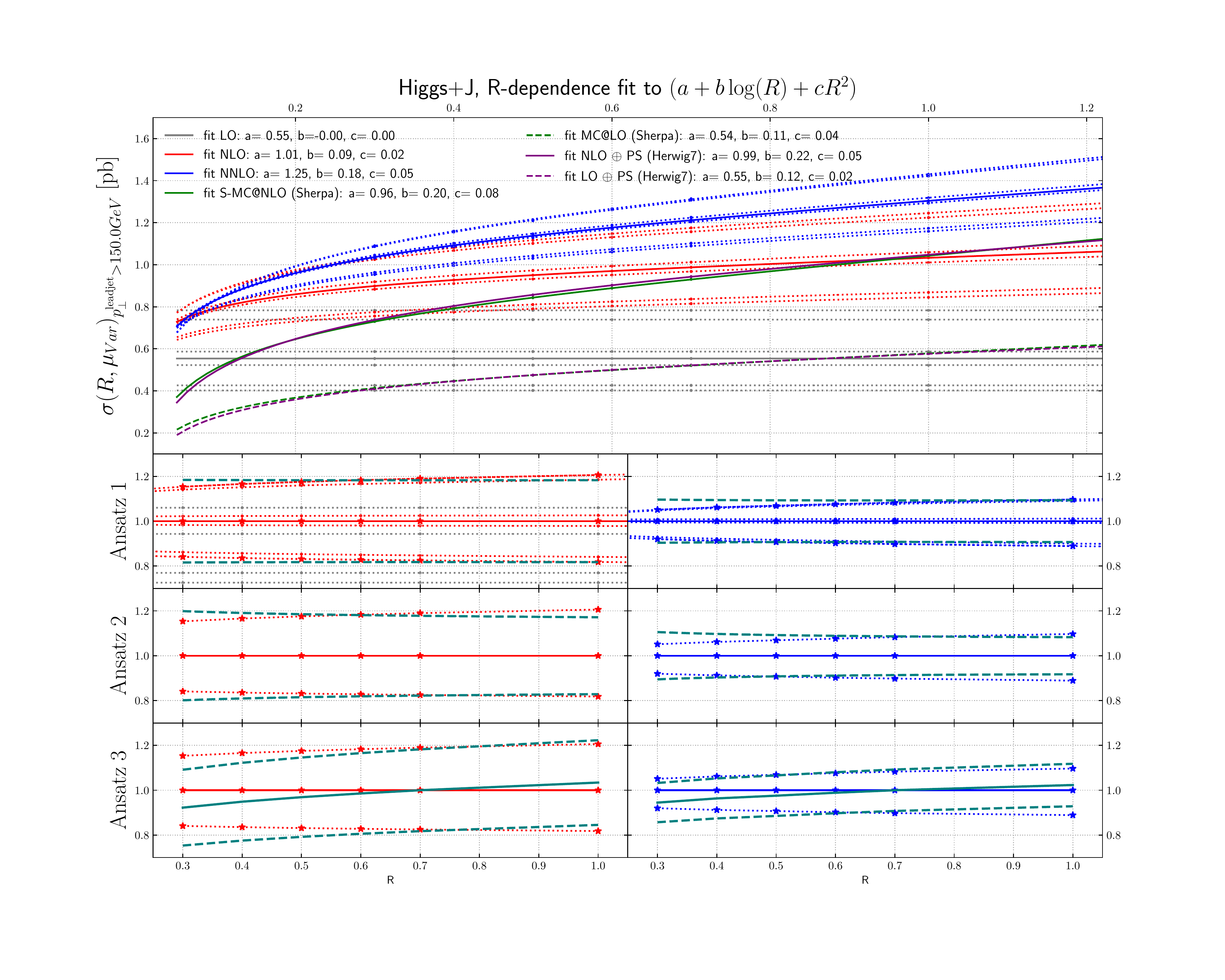}}
\caption{The $R$-dependence of the cross sections at NLO, NNLO and NLO+PS are
shown, for particular scale values, as a function of the jet radius, for $H
+\ge1$ jet production, for leading jet transverse momenta above 150~GeV.
\label{fig:AccidentalScaleComp_H}} 
\end{figure}

\begin{figure}[t]
\centerline{\includegraphics[width=\textwidth]{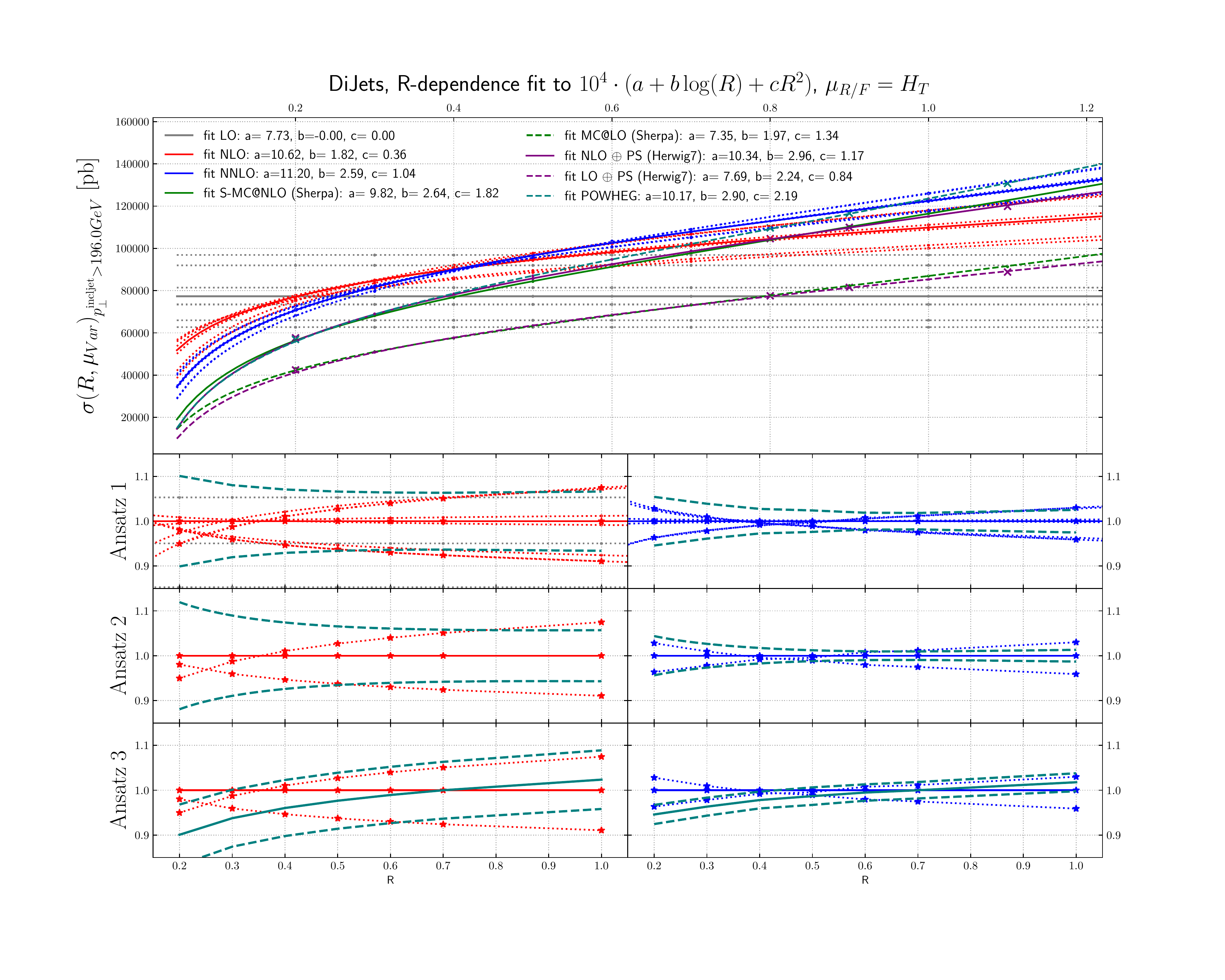}}
\caption{The $R$-dependence of the cross sections for inclusive jet production
at LO, NLO, NNLO and NLO+PS are shown, for scale variations around a central
scale of $H_T$, as a function of jet radius, for dijet production, for leading
jet transverse momenta above 196 GeV. 
\label{fig:AccidentalScaleComp_dijet}}
\end{figure}

\begin{figure}[t]
\centerline{\includegraphics[width=\textwidth]{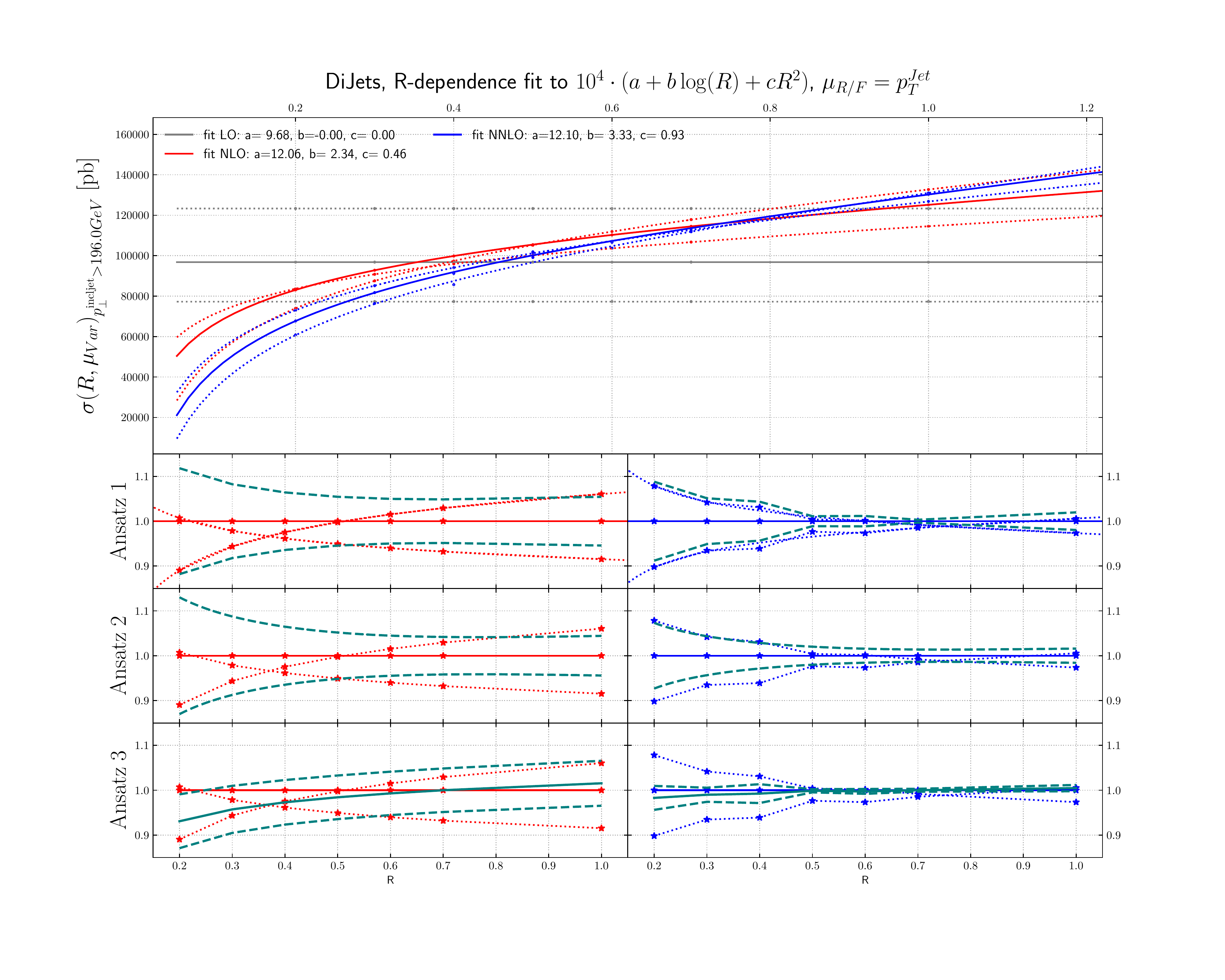}}
\caption{The $R$-dependence of the cross sections for inclusive jet production
at LO, NLO, NNLO and NLO+PS are shown, for scale variations around a central
scale of $p_T^\mathrm{jet}$, as a function of jet radius, for dijet production,
for leading jet transverse momenta above 196 GeV.
\label{fig:AccidentalScaleComp_dijet2}} 
\end{figure}

\begin{figure}[t]
\centerline{\includegraphics[width=\textwidth]{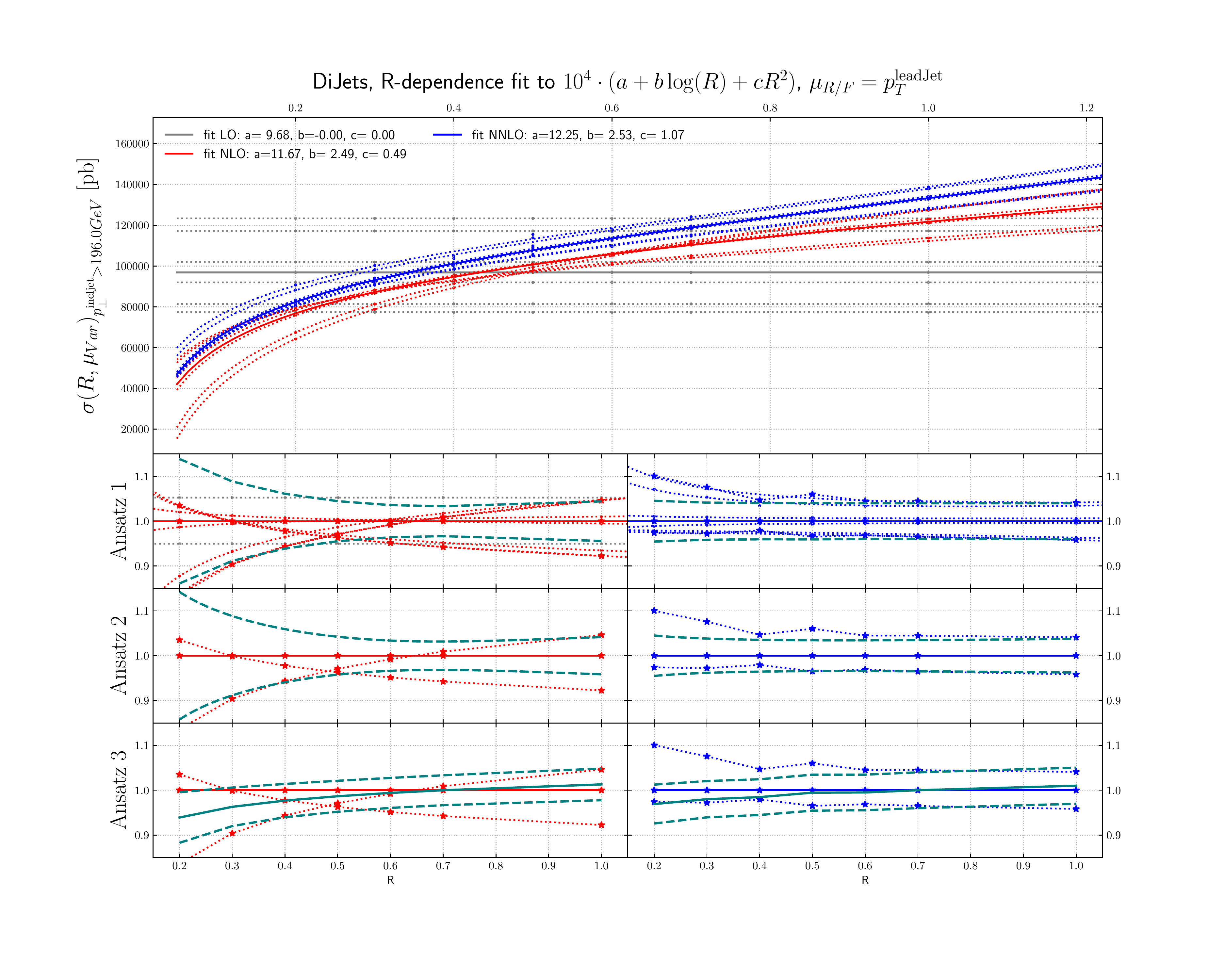}}
\caption{The $R$-dependence of the cross sections for inclusive jet production
at LO, NLO, NNLO and NLO+PS are shown, for scale variations around a central
scale of $\mu_{R/F}=p^{\mathrm{lead\;jet}}_T$, as a function of jet radius, for
dijet production, for leading jet transverse momenta above 196 GeV.
\label{fig:AccidentalScaleComp_dijet3}} 
\end{figure}

Fig.~\ref{fig:AccidentalScaleComp_H} shows the leading jet $p_T$ cross sections
for $H +\ge1$ jet production for the different scale choices, at LO, NLO and
NNLO, as a function of the jet size $R$. In this case, a minimum transverse
momentum requirement of 150~GeV has been placed on the leading jet. We assume
this scale to be large enough to replace $M_H$ as the largest scale in the
process, see discussion of Fig. \ref{fig:SM_Higgs_jet_R:ScaleVariationsFO}.
The dots for each scale choice have been fit to a functional form motivated by
the expected behavior for jet cross sections. We assume the leading functional
form~\cite{Ellis:1992qq}: 
\begin{equation} 
f(R)=a+b\log(R)+c R^2
\label{eq:SM_Higgs_jet_R:fit} 
\end{equation} 
which is motivated by the
logarithmic behaviour scaling of the cross section with the jet size $R$ and an
area-dependent contribution from initial-state radiation. The lines in
Fig.~\ref{fig:AccidentalScaleComp_H} are then interpolations with
Eq.~\eqref{eq:SM_Higgs_jet_R:fit} and the fitted values. 
Again, the scale variation band is given by the upper and lower curves at each
order. It is notable that the scale uncertainty bands shrink as the jet size
decreases, as mentioned earlier. For very low values of $R$, this improvement
in the uncertainty can be regarded at least partially due to accidental
cancellations that stem from the restrictions in phase space.
It can also be observed that for each particular scale, the slope is greater at
NNLO than at NLO. The NLO+PS predictions are also plotted in the figure, and
can be observed to have a greater slope than even the NNLO predictions. This
can be seen as an effect of either including (at large $R$) or not excluding
(at small $R$) additional semi-hard real emissions, which have a leading-order
scale dependence and therefore induce a large change in the cross section. The
ratio panels of Fig.~\ref{fig:AccidentalScaleComp_H} to
Fig.~\ref{fig:AccidentalScaleComp_dijet3} will be discussed in
Sec.~\ref{sec:uncertainties} in the context of improved scale uncertainties. 

Fig.~\ref{fig:AccidentalScaleComp_Z} shows the leading jet $p_T$ cross sections
for $Z +\ge1$ jet production for the different scale choices, at LO, NLO and
NNLO, as a function of the jet size $R$, and again with a minimum transverse
momentum requirement of 150~GeV placed on the leading jet. The behavior at NLO
is similar to what was observed for Higgs + jet. As for Higgs +jet, there is a
large decrease of the scale uncertainty at NNLO at all $R$ values. In fact, the
scale uncertainty decreases to zero at $R$=0.3, emphasizing the accidental
cancellations noted for Higgs+jet. This may indicate that the especially small
scale uncertainties for $R$=0.4, as observed for example in
Fig.~\ref{fig:SM_Higgs_jet_R:ScaleVariationsFO}, may be underestimated.

Figures~\ref{fig:AccidentalScaleComp_dijet},
\ref{fig:AccidentalScaleComp_dijet2} and \ref{fig:AccidentalScaleComp_dijet3}
show the inclusive jet cross section from dijet production, again at LO, NLO
and NNLO, as a function of $R$, using scale variations around a central scales
of $H_T$, $p_T^\mathrm{jet}$ and $\mu_{R/F}=p^{\mathrm{lead\;jet}}_T$,
respectively. Here, the behavior is in some sense more extreme in that e.g. for
the scale choice $H_T$ the jet $R$ value for essentially zero scale uncertainty
is at $R$=0.4 which is one of the jet sizes that is commonly used at the
LHC\footnote{Note that this statement is dependent on the functional form used
for the scale.}. 

\begin{figure}[t]
\centerline{\includegraphics[width=\textwidth]{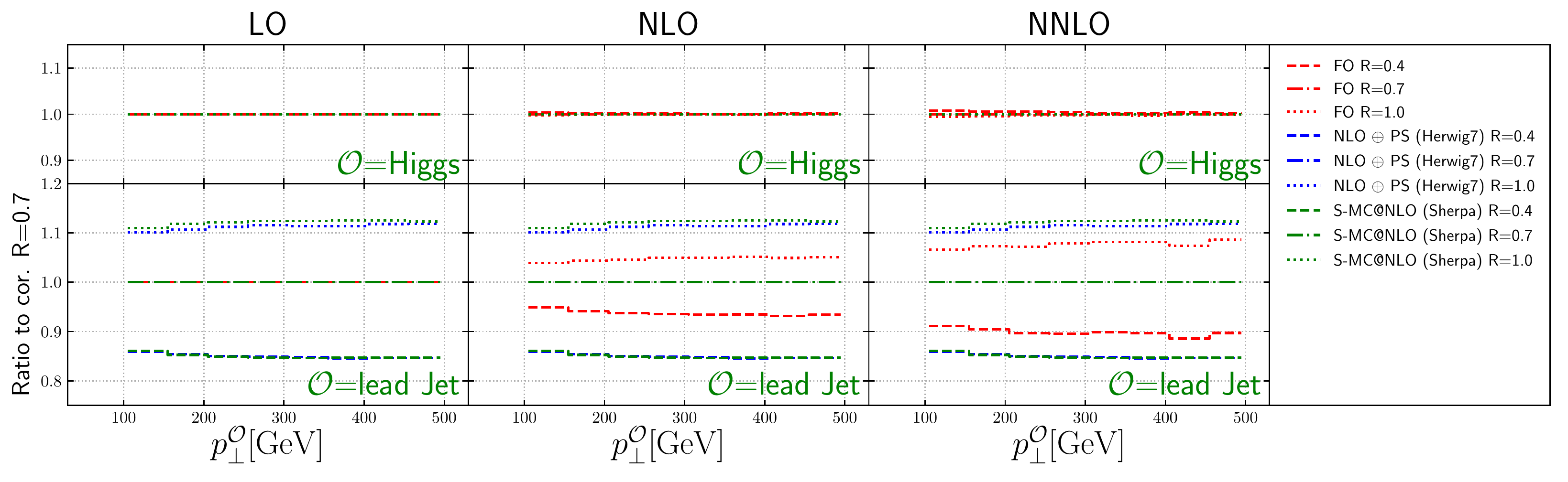}}
\centerline{\includegraphics[width=\textwidth]{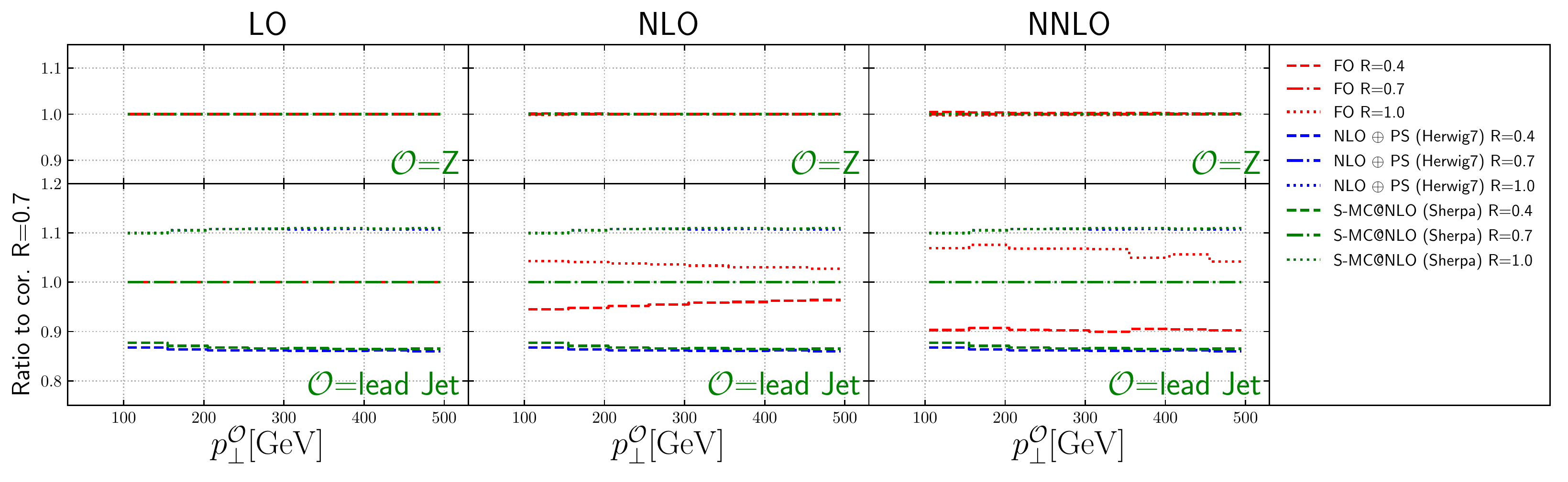}}
\caption{The ratio of each cross section (either Higgs($Z$) $p_T$ or lead jet
$p_T$) for specific jet sizes, scaled to the cross section for each prediction
for a jet size of $R=0.7$. \label{fig:SM_Higgs_jet_R:multiratios_vs_NLO_07}}
\end{figure}

Fig.~\ref{fig:SM_Higgs_jet_R:multiratios_vs_NLO_07} shows the cross sections
for the Higgs(Z) transverse momentum and leading jet transverse momentum for
several different jet sizes, at LO, NLO and NNLO (from \nnlojet) and from the
two NLO+PS predictions. All cross sections have been scaled to their respective
value for the reference jet size of $R=0.7$. Near this value we observe the
best agreement between fixed-order and NLO matched results, save for an overall
normalization which can be extracted from the Higgs($Z$) transverse momentum
spectrum, cf.\ Fig.~\ref{fig:K-factors}. 

\begin{figure}[t]
\centerline{\includegraphics[width=\textwidth]{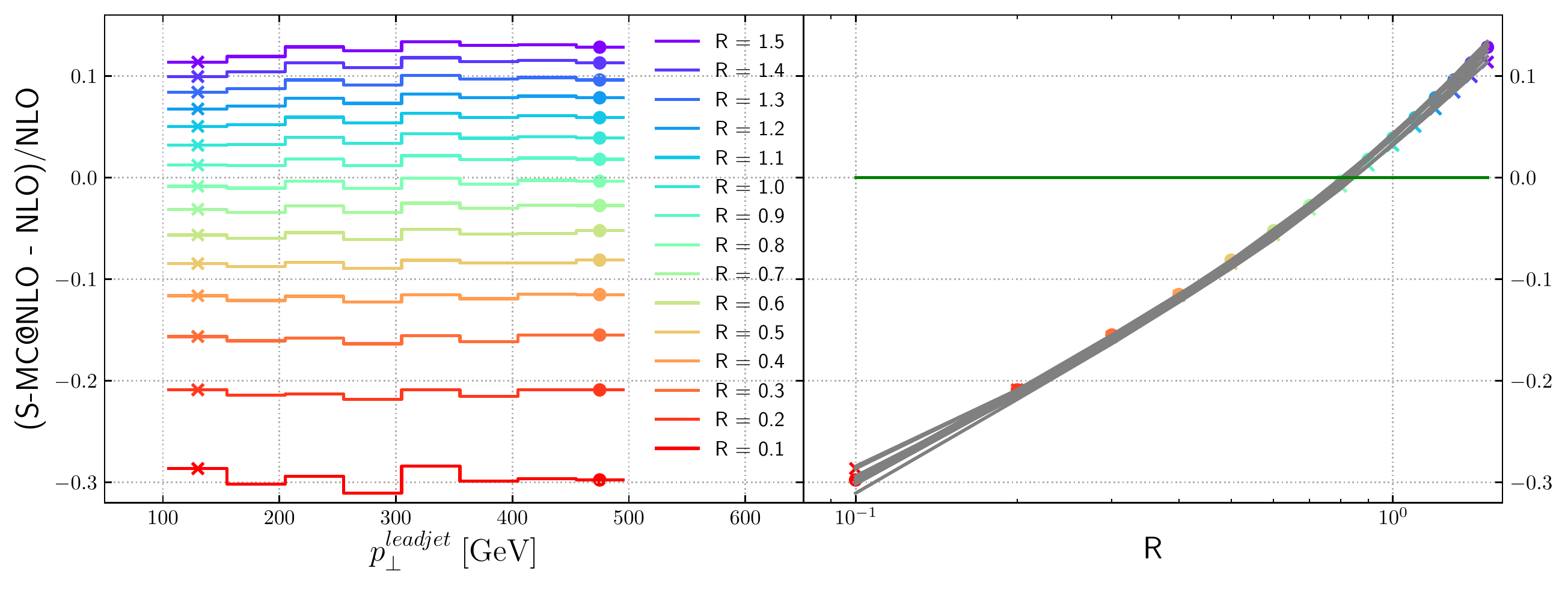}}
\caption{Relative difference between the NLO-matched prediction and the
fixed-order result as a function of the leading jet transverse momentum and the
jet radius, for $H+\ge1$ jet.\label{fig:SM_Higgs_jet_R:ps_vs_fo_rpt_higgs}}
\end{figure}

The absolute value of the difference between the fixed-order and the NLO
matched predictions away from $R=0.7$ increases roughly proportional to $\log
(R/0.7)$ (cf.\ Fig.~\ref{fig:SM_Higgs_jet_R:ps_vs_fo_rpt_higgs}), which is
expected due to the higher-multiplicity emissions included in the PS
simulations. Depending on kinematics they either enhance (at $R>0.7$) or reduce
(at $R<0.7$) the cross section. 

The differences between the NLO+PS predictions and those from \nnlojet decrease
as the order is raised from NLO to NNLO for both Higgs and $Z$-boson
production. The difference for Higgs boson production is of the order of 5-10\%
for $R=0.4$ at NLO and of the order of less than 5\% at NNLO, relatively flat
with $p_T$. For $Z$-boson production, the differences between the NLO+PS
predictions and those from \nnlojet at NLO slightly increase with increasing
$p_T$, and are relatively flat and small at NNLO.

\begin{figure}[t]
\centerline{\includegraphics[width=\textwidth]{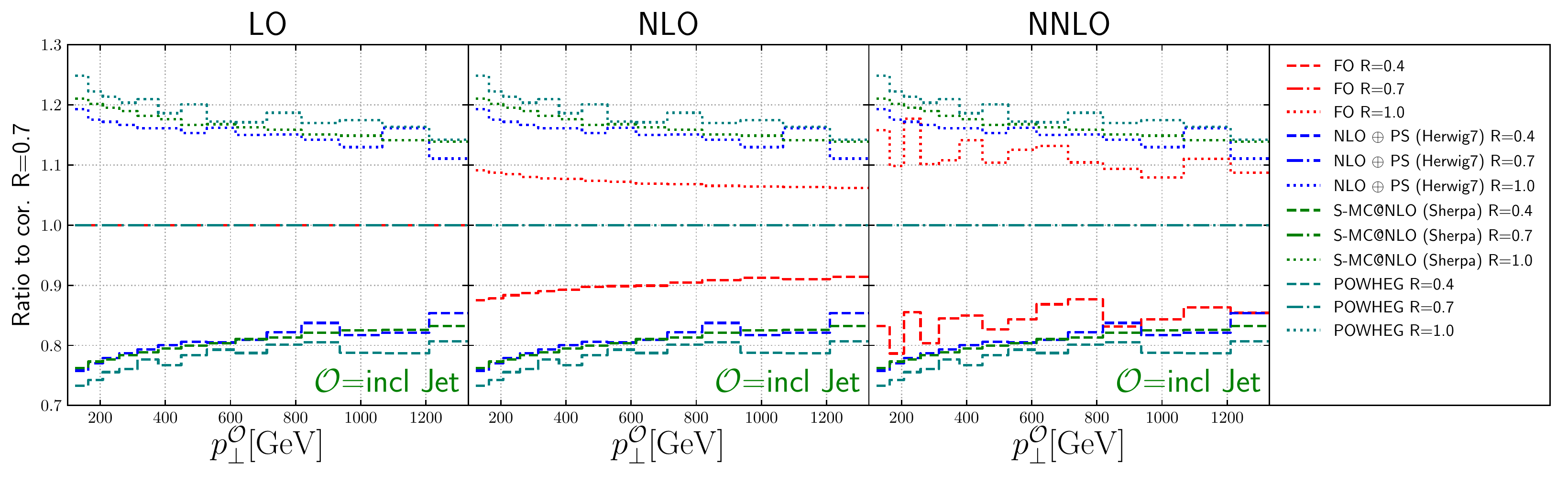}}
\caption{The ratio of the inclusive jet $p_T$ cross section for specific jet
sizes, scaled to the cross section for each prediction for a jet size of
$R=0.7$. \label{fig:SM_Higgs_jet_R:multiratios_vs_NLO_07_jet}} 
\end{figure}

Fig.~\ref{fig:SM_Higgs_jet_R:multiratios_vs_NLO_07_jet} shows the cross
sections for the inclusive jet $p_T$ distribution for several different jet
sizes, at LO, NLO and NNLO (from \nnlojet) and from the two NLO+PS predictions.
The ratios to $R$= 0.7 decrease as a function of increasing jet $p_T$ at all
orders. The differences between the three NLO+PS predictions and those from
\nnlojet are of the order of 10\% at NLO and of the order of 5\% or less at
NNLO. 

Given the better description of the jet shape provided by the NLO+PS
predictions, this is an indication of the theoretical uncertainty associated
with the truncation of the perturbative series. The uncertainty is reduced at
NNLO as expected. It is noteworthy that the ratios in
Fig.~\ref{fig:SM_Higgs_jet_R:multiratios_vs_NLO_07} are relatively flat as a
function of the transverse momenta.

Fig.~\ref{fig:SM_Higgs_jet_R:ps_vs_fo_rpt_higgs} shows the dependence of the
relative difference between a NLO-matched prediction from Sherpa and the NLO
fixed-order result for $H+\ge1$ jet production, as a function of the leading
jet transverse momentum for varying jet radii. The ratio is flat as a function
of the leading jet $p_T$. In Fig.~\ref{fig:AccidentalScaleComp_H} we compared
integrated cross sections, while here we observe interestingly a similar
behaviour for the differential cross sections. In the right plot, the
projection is with respect to the radius, and displays, in grey, the various
transverse momentum intervals and, in coloured, the lowest and highest
energies. Assuming the leading behaviour is given by
Eq.~\eqref{eq:SM_Higgs_jet_R:fit}, and with the flatness in the leading jet
transverse momentum, the linear, (but slightly quadratic) behaviour in the
logarithmic plot is expected. We note the zero crossing of the curve on the
right-hand side, which corresponds to the best agreement between fixed-order
and NLO-matched result, is located at $R \approx$0.8 (see the discussion of
Fig.~\ref{fig:SM_Higgs_jet_R:multiratios_vs_NLO_07}). In configurations where
the jet rapidity is zero, this corresponds to a roughly equal partitioning of
the rapidity phase-space into collinear sectors for color dipoles spanned
between the initial-state partons and the final-state jet, and thus to a
roughly equal partitioning of soft-enhanced radiation. This geometric argument
favours the commonly used $R=0.7$ with respect to smaller values when
experimental data is compared to fixed order calculations, although the precise
value will depend on the color structure of the process and on the parton
luminosity.

\begin{figure}[t]
\centerline{\includegraphics[width=\textwidth]{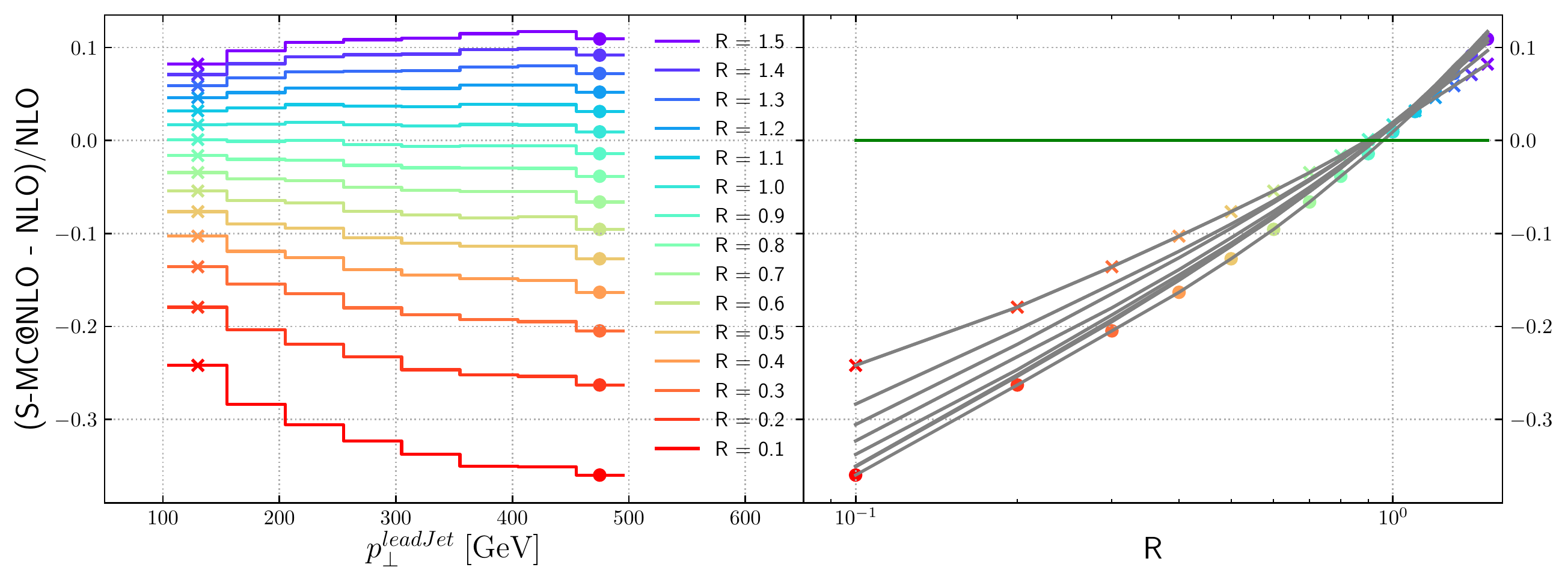}}
\caption{Relative difference between the NLO-matched prediction and the
fixed-order result as a function of the leading jet transverse momentum and the
jet radius, for $Z+\ge1$ jet.\label{fig:SM_Z_jet_R:ps_vs_fo_rpt_Z}}
\end{figure}

Fig.~\ref{fig:SM_Z_jet_R:ps_vs_fo_rpt_Z} shows the dependence of the relative
difference between a NLO-matched prediction from Sherpa and the NLO fixed-order
result, for $Z+\ge1$ jet production, as a function of the leading jet
transverse momentum for varying jet radii. In contrast to the Higgs boson case,
the distributions are not flat as a function of lead jet transverse momentum
for small-$R$ jets and for large-$R$ jets. Note also that the zero-crossing for
the curves on the right-hand side is closer to $R=0.9$ than to $R=0.8$. 

\begin{figure}[t]
\centerline{\includegraphics[width=\textwidth]{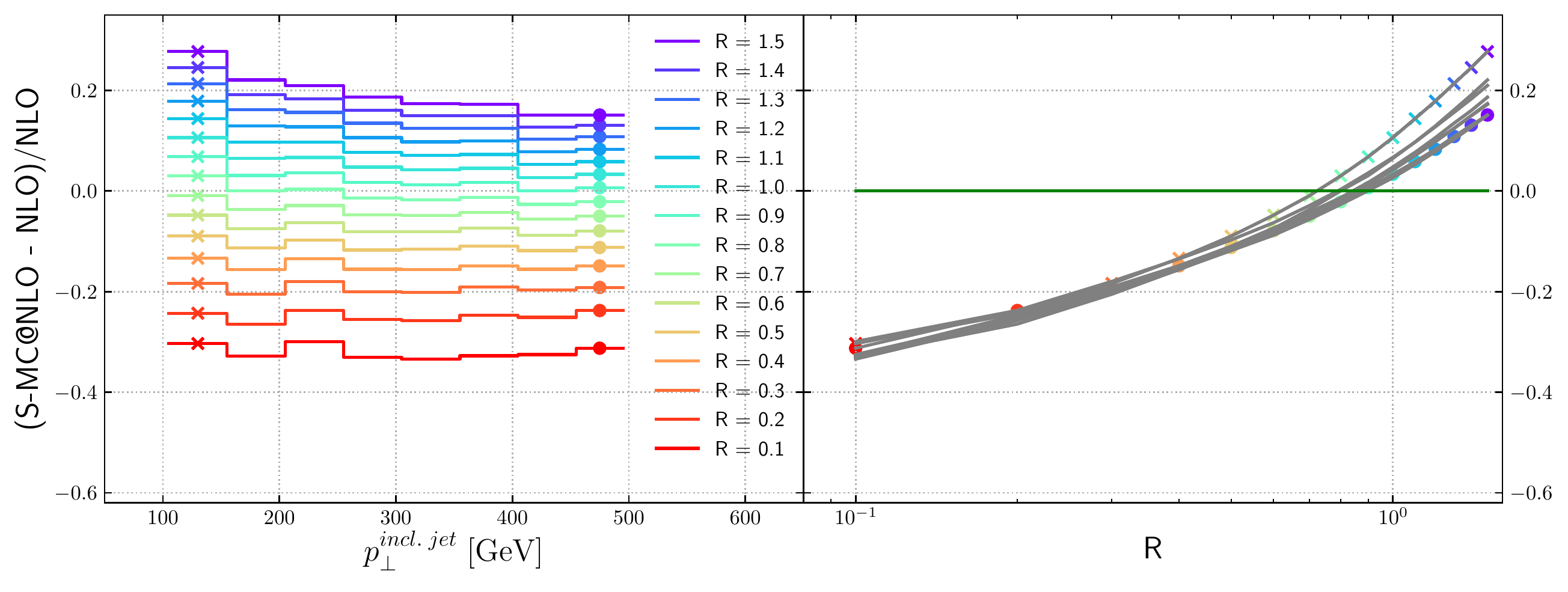}}
\caption{Relative difference between the NLO-matched prediction and the
fixed-order result as a function of the inclusive jet transverse momentum and
the jet radius, for dijet
production.\label{fig:SM_incl_jet_R:ps_vs_fo_rpt_incl_jet}} 
\end{figure}

Fig.~\ref{fig:SM_incl_jet_R:ps_vs_fo_rpt_incl_jet} shows the dependence of the
relative difference between a NLO-matched prediction from Sherpa and the NLO
fixed-order result, for the inclusive jet transverse momentum, for dijet
production, as a function of the inclusive jet transverse momentum for varying
jet radii. The curves are relatively flat as a function of inclusive jet
transverse momentum, for jet $R$ values less than 0.7, but fall more steeply
for larger $R$ values. Note that the zero-crossings for the curves on the
right-hand sides of the figures (for lead jet and inclusive jet) are around
$R=0.8$.

\begin{figure}[t]
\centerline{\includegraphics[width=\textwidth]{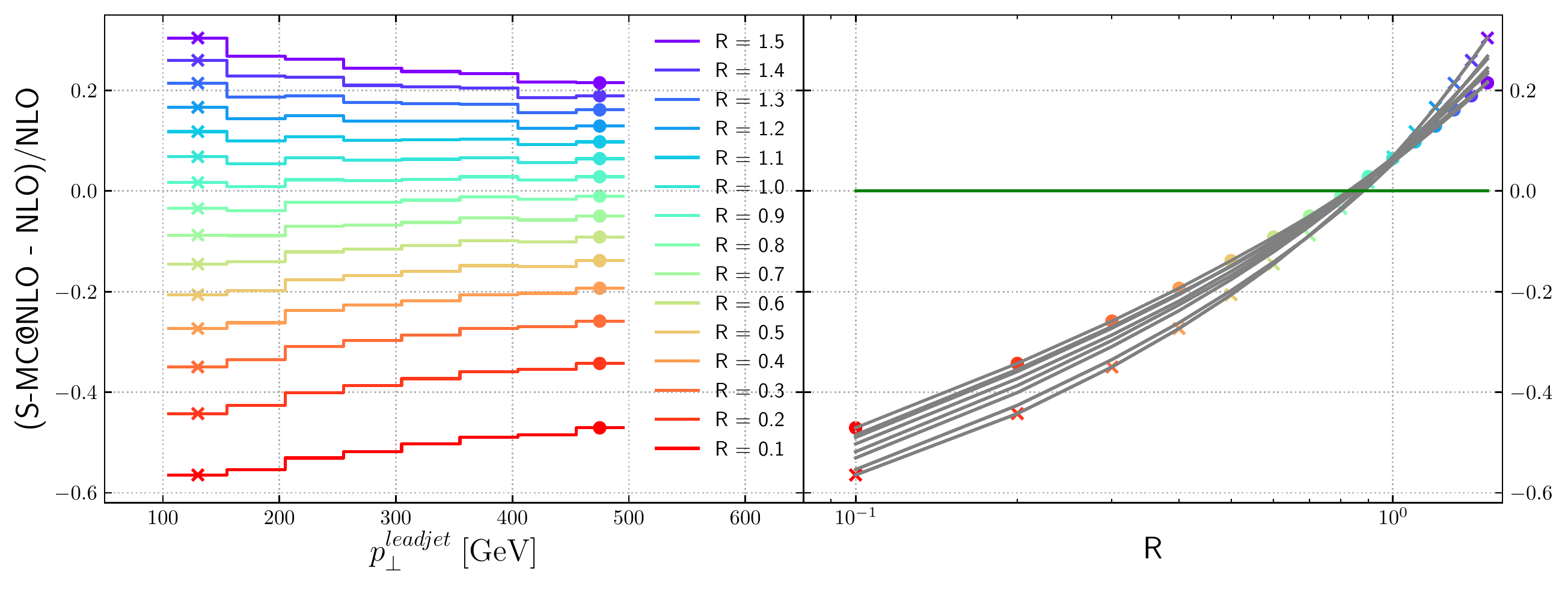}}
\caption{Relative difference between the NLO-matched prediction and the
fixed-order result as a function of the leading jet transverse momentum and the
jet radius, for dijet production.\label{fig:SM_incl_jet_R:ps_vs_fo_rpt_jet}}
\end{figure}

Fig.~\ref{fig:SM_incl_jet_R:ps_vs_fo_rpt_jet} shows the dependence of the
relative difference between a NLO-matched prediction from Sherpa and the NLO
fixed-order result, for the lead jet transverse momentum, for dijet production,
as a function of the leading jet transverse momentum for varying jet
radii~\footnote{Note that for this distribution the $R$-dependence is larger
than for any of the other distributions.}. The curves are relatively flat as a
function of lead jet transverse momentum, for jet $R$ values around 0.5, but
fall (rise) more steeply for larger (smaller) $R$ values. 

Comparing the Figs. \ref{fig:SM_Higgs_jet_R:ps_vs_fo_rpt_higgs}
-~\ref{fig:SM_incl_jet_R:ps_vs_fo_rpt_jet} we note the relative narrow
distribution of grey lines that sample the different $p_T$ bins in the case of
Higgs production. One might expect that this behavior is due to the Higgs
production process being gluon-initiated. However, the decomposition into
flavor channels shows that initial state quarks do play an important role and
that quark-gluon initiated processes start to dominate for high transverse
momenta. This diverse flavour composition of initial and final state does not
allow us to make a definite statement without further studies.

\section{Uncertainty estimates in processes with final-state jets}
\label{sec:uncertainties}

The reduction of scale uncertainties achievable at NNLO is remarkable.
However, the $R$-dependence of the uncertainty discussed in
Sec.~\ref{sec:results} indicates that some of the improvements may be due to
accidental cancellations. It is well known, that the scale variation for
exclusive cross sections is prone to the accidental compensation of
logarithmically enhanced higher-order corrections that appear both as a result
of scale variations and as a result of the phase-space restrictions. The very
definition of a final-state jet implies an exclusive measurement and
effectively acts as a veto on real-radiative corrections that fall outside the
jet area. This effect has been studied in different contexts
\cite{Dasgupta:2016bnd,Moch:2018hgy} An accurate assessment of the perturbative
uncertainties is important for inclusive jet production (and to a lesser extent
for $Z$+jet production), as the PDF fitting groups are working to incorporate
scale uncertainties in their analyses, and jet production serves as one of the
major constraints on the gluon distribution, especially at large $x$. The
impact is also especially important for smaller jet sizes ($R$=0.4), commonly
used for many measurements at the LHC, such as Higgs + jet production. The
accidental cancellations can also be an issue at NLO, but it is less
noticeable, given the larger intrinsic uncertainties at that order.

The \textit{ansatz} advocated in \cite{Dasgupta:2016bnd} is to view the
differential cross section as a combination of a fixed-order term and the
normalized all-orders resummed result. The two are then combined through
multiplicative matching, and their perturbative uncertainties are added in
quadrature. Upon re-expanding this result to fixed-order, one obtains the
$\mathrm{N}^n\mathrm{LO}$-mult prescription given in \cite{Dasgupta:2016bnd},
Eqs.~(3.5) and~(4.3). The result can be written as 
\begin{equation}
\sigma(R)=\sigma(R_0)\frac{\sigma(R)}{\sigma(R_0)}\approx
\sigma(R_0)\cdot\left(1+\alpha_S\, \partial_{\alpha_S}
\frac{\sigma(R)}{\sigma(R_0)}\Big|_{\alpha_S=0}+\alpha^2_S\,
\partial^2_{\alpha_S} \frac{\sigma(R)}{\sigma(R_0)}\Big|_{\alpha_S=0}\right)
\label{eq:expand}\;. 
\end{equation}

Clearly there are several possible choices in regards to the implementation of
the factorization of terms on the right-hand side of Eq.~\eqref{eq:expand}.
Results from the original proposal in \cite{Dasgupta:2016bnd} are shown in the
lower panels of
Figs.~\ref{fig:AccidentalScaleComp_H}-\ref{fig:AccidentalScaleComp_dijet2}. We
refer to this technique as ``Ansatz~3''. While the red and blue dotted lines in
Figs.~\ref{fig:AccidentalScaleComp_H}-\ref{fig:AccidentalScaleComp_dijet2}
correspond to typical scale variations the green dashed lines show the ratio of
``Ansatz~3'' (and other choices explained in the following) to the central
scale prediction.
The uncertainty of ``Ansatz~3'' has a more realistic-seeming value for all $R$,
but the central value of the prediction is modified at small $R$, in some cases
leading to the resultant uncertainty not encompassing the central value of the
original NNLO prediction. We therefore investigate two alternative approaches.
In the first (``Ansatz~1''), the ratio $\sigma(R)/\sigma(R_0)$ on the
right-hand side of Eq.~\eqref{eq:expand} is not expanded, and we combine the
uncertainties from the ratio and the seed cross section $\sigma(R_0)$ in
quadrature. The results of this procedure are shown in the top ratio panels of
Figs.~\ref{fig:AccidentalScaleComp_H}-\ref{fig:AccidentalScaleComp_dijet2}.
Our second alternative method (``Ansatz~2''), is based on the parametrization
of the cross section as a function of $R$ according to
Eq.~\eqref{eq:SM_Higgs_jet_R:fit}. We then determine the scale uncertainties of
the fit coefficients $a$, $b$ and $c$ and combine them in quadrature to arrive
at the full uncertainty. It can be seen in comparison between the top and
middle ratio panels of
Figs.~\ref{fig:AccidentalScaleComp_H}-\ref{fig:AccidentalScaleComp_dijet2}.
that Ansatz~1 and Ansatz~2 give similar results, and both preserve the central
value of the original NNLO fixed-order result. Although larger than the
original uncertainties, the perturbative scale variations determined in this
way are still smaller than the uncertainties observed at NLO, as would be
expected from a higher order calculation.

It is important to note that all of the aforementioned approaches of estimating
the theoretical uncertainty from missing higher-order corrections have an
intrinsic dependence on the arbitrary reference value $R_0$. By varying $R_0$
it is possible to create again a situation where the logarithmic corrections
due to higher-order effects and due to phase-space restrictions compensate each
other and the scale uncertainty is reduced to nearly zero. Based on the
analysis in Secs.~\ref{sec:shapes}-\ref{sec:results} we advocate to fix the
reference radius $R_0$ by comparing the higher-order result to a parton-shower
matched calculation and choose the reference point where the two
(approximately) agree. As discussed in Sec.~\ref{sec:results}, this corresponds
to selecting a reference radius where large logarithmic higher-order
corrections are minimized. Here we choose $R_0=0.7$ for all uncertainty
ans\"atze.

\section{Hadronization corrections and uncertainties} 
\label{sec:hadcorr} 

In this section we examine the non-perturbative corrections on the predictions
presented before. We determine the hadronization uncertainties by taking the
difference between NLO matched and hadronized results from Sherpa, using either
the cluster fragmentation model as implemented in Sherpa~\cite{Winter:2003tt}
or an interface to the Lund string fragmentation model as implemented in
Pythia~\cite{Sjostrand:2006za}.

\begin{figure}[p]
\centerline{\includegraphics[width=\textwidth]{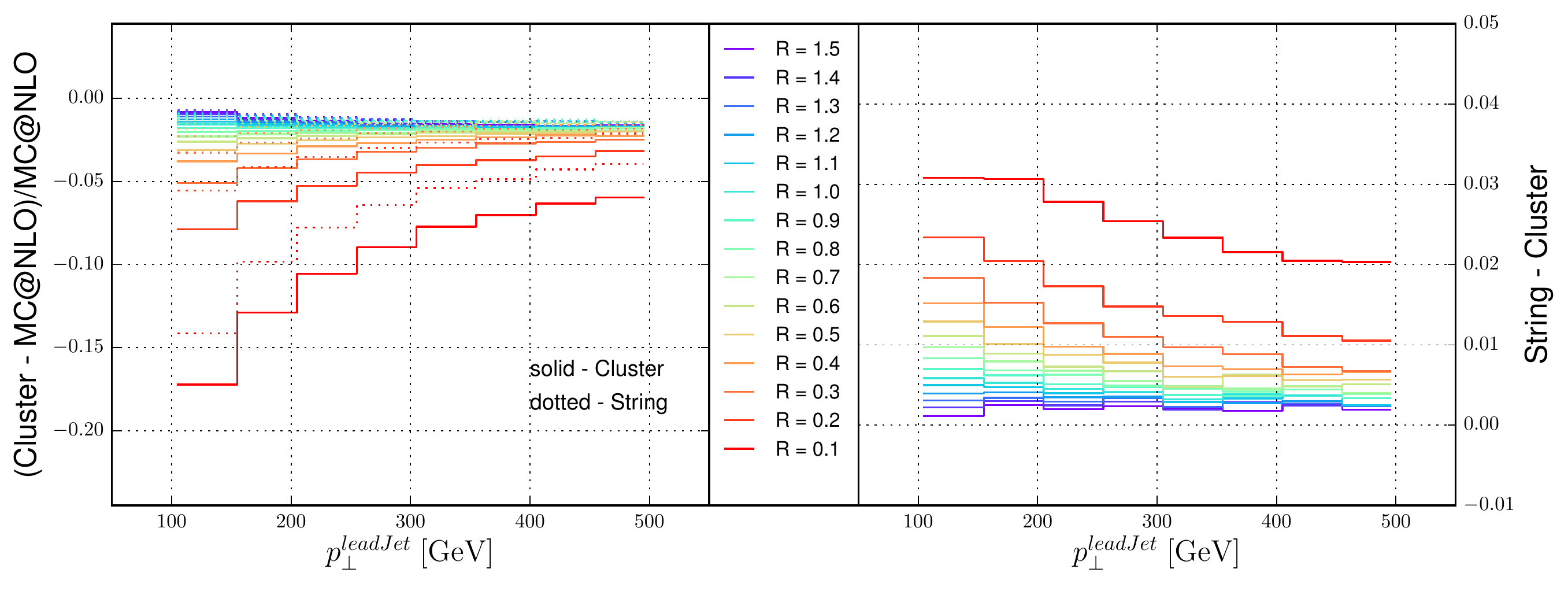}}
\centerline{\includegraphics[width=\textwidth]{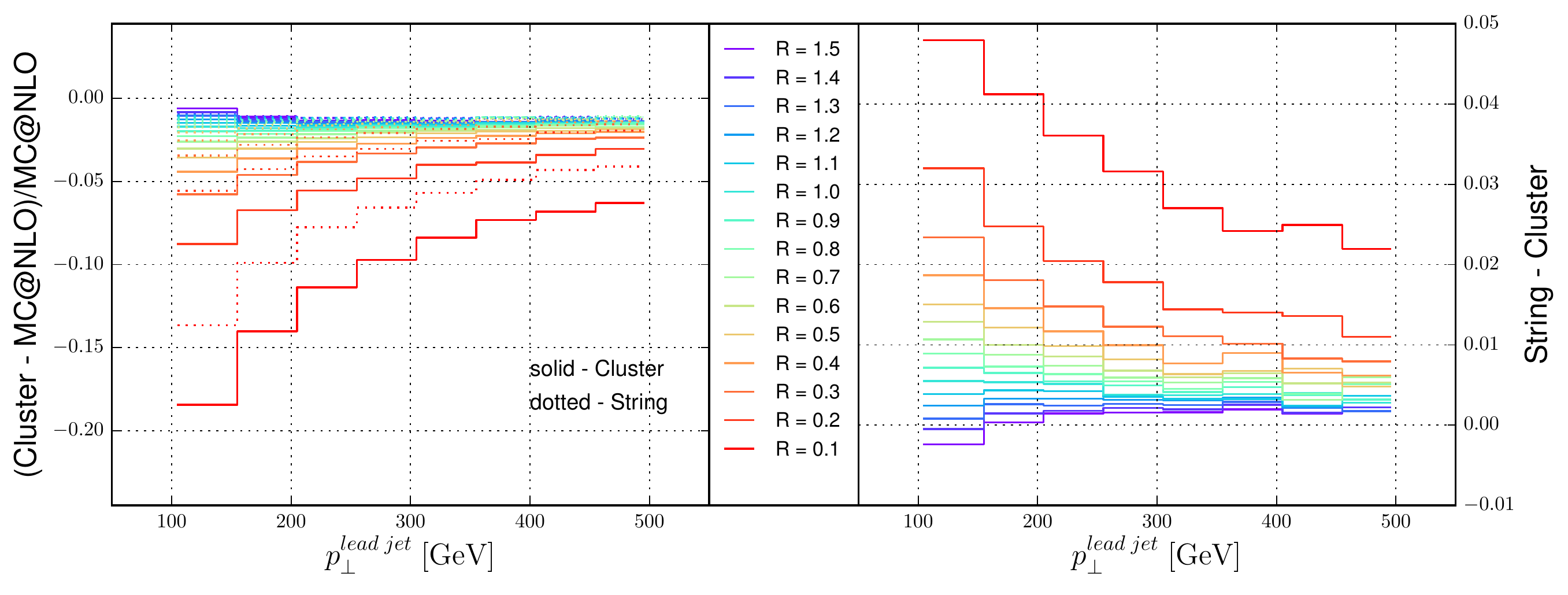}}
\centerline{\includegraphics[width=\textwidth]{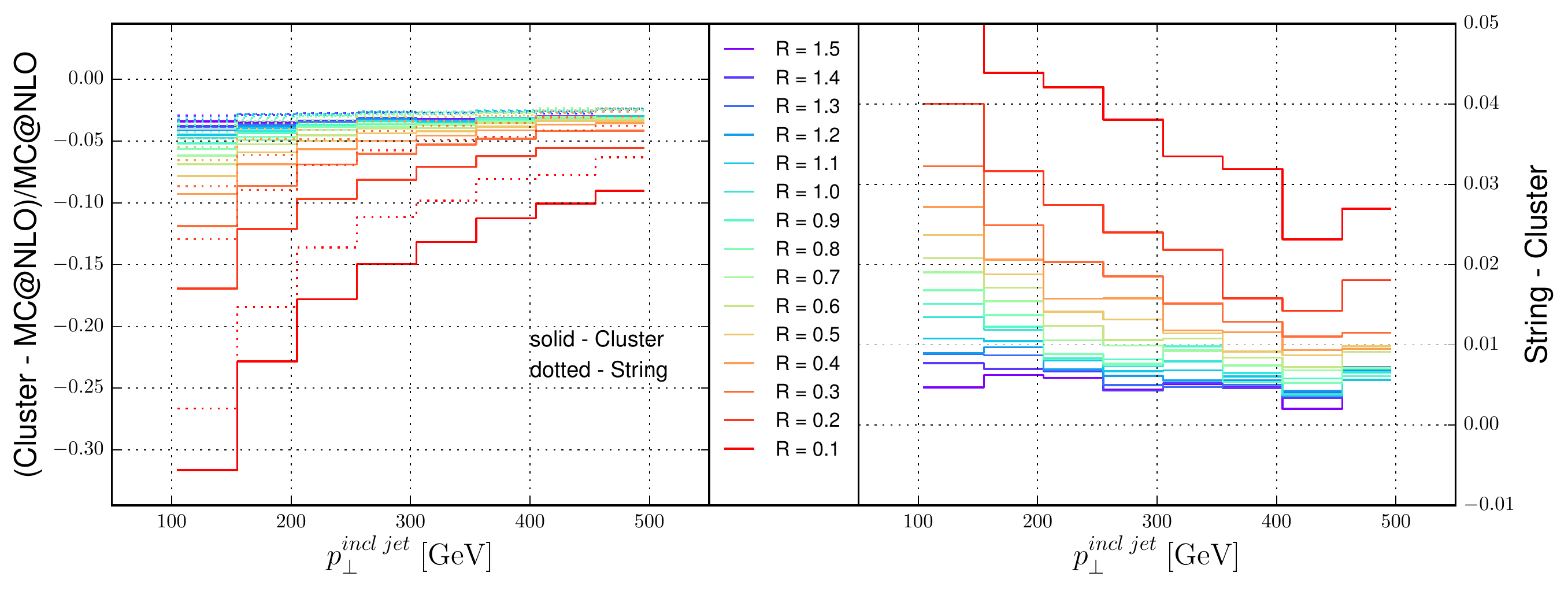}}
\caption{Hadronization corrections and uncertainties for Higgs+jets (top),
Z+jets (middle) and inclusive jets (bottom).\label{fig:hadcor}} 
\end{figure}

Fig.~\ref{fig:hadcor} (top and middle) shows that the hadronization corrections
for the lead jet in $H+\ge1$ jet and in $Z+\ge1$ jet are very similar. For the
commonly used jet size $R$=0.4, the corrections are of the order of 5\% or
less. String fragmentation leads to slightly larger corrections, but the
differences between the cluster and string fragmentation models are
significantly smaller than the magnitudes of the corrections, on the order of
2\% or less for $R$=0.4 and decreasing for larger $R$, as expected.

The pattern is similar for the inclusive jet transverse momentum spectrum for
dijet production, as shown in Fig.~\ref{fig:hadcor} (bottom), although the
impacts are magnified given the dijet final state at Born level. For $R$=0.4,
the difference between cluster and string fragmentation is of the order of
2.5\% or less. 

\begin{figure}[p]
\centerline{\includegraphics[width=\textwidth]{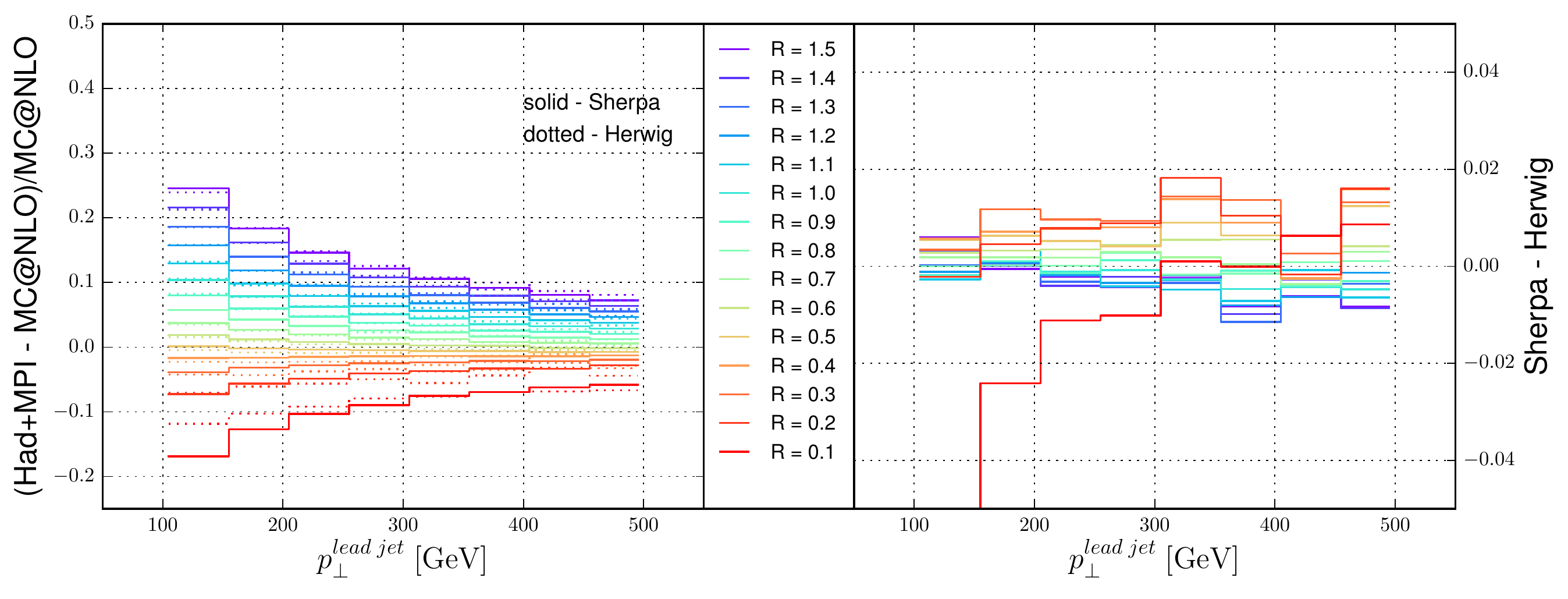}}
\centerline{\includegraphics[width=\textwidth]{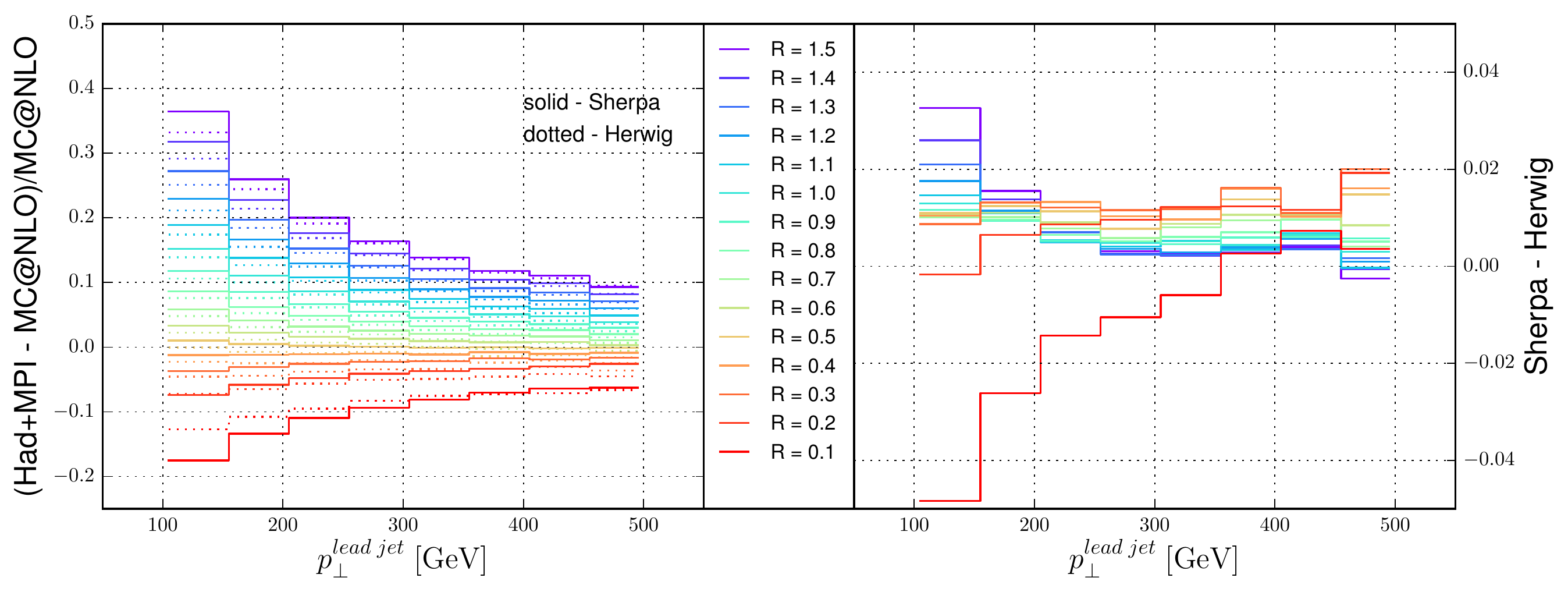}}
\centerline{\includegraphics[width=\textwidth]{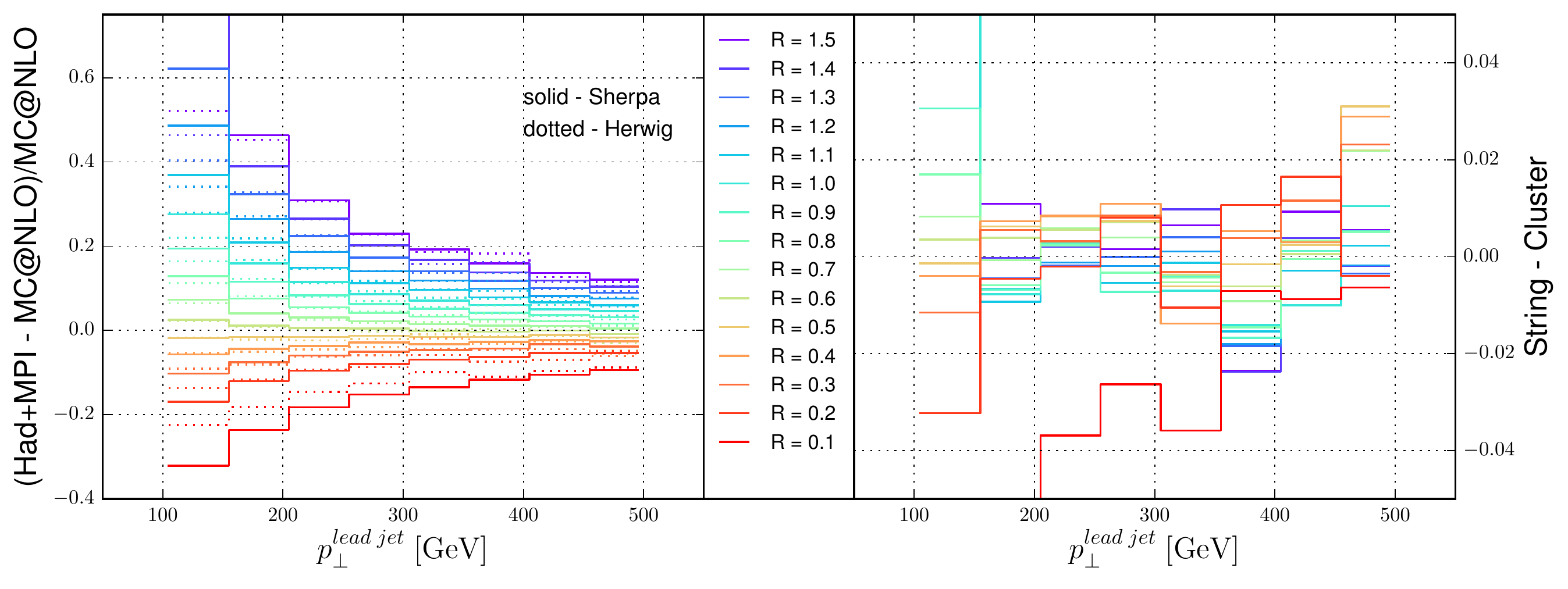}}
\caption{Hadronization plus MPI corrections for Higgs+jets (top), Z+jets
(middle) and inclusive jets (bottom).\label{fig:hadcor_mpi}} 
\end{figure}

The combined corrections from hadronization and the underlying event, modeled
through multiple parton interactions (MPI) are shown in
Fig.~\ref{fig:hadcor_mpi}. As the two corrections are in opposite directions,
and are of similar magnitude for jet sizes of the order of 0.4, the combined
correction is small, of the order of 2\% or less for $R$=0.4, except for dijet
production, where the combined correction can be as large as 5\%. The related
uncertainties shown on the right-hand side of Fig.~\ref{fig:hadcor_mpi} are
determined by taking the difference between predictions from Sherpa and Herwig,
both using their default MPI tunes and Cluster fragmentation.

\section{Conclusion and Outlook} 
\label{sec:outlook}

Searches for new physics, as well as a better understanding of standard model
physics, require an increasing level of precision, both for measurement and for
theory. For differential distributions, the highest level of precision is
obtained with NNLO calculations. Matched NLO plus parton shower predictions
(NLO+PS) start form less accurate fixed-order results, but provide a more
complete description of the event structure, including resummation effects at
leading logarithmic accuracy. Most physics measurements at the LHC make use of
relatively small jet sizes (anti-$k_T$ with $R=0.4$), and $H(Z)+\ge1$ jet
production and dijet production are no exception. There can be differences
between fixed order and NLO+PS predictions for the same observable just due to
the different estimates of the amount of jet energy contained in a jet of
radius $R$. These differences can be comparable to the size of the scale
uncertainty for the cross section at that order. 

In this contribution, we have reported on an investigation of the impact of
different jet sizes on Higgs plus jet, $Z$ boson plus jet, and dijet physics at
the LHC, paying close attention to the impact of the jet size on $K$-factors,
on scale uncertainties, and on differences between fixed order and NLO+PS
predictions. Better understanding of the issues described here may allow an
improvement in the accuracy, and precision, of such predictions at the LHC.

Our comparisons of the jet shapes for the three processes at fixed-order, full
parton shower level, and truncated parton shower level indicated that the
differences observed between fixed-order and NLO-matched results are due to
higher parton multiplicity final states. We have observed the best agreement,
for predictions involving jets, between fixed-order and NLO-matched
predictions, occur when the jet size R is relatively large, and/or the
fixed-order prediction is at NNLO compared to NLO. In the former, the jet shape
is not as critical, and in the latter the jet shape is better described. We
have found excellent agreement among the NLO-matched predictions for all
observables. 

The scale uncertainty naturally decreases in going from LO to NLO to NNLO, and
also tends to decrease as the jet size decreases. We have observed that the
(suitably normalized) NLO-matched results are within the fixed-order scale
uncertainty bands at LO, NLO and NNLO, for $H+\ge1$ jet for all jet sizes, but
are typically outside the NNLO scale uncertainty bands for $Z+\ge1$ jet, due to
the very small values of the scale uncertainties for this process at this
order. The scale uncertainties at NNLO can in fact be at or near zero for small
jet sizes, indicating that the standard scale uncertainty paradigm does not
provide an accurate description of the uncertainty of the calculation. These
small uncertainties are due to accidental cancellations arising from the
restriction of the phase space for small-R jets. We have constructed several
ways of providing more robust determinations of the scale uncertainties. 

Lastly, we have compared the non-perturbative predictions for all three
processes as a function of jet size R and jet transverse momentum, and have
found very good agreement between string and cluster fragmentation, and between
the full non-perturbative corrections, fragmentation plus MPI, between Sherpa
and Herwig. 

In summary, we expect parton-shower matched predictions to differ from the
underlying fixed-order results in regions where (1) there is a large
sensitivity to jet shapes (typically small $R$ jets), (2) there is another
restriction in phase space such that soft gluon resummation effects become
important, (3) the observable contains multiple, disparate scales, (4) the
observable is sensitive to higher multiplicity final states than those
described by the fixed-order calculation. Such differences should be smaller at
NNLO than at NLO. Large parton shower effects in the absence of large
higher-order corrections of type (1)-(4) should be viewed with suspicion, as
should large differences between parton shower predictions in general.

\section*{Acknowledgements} 

We thank the University of Zurich S3IT and CSCS
Lugano for providing the computational resources for \nnlojet. The work related
to \nnlojet was supported in part by the UK Science and Technology Facilities
Council, by the Swiss National Science Foundation (SNF) under contracts
200020-175595 and 200021-172478, by the ERC Consolidator Grant HICCUP
(No.614577), by the Swiss National Supercomputing Centre (CSCS) under project
ID UZH10, by the Research Executive Agency (REA) of the European Union under
the ERC Advanced Grant MC@NNLO(340983) and by the Funda\c{c}\~{a}o para a
Ci\^{e}ncia e Tecnologia (FCT-Portugal), project UID/FIS/00777/2019. This
research was supported by the U.S. Department of Energy under contract
DE-AC02-76SF00515. It used resources of the Fermi National Accelerator
Laboratory (Fermilab), a U.S. Department of Energy, Office of Science, HEP User
Facility. Fermilab is managed by Fermi Research Alliance, LLC (FRA), acting
under Contract No. DE-AC02-07CH11359. The work of AB was partially supported by
Royal Society University Research Fellowship number UF160548, and by the Marie
Sklodowska-Curie Action Innovative Training Network MCnetITN3. The work of ER
was partially supported by a Marie Sklodowska-Curie Individual Fellowship of
the European Commission's Horizon 2020 Programme under contract number 659147
PrecisionTools4LHC. This project has received funding from the European
Research Council (ERC) under the European Union's Horizon 2020 research and
innovation programme, grant agreement No 668679.

\bibliography{journal.bib}

\begin{thebibliography}{108}
\expandafter\ifx\csname natexlab\endcsname\relax\def\natexlab#1{#1}\fi
\expandafter\ifx\csname bibnamefont\endcsname\relax
  \def\bibnamefont#1{#1}\fi
\expandafter\ifx\csname bibfnamefont\endcsname\relax
  \def\bibfnamefont#1{#1}\fi
\expandafter\ifx\csname citenamefont\endcsname\relax
  \def\citenamefont#1{#1}\fi
\expandafter\ifx\csname url\endcsname\relax
  \def\url#1{\texttt{#1}}\fi
\expandafter\ifx\csname urlprefix\endcsname\relax\def\urlprefix{URL }\fi
\providecommand{\bibinfo}[2]{#2}
\providecommand{\eprint}[2][]{\url{#2}}

\bibitem[{\citenamefont{Gehrmann-De~Ridder
  et~al.}(2016{\natexlab{a}})\citenamefont{Gehrmann-De~Ridder, Gehrmann,
  Glover, Huss, and Morgan}}]{Ridder:2015dxa}
\bibinfo{author}{\bibfnamefont{A.}~\bibnamefont{Gehrmann-De~Ridder}},
  \bibinfo{author}{\bibfnamefont{T.}~\bibnamefont{Gehrmann}},
  \bibinfo{author}{\bibfnamefont{E.~W.~N.} \bibnamefont{Glover}},
  \bibinfo{author}{\bibfnamefont{A.}~\bibnamefont{Huss}}, \bibnamefont{and}
  \bibinfo{author}{\bibfnamefont{T.~A.} \bibnamefont{Morgan}},
  \bibinfo{journal}{Phys. Rev. Lett.} \textbf{\bibinfo{volume}{117}},
  \bibinfo{pages}{022001} (\bibinfo{year}{2016}{\natexlab{a}}),
  \eprint{1507.02850}.

\bibitem[{\citenamefont{Boughezal et~al.}(2016)\citenamefont{Boughezal,
  Campbell, Ellis, Focke, Giele, Liu, and Petriello}}]{Boughezal:2015ded}
\bibinfo{author}{\bibfnamefont{R.}~\bibnamefont{Boughezal}},
  \bibinfo{author}{\bibfnamefont{J.~M.} \bibnamefont{Campbell}},
  \bibinfo{author}{\bibfnamefont{R.~K.} \bibnamefont{Ellis}},
  \bibinfo{author}{\bibfnamefont{C.}~\bibnamefont{Focke}},
  \bibinfo{author}{\bibfnamefont{W.~T.} \bibnamefont{Giele}},
  \bibinfo{author}{\bibfnamefont{X.}~\bibnamefont{Liu}}, \bibnamefont{and}
  \bibinfo{author}{\bibfnamefont{F.}~\bibnamefont{Petriello}},
  \bibinfo{journal}{Phys. Rev. Lett.} \textbf{\bibinfo{volume}{116}},
  \bibinfo{pages}{152001} (\bibinfo{year}{2016}), \eprint{1512.01291}.

\bibitem[{\citenamefont{Gehrmann-De~Ridder
  et~al.}(2016{\natexlab{b}})\citenamefont{Gehrmann-De~Ridder, Gehrmann,
  Glover, Huss, and Morgan}}]{Gehrmann-DeRidder:2016jns}
\bibinfo{author}{\bibfnamefont{A.}~\bibnamefont{Gehrmann-De~Ridder}},
  \bibinfo{author}{\bibfnamefont{T.}~\bibnamefont{Gehrmann}},
  \bibinfo{author}{\bibfnamefont{E.~W.~N.} \bibnamefont{Glover}},
  \bibinfo{author}{\bibfnamefont{A.}~\bibnamefont{Huss}}, \bibnamefont{and}
  \bibinfo{author}{\bibfnamefont{T.~A.} \bibnamefont{Morgan}},
  \bibinfo{journal}{JHEP} \textbf{\bibinfo{volume}{11}}, \bibinfo{pages}{094}
  (\bibinfo{year}{2016}{\natexlab{b}}), \bibinfo{note}{[Erratum:
  JHEP10,126(2018)]}, \eprint{1610.01843}.

\bibitem[{\citenamefont{Gehrmann-De~Ridder
  et~al.}(2016{\natexlab{c}})\citenamefont{Gehrmann-De~Ridder, Gehrmann,
  Glover, Huss, and Morgan}}]{Ridder:2016nkl}
\bibinfo{author}{\bibfnamefont{A.}~\bibnamefont{Gehrmann-De~Ridder}},
  \bibinfo{author}{\bibfnamefont{T.}~\bibnamefont{Gehrmann}},
  \bibinfo{author}{\bibfnamefont{E.~W.~N.} \bibnamefont{Glover}},
  \bibinfo{author}{\bibfnamefont{A.}~\bibnamefont{Huss}}, \bibnamefont{and}
  \bibinfo{author}{\bibfnamefont{T.~A.} \bibnamefont{Morgan}},
  \bibinfo{journal}{JHEP} \textbf{\bibinfo{volume}{07}}, \bibinfo{pages}{133}
  (\bibinfo{year}{2016}{\natexlab{c}}), \eprint{1605.04295}.

\bibitem[{\citenamefont{Gehrmann-De~Ridder
  et~al.}(2018)\citenamefont{Gehrmann-De~Ridder, Gehrmann, Glover, Huss, and
  Walker}}]{Gehrmann-DeRidder:2017mvr}
\bibinfo{author}{\bibfnamefont{A.}~\bibnamefont{Gehrmann-De~Ridder}},
  \bibinfo{author}{\bibfnamefont{T.}~\bibnamefont{Gehrmann}},
  \bibinfo{author}{\bibfnamefont{E.~W.~N.} \bibnamefont{Glover}},
  \bibinfo{author}{\bibfnamefont{A.}~\bibnamefont{Huss}}, \bibnamefont{and}
  \bibinfo{author}{\bibfnamefont{D.~M.} \bibnamefont{Walker}},
  \bibinfo{journal}{Phys. Rev. Lett.} \textbf{\bibinfo{volume}{120}},
  \bibinfo{pages}{122001} (\bibinfo{year}{2018}), \eprint{1712.07543}.

\bibitem[{\citenamefont{Boughezal et~al.}(2013)\citenamefont{Boughezal, Caola,
  Melnikov, Petriello, and Schulze}}]{Boughezal:2013uia}
\bibinfo{author}{\bibfnamefont{R.}~\bibnamefont{Boughezal}},
  \bibinfo{author}{\bibfnamefont{F.}~\bibnamefont{Caola}},
  \bibinfo{author}{\bibfnamefont{K.}~\bibnamefont{Melnikov}},
  \bibinfo{author}{\bibfnamefont{F.}~\bibnamefont{Petriello}},
  \bibnamefont{and} \bibinfo{author}{\bibfnamefont{M.}~\bibnamefont{Schulze}},
  \bibinfo{journal}{JHEP} \textbf{\bibinfo{volume}{06}}, \bibinfo{pages}{072}
  (\bibinfo{year}{2013}), \eprint{1302.6216}.

\bibitem[{\citenamefont{Chen et~al.}(2015)\citenamefont{Chen, Gehrmann, Glover,
  and Jaquier}}]{Chen:2014gva}
\bibinfo{author}{\bibfnamefont{X.}~\bibnamefont{Chen}},
  \bibinfo{author}{\bibfnamefont{T.}~\bibnamefont{Gehrmann}},
  \bibinfo{author}{\bibfnamefont{E.~W.~N.} \bibnamefont{Glover}},
  \bibnamefont{and} \bibinfo{author}{\bibfnamefont{M.}~\bibnamefont{Jaquier}},
  \bibinfo{journal}{Phys. Lett.} \textbf{\bibinfo{volume}{B740}},
  \bibinfo{pages}{147} (\bibinfo{year}{2015}), \eprint{1408.5325}.

\bibitem[{\citenamefont{Boughezal et~al.}(2015)\citenamefont{Boughezal, Caola,
  Melnikov, Petriello, and Schulze}}]{Boughezal:2015dra}
\bibinfo{author}{\bibfnamefont{R.}~\bibnamefont{Boughezal}},
  \bibinfo{author}{\bibfnamefont{F.}~\bibnamefont{Caola}},
  \bibinfo{author}{\bibfnamefont{K.}~\bibnamefont{Melnikov}},
  \bibinfo{author}{\bibfnamefont{F.}~\bibnamefont{Petriello}},
  \bibnamefont{and} \bibinfo{author}{\bibfnamefont{M.}~\bibnamefont{Schulze}},
  \bibinfo{journal}{Phys. Rev. Lett.} \textbf{\bibinfo{volume}{115}},
  \bibinfo{pages}{082003} (\bibinfo{year}{2015}), \eprint{1504.07922}.

\bibitem[{\citenamefont{Chen et~al.}(2016{\natexlab{a}})\citenamefont{Chen,
  Cruz-Martinez, Gehrmann, Glover, and Jaquier}}]{Chen:2016zka}
\bibinfo{author}{\bibfnamefont{X.}~\bibnamefont{Chen}},
  \bibinfo{author}{\bibfnamefont{J.}~\bibnamefont{Cruz-Martinez}},
  \bibinfo{author}{\bibfnamefont{T.}~\bibnamefont{Gehrmann}},
  \bibinfo{author}{\bibfnamefont{E.~W.~N.} \bibnamefont{Glover}},
  \bibnamefont{and} \bibinfo{author}{\bibfnamefont{M.}~\bibnamefont{Jaquier}},
  \bibinfo{journal}{JHEP} \textbf{\bibinfo{volume}{10}}, \bibinfo{pages}{066}
  (\bibinfo{year}{2016}{\natexlab{a}}), \eprint{1607.08817}.

\bibitem[{\citenamefont{Chen et~al.}(2016{\natexlab{b}})\citenamefont{Chen,
  Gehrmann, Glover, and Jaquier}}]{Chen:2016vqn}
\bibinfo{author}{\bibfnamefont{X.}~\bibnamefont{Chen}},
  \bibinfo{author}{\bibfnamefont{T.}~\bibnamefont{Gehrmann}},
  \bibinfo{author}{\bibfnamefont{N.}~\bibnamefont{Glover}}, \bibnamefont{and}
  \bibinfo{author}{\bibfnamefont{M.}~\bibnamefont{Jaquier}},
  \bibinfo{journal}{PoS} \textbf{\bibinfo{volume}{RADCOR2015}},
  \bibinfo{pages}{056} (\bibinfo{year}{2016}{\natexlab{b}}),
  \eprint{1604.04085}.

\bibitem[{\citenamefont{Gehrmann-De~Ridder
  et~al.}(2013)\citenamefont{Gehrmann-De~Ridder, Gehrmann, Glover, and
  Pires}}]{Ridder:2013mf}
\bibinfo{author}{\bibfnamefont{A.}~\bibnamefont{Gehrmann-De~Ridder}},
  \bibinfo{author}{\bibfnamefont{T.}~\bibnamefont{Gehrmann}},
  \bibinfo{author}{\bibfnamefont{E.}~\bibnamefont{Glover}}, \bibnamefont{and}
  \bibinfo{author}{\bibfnamefont{J.}~\bibnamefont{Pires}},
  \bibinfo{journal}{Phys.Rev.Lett.} \textbf{\bibinfo{volume}{110}},
  \bibinfo{pages}{162003} (\bibinfo{year}{2013}), \eprint{1301.7310}.

\bibitem[{\citenamefont{Currie et~al.}(2014)\citenamefont{Currie,
  Gehrmann-De~Ridder, Glover, and Pires}}]{Currie:2013dwa}
\bibinfo{author}{\bibfnamefont{J.}~\bibnamefont{Currie}},
  \bibinfo{author}{\bibfnamefont{A.}~\bibnamefont{Gehrmann-De~Ridder}},
  \bibinfo{author}{\bibfnamefont{E.}~\bibnamefont{Glover}}, \bibnamefont{and}
  \bibinfo{author}{\bibfnamefont{J.}~\bibnamefont{Pires}},
  \bibinfo{journal}{JHEP} \textbf{\bibinfo{volume}{1401}}, \bibinfo{pages}{110}
  (\bibinfo{year}{2014}), \eprint{1310.3993}.

\bibitem[{\citenamefont{Currie et~al.}(2017{\natexlab{a}})\citenamefont{Currie,
  Glover, and Pires}}]{Currie:2016bfm}
\bibinfo{author}{\bibfnamefont{J.}~\bibnamefont{Currie}},
  \bibinfo{author}{\bibfnamefont{E.~W.~N.} \bibnamefont{Glover}},
  \bibnamefont{and} \bibinfo{author}{\bibfnamefont{J.}~\bibnamefont{Pires}},
  \bibinfo{journal}{Phys. Rev. Lett.} \textbf{\bibinfo{volume}{118}},
  \bibinfo{pages}{072002} (\bibinfo{year}{2017}{\natexlab{a}}),
  \eprint{1611.01460}.

\bibitem[{\citenamefont{Currie et~al.}(2017{\natexlab{b}})\citenamefont{Currie,
  Gehrmann-De~Ridder, Gehrmann, Glover, Huss, and Pires}}]{Currie:2017eqf}
\bibinfo{author}{\bibfnamefont{J.}~\bibnamefont{Currie}},
  \bibinfo{author}{\bibfnamefont{A.}~\bibnamefont{Gehrmann-De~Ridder}},
  \bibinfo{author}{\bibfnamefont{T.}~\bibnamefont{Gehrmann}},
  \bibinfo{author}{\bibfnamefont{E.~W.~N.} \bibnamefont{Glover}},
  \bibinfo{author}{\bibfnamefont{A.}~\bibnamefont{Huss}}, \bibnamefont{and}
  \bibinfo{author}{\bibfnamefont{J.}~\bibnamefont{Pires}},
  \bibinfo{journal}{Phys. Rev. Lett.} \textbf{\bibinfo{volume}{119}},
  \bibinfo{pages}{152001} (\bibinfo{year}{2017}{\natexlab{b}}),
  \eprint{1705.10271}.

\bibitem[{\citenamefont{Currie et~al.}(2018)\citenamefont{Currie,
  Gehrmann-De~Ridder, Gehrmann, Glover, Huss, and Pires}}]{Currie:2018xkj}
\bibinfo{author}{\bibfnamefont{J.}~\bibnamefont{Currie}},
  \bibinfo{author}{\bibfnamefont{A.}~\bibnamefont{Gehrmann-De~Ridder}},
  \bibinfo{author}{\bibfnamefont{T.}~\bibnamefont{Gehrmann}},
  \bibinfo{author}{\bibfnamefont{E.~W.~N.} \bibnamefont{Glover}},
  \bibinfo{author}{\bibfnamefont{A.}~\bibnamefont{Huss}}, \bibnamefont{and}
  \bibinfo{author}{\bibfnamefont{J.}~\bibnamefont{Pires}},
  \bibinfo{journal}{JHEP} \textbf{\bibinfo{volume}{10}}, \bibinfo{pages}{155}
  (\bibinfo{year}{2018}), \eprint{1807.03692}.

\bibitem[{\citenamefont{Ellis et~al.}(1983)\citenamefont{Ellis, Martinelli, and
  Petronzio}}]{Ellis:1981hk}
\bibinfo{author}{\bibfnamefont{R.~K.} \bibnamefont{Ellis}},
  \bibinfo{author}{\bibfnamefont{G.}~\bibnamefont{Martinelli}},
  \bibnamefont{and}
  \bibinfo{author}{\bibfnamefont{R.}~\bibnamefont{Petronzio}},
  \bibinfo{journal}{Nucl. Phys.} \textbf{\bibinfo{volume}{B211}},
  \bibinfo{pages}{106} (\bibinfo{year}{1983}).

\bibitem[{\citenamefont{Altarelli et~al.}(1984)\citenamefont{Altarelli, Ellis,
  Greco, and Martinelli}}]{Altarelli:1984pt}
\bibinfo{author}{\bibfnamefont{G.}~\bibnamefont{Altarelli}},
  \bibinfo{author}{\bibfnamefont{R.~K.} \bibnamefont{Ellis}},
  \bibinfo{author}{\bibfnamefont{M.}~\bibnamefont{Greco}}, \bibnamefont{and}
  \bibinfo{author}{\bibfnamefont{G.}~\bibnamefont{Martinelli}},
  \bibinfo{journal}{Nucl. Phys.} \textbf{\bibinfo{volume}{B246}},
  \bibinfo{pages}{12} (\bibinfo{year}{1984}).

\bibitem[{\citenamefont{Ellis and Sexton}(1986)}]{Ellis:1985er}
\bibinfo{author}{\bibfnamefont{R.~K.} \bibnamefont{Ellis}} \bibnamefont{and}
  \bibinfo{author}{\bibfnamefont{J.~C.} \bibnamefont{Sexton}},
  \bibinfo{journal}{Nucl. Phys.} \textbf{\bibinfo{volume}{B269}},
  \bibinfo{pages}{445} (\bibinfo{year}{1986}).

\bibitem[{\citenamefont{Gleisberg and Krauss}(2008)}]{Gleisberg:2007md}
\bibinfo{author}{\bibfnamefont{T.}~\bibnamefont{Gleisberg}} \bibnamefont{and}
  \bibinfo{author}{\bibfnamefont{F.}~\bibnamefont{Krauss}},
  \bibinfo{journal}{Eur. Phys. J.} \textbf{\bibinfo{volume}{C53}},
  \bibinfo{pages}{501} (\bibinfo{year}{2008}), \eprint{0709.2881}.

\bibitem[{\citenamefont{Frederix et~al.}(2008)\citenamefont{Frederix, Gehrmann,
  and Greiner}}]{Frederix:2008hu}
\bibinfo{author}{\bibfnamefont{R.}~\bibnamefont{Frederix}},
  \bibinfo{author}{\bibfnamefont{T.}~\bibnamefont{Gehrmann}}, \bibnamefont{and}
  \bibinfo{author}{\bibfnamefont{N.}~\bibnamefont{Greiner}},
  \bibinfo{journal}{JHEP} \textbf{\bibinfo{volume}{09}}, \bibinfo{pages}{122}
  (\bibinfo{year}{2008}), \eprint{0808.2128}.

\bibitem[{\citenamefont{Frederix et~al.}(2009)\citenamefont{Frederix, Frixione,
  Maltoni, and Stelzer}}]{Frederix:2009yq}
\bibinfo{author}{\bibfnamefont{R.}~\bibnamefont{Frederix}},
  \bibinfo{author}{\bibfnamefont{S.}~\bibnamefont{Frixione}},
  \bibinfo{author}{\bibfnamefont{F.}~\bibnamefont{Maltoni}}, \bibnamefont{and}
  \bibinfo{author}{\bibfnamefont{T.}~\bibnamefont{Stelzer}},
  \bibinfo{journal}{JHEP} \textbf{\bibinfo{volume}{10}}, \bibinfo{pages}{003}
  (\bibinfo{year}{2009}), \eprint{0908.4272}.

\bibitem[{\citenamefont{Cascioli et~al.}(2012)\citenamefont{Cascioli,
  Maierh{\"o}fer, and Pozzorini}}]{Cascioli:2011va}
\bibinfo{author}{\bibfnamefont{F.}~\bibnamefont{Cascioli}},
  \bibinfo{author}{\bibfnamefont{P.}~\bibnamefont{Maierh{\"o}fer}},
  \bibnamefont{and}
  \bibinfo{author}{\bibfnamefont{S.}~\bibnamefont{Pozzorini}},
  \bibinfo{journal}{Phys.Rev.Lett.} \textbf{\bibinfo{volume}{108}},
  \bibinfo{pages}{111601} (\bibinfo{year}{2012}), \eprint{1111.5206}.

\bibitem[{\citenamefont{Hirschi et~al.}(2011)\citenamefont{Hirschi, Frederix,
  Frixione, Garzelli, Maltoni, and Pittau}}]{Hirschi:2011pa}
\bibinfo{author}{\bibfnamefont{V.}~\bibnamefont{Hirschi}},
  \bibinfo{author}{\bibfnamefont{R.}~\bibnamefont{Frederix}},
  \bibinfo{author}{\bibfnamefont{S.}~\bibnamefont{Frixione}},
  \bibinfo{author}{\bibfnamefont{M.~V.} \bibnamefont{Garzelli}},
  \bibinfo{author}{\bibfnamefont{F.}~\bibnamefont{Maltoni}}, \bibnamefont{and}
  \bibinfo{author}{\bibfnamefont{R.}~\bibnamefont{Pittau}},
  \bibinfo{journal}{JHEP} \textbf{\bibinfo{volume}{05}}, \bibinfo{pages}{044}
  (\bibinfo{year}{2011}), \eprint{1103.0621}.

\bibitem[{\citenamefont{Cullen et~al.}(2012)\citenamefont{Cullen, Greiner,
  Heinrich, Luisoni, Mastrolia, Ossola, Reiter, and
  Tramontano}}]{Cullen:2011ac}
\bibinfo{author}{\bibfnamefont{G.}~\bibnamefont{Cullen}},
  \bibinfo{author}{\bibfnamefont{N.}~\bibnamefont{Greiner}},
  \bibinfo{author}{\bibfnamefont{G.}~\bibnamefont{Heinrich}},
  \bibinfo{author}{\bibfnamefont{G.}~\bibnamefont{Luisoni}},
  \bibinfo{author}{\bibfnamefont{P.}~\bibnamefont{Mastrolia}},
  \bibinfo{author}{\bibfnamefont{G.}~\bibnamefont{Ossola}},
  \bibinfo{author}{\bibfnamefont{T.}~\bibnamefont{Reiter}}, \bibnamefont{and}
  \bibinfo{author}{\bibfnamefont{F.}~\bibnamefont{Tramontano}},
  \bibinfo{journal}{Eur.Phys.J.} \textbf{\bibinfo{volume}{C72}},
  \bibinfo{pages}{1889} (\bibinfo{year}{2012}), \eprint{1111.2034}.

\bibitem[{\citenamefont{Pl{\"a}tzer and Gieseke}(2012)}]{Platzer:2011bc}
\bibinfo{author}{\bibfnamefont{S.}~\bibnamefont{Pl{\"a}tzer}} \bibnamefont{and}
  \bibinfo{author}{\bibfnamefont{S.}~\bibnamefont{Gieseke}},
  \bibinfo{journal}{Eur.Phys.J.} \textbf{\bibinfo{volume}{C72}},
  \bibinfo{pages}{2187} (\bibinfo{year}{2012}), \eprint{1109.6256}.

\bibitem[{\citenamefont{Cullen et~al.}(2014)\citenamefont{Cullen, van Deurzen,
  Greiner, Heinrich, Luisoni et~al.}}]{Cullen:2014yla}
\bibinfo{author}{\bibfnamefont{G.}~\bibnamefont{Cullen}},
  \bibinfo{author}{\bibfnamefont{H.}~\bibnamefont{van Deurzen}},
  \bibinfo{author}{\bibfnamefont{N.}~\bibnamefont{Greiner}},
  \bibinfo{author}{\bibfnamefont{G.}~\bibnamefont{Heinrich}},
  \bibinfo{author}{\bibfnamefont{G.}~\bibnamefont{Luisoni}},
  \bibnamefont{et~al.}, \bibinfo{journal}{Eur.Phys.J.}
  \textbf{\bibinfo{volume}{C74}}, \bibinfo{pages}{3001} (\bibinfo{year}{2014}),
  \eprint{1404.7096}.

\bibitem[{\citenamefont{Alwall et~al.}(2014)\citenamefont{Alwall, Frederix,
  Frixione, Hirschi, Maltoni, Mattelaer, Shao, Stelzer, Torrielli, and
  Zaro}}]{Alwall:2014hca}
\bibinfo{author}{\bibfnamefont{J.}~\bibnamefont{Alwall}},
  \bibinfo{author}{\bibfnamefont{R.}~\bibnamefont{Frederix}},
  \bibinfo{author}{\bibfnamefont{S.}~\bibnamefont{Frixione}},
  \bibinfo{author}{\bibfnamefont{V.}~\bibnamefont{Hirschi}},
  \bibinfo{author}{\bibfnamefont{F.}~\bibnamefont{Maltoni}},
  \bibinfo{author}{\bibfnamefont{O.}~\bibnamefont{Mattelaer}},
  \bibinfo{author}{\bibfnamefont{H.-S.} \bibnamefont{Shao}},
  \bibinfo{author}{\bibfnamefont{T.}~\bibnamefont{Stelzer}},
  \bibinfo{author}{\bibfnamefont{P.}~\bibnamefont{Torrielli}},
  \bibnamefont{and} \bibinfo{author}{\bibfnamefont{M.}~\bibnamefont{Zaro}},
  \bibinfo{journal}{JHEP} \textbf{\bibinfo{volume}{07}}, \bibinfo{pages}{079}
  (\bibinfo{year}{2014}), \eprint{1405.0301}.

\bibitem[{\citenamefont{Frixione and Webber}(2002)}]{Frixione:2002ik}
\bibinfo{author}{\bibfnamefont{S.}~\bibnamefont{Frixione}} \bibnamefont{and}
  \bibinfo{author}{\bibfnamefont{B.~R.} \bibnamefont{Webber}},
  \bibinfo{journal}{JHEP} \textbf{\bibinfo{volume}{06}}, \bibinfo{pages}{029}
  (\bibinfo{year}{2002}), \eprint{hep-ph/0204244}.

\bibitem[{\citenamefont{Nason}(2004)}]{Nason:2004rx}
\bibinfo{author}{\bibfnamefont{P.}~\bibnamefont{Nason}},
  \bibinfo{journal}{JHEP} \textbf{\bibinfo{volume}{11}}, \bibinfo{pages}{040}
  (\bibinfo{year}{2004}), \eprint{hep-ph/0409146}.

\bibitem[{\citenamefont{Dasgupta et~al.}(2016)\citenamefont{Dasgupta, Dreyer,
  Salam, and Soyez}}]{Dasgupta:2016bnd}
\bibinfo{author}{\bibfnamefont{M.}~\bibnamefont{Dasgupta}},
  \bibinfo{author}{\bibfnamefont{F.~A.} \bibnamefont{Dreyer}},
  \bibinfo{author}{\bibfnamefont{G.~P.} \bibnamefont{Salam}}, \bibnamefont{and}
  \bibinfo{author}{\bibfnamefont{G.}~\bibnamefont{Soyez}},
  \bibinfo{journal}{JHEP} \textbf{\bibinfo{volume}{06}}, \bibinfo{pages}{057}
  (\bibinfo{year}{2016}), \eprint{1602.01110}.

\bibitem[{\citenamefont{Liu et~al.}(2017)\citenamefont{Liu, Moch, and
  Ringer}}]{Liu:2017pbb}
\bibinfo{author}{\bibfnamefont{X.}~\bibnamefont{Liu}},
  \bibinfo{author}{\bibfnamefont{S.-O.} \bibnamefont{Moch}}, \bibnamefont{and}
  \bibinfo{author}{\bibfnamefont{F.}~\bibnamefont{Ringer}},
  \bibinfo{journal}{Phys. Rev. Lett.} \textbf{\bibinfo{volume}{119}},
  \bibinfo{pages}{212001} (\bibinfo{year}{2017}), \eprint{1708.04641}.

\bibitem[{\citenamefont{Liu et~al.}(2018)\citenamefont{Liu, Moch, and
  Ringer}}]{Liu:2018ktv}
\bibinfo{author}{\bibfnamefont{X.}~\bibnamefont{Liu}},
  \bibinfo{author}{\bibfnamefont{S.-O.} \bibnamefont{Moch}}, \bibnamefont{and}
  \bibinfo{author}{\bibfnamefont{F.}~\bibnamefont{Ringer}},
  \bibinfo{journal}{Phys. Rev.} \textbf{\bibinfo{volume}{D97}},
  \bibinfo{pages}{056026} (\bibinfo{year}{2018}), \eprint{1801.07284}.

\bibitem[{Ben(2018)}]{Bendavid:2018nar}
\emph{\bibinfo{title}{{Les Houches 2017: Physics at TeV Colliders Standard
  Model Working Group Report}}} (\bibinfo{year}{2018}), \eprint{1803.07977}.

\bibitem[{\citenamefont{Cacciari et~al.}(2008)\citenamefont{Cacciari, Salam,
  and Soyez}}]{Cacciari:2008gp}
\bibinfo{author}{\bibfnamefont{M.}~\bibnamefont{Cacciari}},
  \bibinfo{author}{\bibfnamefont{G.~P.} \bibnamefont{Salam}}, \bibnamefont{and}
  \bibinfo{author}{\bibfnamefont{G.}~\bibnamefont{Soyez}},
  \bibinfo{journal}{JHEP} \textbf{\bibinfo{volume}{04}}, \bibinfo{pages}{063}
  (\bibinfo{year}{2008}), \eprint{0802.1189}.

\bibitem[{\citenamefont{Butterworth et~al.}(2016)}]{Butterworth:2015oua}
\bibinfo{author}{\bibfnamefont{J.}~\bibnamefont{Butterworth}}
  \bibnamefont{et~al.}, \bibinfo{journal}{J. Phys.}
  \textbf{\bibinfo{volume}{G43}}, \bibinfo{pages}{023001}
  (\bibinfo{year}{2016}), \eprint{1510.03865}.

\bibitem[{\citenamefont{Andersen et~al.}(2016)}]{Badger:2016bpw}
\bibinfo{author}{\bibfnamefont{J.~R.} \bibnamefont{Andersen}}
  \bibnamefont{et~al.}, in \emph{\bibinfo{booktitle}{{9th Les Houches Workshop
  on Physics at TeV Colliders (PhysTeV 2015) Les Houches, France, June 1-19,
  2015}}} (\bibinfo{year}{2016}), \eprint{1605.04692}.

\bibitem[{\citenamefont{Buckley et~al.}(2013)\citenamefont{Buckley,
  Butterworth, L{\"o}nnblad, Grellscheid, Hoeth et~al.}}]{Buckley:2010ar}
\bibinfo{author}{\bibfnamefont{A.}~\bibnamefont{Buckley}},
  \bibinfo{author}{\bibfnamefont{J.}~\bibnamefont{Butterworth}},
  \bibinfo{author}{\bibfnamefont{L.}~\bibnamefont{L{\"o}nnblad}},
  \bibinfo{author}{\bibfnamefont{D.}~\bibnamefont{Grellscheid}},
  \bibinfo{author}{\bibfnamefont{H.}~\bibnamefont{Hoeth}},
  \bibnamefont{et~al.}, \bibinfo{journal}{Comput.Phys.Commun.}
  \textbf{\bibinfo{volume}{184}}, \bibinfo{pages}{2803} (\bibinfo{year}{2013}),
  \eprint{1003.0694}.

\bibitem[{\citenamefont{Khachatryan et~al.}(2016)}]{Khachatryan:2016wdh}
\bibinfo{author}{\bibfnamefont{V.}~\bibnamefont{Khachatryan}}
  \bibnamefont{et~al.} (\bibinfo{collaboration}{CMS}), \bibinfo{journal}{Eur.
  Phys. J.} \textbf{\bibinfo{volume}{C76}}, \bibinfo{pages}{451}
  (\bibinfo{year}{2016}), \eprint{1605.04436}.

\bibitem[{\citenamefont{Del~Duca et~al.}(2004)\citenamefont{Del~Duca, Frizzo,
  and Maltoni}}]{DelDuca:2004wt}
\bibinfo{author}{\bibfnamefont{V.}~\bibnamefont{Del~Duca}},
  \bibinfo{author}{\bibfnamefont{A.}~\bibnamefont{Frizzo}}, \bibnamefont{and}
  \bibinfo{author}{\bibfnamefont{F.}~\bibnamefont{Maltoni}},
  \bibinfo{journal}{JHEP} \textbf{\bibinfo{volume}{05}}, \bibinfo{pages}{064}
  (\bibinfo{year}{2004}), \eprint{hep-ph/0404013}.

\bibitem[{\citenamefont{Dixon et~al.}(2004)\citenamefont{Dixon, Glover, and
  Khoze}}]{Dixon:2004za}
\bibinfo{author}{\bibfnamefont{L.~J.} \bibnamefont{Dixon}},
  \bibinfo{author}{\bibfnamefont{E.~W.~N.} \bibnamefont{Glover}},
  \bibnamefont{and} \bibinfo{author}{\bibfnamefont{V.~V.} \bibnamefont{Khoze}},
  \bibinfo{journal}{JHEP} \textbf{\bibinfo{volume}{12}}, \bibinfo{pages}{015}
  (\bibinfo{year}{2004}), \eprint{hep-th/0411092}.

\bibitem[{\citenamefont{Badger et~al.}(2005)\citenamefont{Badger, Glover, and
  Khoze}}]{Badger:2004ty}
\bibinfo{author}{\bibfnamefont{S.~D.} \bibnamefont{Badger}},
  \bibinfo{author}{\bibfnamefont{E.~W.~N.} \bibnamefont{Glover}},
  \bibnamefont{and} \bibinfo{author}{\bibfnamefont{V.~V.} \bibnamefont{Khoze}},
  \bibinfo{journal}{JHEP} \textbf{\bibinfo{volume}{03}}, \bibinfo{pages}{023}
  (\bibinfo{year}{2005}), \eprint{hep-th/0412275}.

\bibitem[{\citenamefont{Badger et~al.}(2010)\citenamefont{Badger, Nigel~Glover,
  Mastrolia, and Williams}}]{Badger:2009hw}
\bibinfo{author}{\bibfnamefont{S.}~\bibnamefont{Badger}},
  \bibinfo{author}{\bibfnamefont{E.~W.} \bibnamefont{Nigel~Glover}},
  \bibinfo{author}{\bibfnamefont{P.}~\bibnamefont{Mastrolia}},
  \bibnamefont{and} \bibinfo{author}{\bibfnamefont{C.}~\bibnamefont{Williams}},
  \bibinfo{journal}{JHEP} \textbf{\bibinfo{volume}{01}}, \bibinfo{pages}{036}
  (\bibinfo{year}{2010}), \eprint{0909.4475}.

\bibitem[{\citenamefont{Badger et~al.}(2009)\citenamefont{Badger, Campbell,
  Ellis, and Williams}}]{Badger:2009vh}
\bibinfo{author}{\bibfnamefont{S.}~\bibnamefont{Badger}},
  \bibinfo{author}{\bibfnamefont{J.~M.} \bibnamefont{Campbell}},
  \bibinfo{author}{\bibfnamefont{R.~K.} \bibnamefont{Ellis}}, \bibnamefont{and}
  \bibinfo{author}{\bibfnamefont{C.}~\bibnamefont{Williams}},
  \bibinfo{journal}{JHEP} \textbf{\bibinfo{volume}{12}}, \bibinfo{pages}{035}
  (\bibinfo{year}{2009}), \eprint{0910.4481}.

\bibitem[{\citenamefont{Dixon and Sofianatos}(2009)}]{Dixon:2009uk}
\bibinfo{author}{\bibfnamefont{L.~J.} \bibnamefont{Dixon}} \bibnamefont{and}
  \bibinfo{author}{\bibfnamefont{Y.}~\bibnamefont{Sofianatos}},
  \bibinfo{journal}{JHEP} \textbf{\bibinfo{volume}{08}}, \bibinfo{pages}{058}
  (\bibinfo{year}{2009}), \eprint{0906.0008}.

\bibitem[{\citenamefont{Gehrmann et~al.}(2012)\citenamefont{Gehrmann, Jaquier,
  Glover, and Koukoutsakis}}]{Gehrmann:2011aa}
\bibinfo{author}{\bibfnamefont{T.}~\bibnamefont{Gehrmann}},
  \bibinfo{author}{\bibfnamefont{M.}~\bibnamefont{Jaquier}},
  \bibinfo{author}{\bibfnamefont{E.~W.~N.} \bibnamefont{Glover}},
  \bibnamefont{and}
  \bibinfo{author}{\bibfnamefont{A.}~\bibnamefont{Koukoutsakis}},
  \bibinfo{journal}{JHEP} \textbf{\bibinfo{volume}{02}}, \bibinfo{pages}{056}
  (\bibinfo{year}{2012}), \eprint{1112.3554}.

\bibitem[{\citenamefont{Hagiwara and Zeppenfeld}(1989)}]{Hagiwara:1988pp}
\bibinfo{author}{\bibfnamefont{K.}~\bibnamefont{Hagiwara}} \bibnamefont{and}
  \bibinfo{author}{\bibfnamefont{D.}~\bibnamefont{Zeppenfeld}},
  \bibinfo{journal}{Nucl. Phys.} \textbf{\bibinfo{volume}{B313}},
  \bibinfo{pages}{560} (\bibinfo{year}{1989}).

\bibitem[{\citenamefont{Berends et~al.}(1989)\citenamefont{Berends, Giele, and
  Kuijf}}]{Berends:1988yn}
\bibinfo{author}{\bibfnamefont{F.~A.} \bibnamefont{Berends}},
  \bibinfo{author}{\bibfnamefont{W.~T.} \bibnamefont{Giele}}, \bibnamefont{and}
  \bibinfo{author}{\bibfnamefont{H.}~\bibnamefont{Kuijf}},
  \bibinfo{journal}{Nucl. Phys.} \textbf{\bibinfo{volume}{B321}},
  \bibinfo{pages}{39} (\bibinfo{year}{1989}).

\bibitem[{\citenamefont{Falck et~al.}(1989)\citenamefont{Falck, Graudenz, and
  Kramer}}]{Falck:1989uz}
\bibinfo{author}{\bibfnamefont{N.~K.} \bibnamefont{Falck}},
  \bibinfo{author}{\bibfnamefont{D.}~\bibnamefont{Graudenz}}, \bibnamefont{and}
  \bibinfo{author}{\bibfnamefont{G.}~\bibnamefont{Kramer}},
  \bibinfo{journal}{Nucl. Phys.} \textbf{\bibinfo{volume}{B328}},
  \bibinfo{pages}{317} (\bibinfo{year}{1989}).

\bibitem[{\citenamefont{Nagy and Trocsanyi}(1999)}]{Nagy:1998bb}
\bibinfo{author}{\bibfnamefont{Z.}~\bibnamefont{Nagy}} \bibnamefont{and}
  \bibinfo{author}{\bibfnamefont{Z.}~\bibnamefont{Trocsanyi}},
  \bibinfo{journal}{Phys. Rev.} \textbf{\bibinfo{volume}{D59}},
  \bibinfo{pages}{014020} (\bibinfo{year}{1999}), \bibinfo{note}{[Erratum:
  Phys. Rev.D62,099902(2000)]}, \eprint{hep-ph/9806317}.

\bibitem[{\citenamefont{Glover and Miller}(1997)}]{Glover:1996eh}
\bibinfo{author}{\bibfnamefont{E.~W.~N.} \bibnamefont{Glover}}
  \bibnamefont{and} \bibinfo{author}{\bibfnamefont{D.~J.}
  \bibnamefont{Miller}}, \bibinfo{journal}{Phys. Lett.}
  \textbf{\bibinfo{volume}{B396}}, \bibinfo{pages}{257} (\bibinfo{year}{1997}),
  \eprint{hep-ph/9609474}.

\bibitem[{\citenamefont{Bern et~al.}(1997)\citenamefont{Bern, Dixon, Kosower,
  and Weinzierl}}]{Bern:1996ka}
\bibinfo{author}{\bibfnamefont{Z.}~\bibnamefont{Bern}},
  \bibinfo{author}{\bibfnamefont{L.~J.} \bibnamefont{Dixon}},
  \bibinfo{author}{\bibfnamefont{D.~A.} \bibnamefont{Kosower}},
  \bibnamefont{and}
  \bibinfo{author}{\bibfnamefont{S.}~\bibnamefont{Weinzierl}},
  \bibinfo{journal}{Nucl. Phys.} \textbf{\bibinfo{volume}{B489}},
  \bibinfo{pages}{3} (\bibinfo{year}{1997}), \eprint{hep-ph/9610370}.

\bibitem[{\citenamefont{Campbell et~al.}(1997)\citenamefont{Campbell, Glover,
  and Miller}}]{Campbell:1997tv}
\bibinfo{author}{\bibfnamefont{J.~M.} \bibnamefont{Campbell}},
  \bibinfo{author}{\bibfnamefont{E.~W.~N.} \bibnamefont{Glover}},
  \bibnamefont{and} \bibinfo{author}{\bibfnamefont{D.~J.}
  \bibnamefont{Miller}}, \bibinfo{journal}{Phys. Lett.}
  \textbf{\bibinfo{volume}{B409}}, \bibinfo{pages}{503} (\bibinfo{year}{1997}),
  \eprint{hep-ph/9706297}.

\bibitem[{\citenamefont{Bern et~al.}(1998)\citenamefont{Bern, Dixon, and
  Kosower}}]{Bern:1997sc}
\bibinfo{author}{\bibfnamefont{Z.}~\bibnamefont{Bern}},
  \bibinfo{author}{\bibfnamefont{L.~J.} \bibnamefont{Dixon}}, \bibnamefont{and}
  \bibinfo{author}{\bibfnamefont{D.~A.} \bibnamefont{Kosower}},
  \bibinfo{journal}{Nucl. Phys.} \textbf{\bibinfo{volume}{B513}},
  \bibinfo{pages}{3} (\bibinfo{year}{1998}), \eprint{hep-ph/9708239}.

\bibitem[{\citenamefont{Moch et~al.}(2002)\citenamefont{Moch, Uwer, and
  Weinzierl}}]{Moch:2002hm}
\bibinfo{author}{\bibfnamefont{S.}~\bibnamefont{Moch}},
  \bibinfo{author}{\bibfnamefont{P.}~\bibnamefont{Uwer}}, \bibnamefont{and}
  \bibinfo{author}{\bibfnamefont{S.}~\bibnamefont{Weinzierl}},
  \bibinfo{journal}{Phys. Rev.} \textbf{\bibinfo{volume}{D66}},
  \bibinfo{pages}{114001} (\bibinfo{year}{2002}), \eprint{hep-ph/0207043}.

\bibitem[{\citenamefont{Garland
  et~al.}(2002{\natexlab{a}})\citenamefont{Garland, Gehrmann, Glover,
  Koukoutsakis, and Remiddi}}]{Garland:2001tf}
\bibinfo{author}{\bibfnamefont{L.~W.} \bibnamefont{Garland}},
  \bibinfo{author}{\bibfnamefont{T.}~\bibnamefont{Gehrmann}},
  \bibinfo{author}{\bibfnamefont{E.~W.~N.} \bibnamefont{Glover}},
  \bibinfo{author}{\bibfnamefont{A.}~\bibnamefont{Koukoutsakis}},
  \bibnamefont{and} \bibinfo{author}{\bibfnamefont{E.}~\bibnamefont{Remiddi}},
  \bibinfo{journal}{Nucl. Phys.} \textbf{\bibinfo{volume}{B627}},
  \bibinfo{pages}{107} (\bibinfo{year}{2002}{\natexlab{a}}),
  \eprint{hep-ph/0112081}.

\bibitem[{\citenamefont{Garland
  et~al.}(2002{\natexlab{b}})\citenamefont{Garland, Gehrmann, Glover,
  Koukoutsakis, and Remiddi}}]{Garland:2002ak}
\bibinfo{author}{\bibfnamefont{L.~W.} \bibnamefont{Garland}},
  \bibinfo{author}{\bibfnamefont{T.}~\bibnamefont{Gehrmann}},
  \bibinfo{author}{\bibfnamefont{E.~W.~N.} \bibnamefont{Glover}},
  \bibinfo{author}{\bibfnamefont{A.}~\bibnamefont{Koukoutsakis}},
  \bibnamefont{and} \bibinfo{author}{\bibfnamefont{E.}~\bibnamefont{Remiddi}},
  \bibinfo{journal}{Nucl. Phys.} \textbf{\bibinfo{volume}{B642}},
  \bibinfo{pages}{227} (\bibinfo{year}{2002}{\natexlab{b}}),
  \eprint{hep-ph/0206067}.

\bibitem[{\citenamefont{Gehrmann and Tancredi}(2012)}]{Gehrmann:2011ab}
\bibinfo{author}{\bibfnamefont{T.}~\bibnamefont{Gehrmann}} \bibnamefont{and}
  \bibinfo{author}{\bibfnamefont{L.}~\bibnamefont{Tancredi}},
  \bibinfo{journal}{JHEP} \textbf{\bibinfo{volume}{02}}, \bibinfo{pages}{004}
  (\bibinfo{year}{2012}), \eprint{1112.1531}.

\bibitem[{\citenamefont{Mangano and Parke}(1991)}]{Mangano:1990by}
\bibinfo{author}{\bibfnamefont{M.~L.} \bibnamefont{Mangano}} \bibnamefont{and}
  \bibinfo{author}{\bibfnamefont{S.~J.} \bibnamefont{Parke}},
  \bibinfo{journal}{Phys. Rept.} \textbf{\bibinfo{volume}{200}},
  \bibinfo{pages}{301} (\bibinfo{year}{1991}), \eprint{hep-th/0509223}.

\bibitem[{\citenamefont{Bern et~al.}(1993)\citenamefont{Bern, Dixon, and
  Kosower}}]{Bern:1993mq}
\bibinfo{author}{\bibfnamefont{Z.}~\bibnamefont{Bern}},
  \bibinfo{author}{\bibfnamefont{L.~J.} \bibnamefont{Dixon}}, \bibnamefont{and}
  \bibinfo{author}{\bibfnamefont{D.~A.} \bibnamefont{Kosower}},
  \bibinfo{journal}{Phys. Rev. Lett.} \textbf{\bibinfo{volume}{70}},
  \bibinfo{pages}{2677} (\bibinfo{year}{1993}), \eprint{hep-ph/9302280}.

\bibitem[{\citenamefont{Bern et~al.}(1995)\citenamefont{Bern, Dixon, and
  Kosower}}]{Bern:1994fz}
\bibinfo{author}{\bibfnamefont{Z.}~\bibnamefont{Bern}},
  \bibinfo{author}{\bibfnamefont{L.~J.} \bibnamefont{Dixon}}, \bibnamefont{and}
  \bibinfo{author}{\bibfnamefont{D.~A.} \bibnamefont{Kosower}},
  \bibinfo{journal}{Nucl. Phys.} \textbf{\bibinfo{volume}{B437}},
  \bibinfo{pages}{259} (\bibinfo{year}{1995}), \eprint{hep-ph/9409393}.

\bibitem[{\citenamefont{Kunszt et~al.}(1994)\citenamefont{Kunszt, Signer, and
  Trocsanyi}}]{Kunszt:1994nq}
\bibinfo{author}{\bibfnamefont{Z.}~\bibnamefont{Kunszt}},
  \bibinfo{author}{\bibfnamefont{A.}~\bibnamefont{Signer}}, \bibnamefont{and}
  \bibinfo{author}{\bibfnamefont{Z.}~\bibnamefont{Trocsanyi}},
  \bibinfo{journal}{Phys. Lett.} \textbf{\bibinfo{volume}{B336}},
  \bibinfo{pages}{529} (\bibinfo{year}{1994}), \eprint{hep-ph/9405386}.

\bibitem[{\citenamefont{Glover et~al.}(2001)\citenamefont{Glover, Oleari, and
  Tejeda-Yeomans}}]{Glover:2001af}
\bibinfo{author}{\bibfnamefont{E.~W.~N.} \bibnamefont{Glover}},
  \bibinfo{author}{\bibfnamefont{C.}~\bibnamefont{Oleari}}, \bibnamefont{and}
  \bibinfo{author}{\bibfnamefont{M.~E.} \bibnamefont{Tejeda-Yeomans}},
  \bibinfo{journal}{Nucl. Phys.} \textbf{\bibinfo{volume}{B605}},
  \bibinfo{pages}{467} (\bibinfo{year}{2001}), \eprint{hep-ph/0102201}.

\bibitem[{\citenamefont{Glover and Tejeda-Yeomans}(2001)}]{Glover:2001rd}
\bibinfo{author}{\bibfnamefont{E.~W.~N.} \bibnamefont{Glover}}
  \bibnamefont{and} \bibinfo{author}{\bibfnamefont{M.~E.}
  \bibnamefont{Tejeda-Yeomans}}, \bibinfo{journal}{JHEP}
  \textbf{\bibinfo{volume}{05}}, \bibinfo{pages}{010} (\bibinfo{year}{2001}),
  \eprint{hep-ph/0104178}.

\bibitem[{\citenamefont{Bern et~al.}(2002)\citenamefont{Bern, De~Freitas, and
  Dixon}}]{Bern:2002tk}
\bibinfo{author}{\bibfnamefont{Z.}~\bibnamefont{Bern}},
  \bibinfo{author}{\bibfnamefont{A.}~\bibnamefont{De~Freitas}},
  \bibnamefont{and} \bibinfo{author}{\bibfnamefont{L.~J.} \bibnamefont{Dixon}},
  \bibinfo{journal}{JHEP} \textbf{\bibinfo{volume}{03}}, \bibinfo{pages}{018}
  (\bibinfo{year}{2002}), \eprint{hep-ph/0201161}.

\bibitem[{\citenamefont{Anastasiou
  et~al.}(2001{\natexlab{a}})\citenamefont{Anastasiou, Glover, Oleari, and
  Tejeda-Yeomans}}]{Anastasiou:2001sv}
\bibinfo{author}{\bibfnamefont{C.}~\bibnamefont{Anastasiou}},
  \bibinfo{author}{\bibfnamefont{E.~W.~N.} \bibnamefont{Glover}},
  \bibinfo{author}{\bibfnamefont{C.}~\bibnamefont{Oleari}}, \bibnamefont{and}
  \bibinfo{author}{\bibfnamefont{M.~E.} \bibnamefont{Tejeda-Yeomans}},
  \bibinfo{journal}{Nucl. Phys.} \textbf{\bibinfo{volume}{B605}},
  \bibinfo{pages}{486} (\bibinfo{year}{2001}{\natexlab{a}}),
  \eprint{hep-ph/0101304}.

\bibitem[{\citenamefont{Anastasiou
  et~al.}(2001{\natexlab{b}})\citenamefont{Anastasiou, Glover, Oleari, and
  Tejeda-Yeomans}}]{Anastasiou:2000mv}
\bibinfo{author}{\bibfnamefont{C.}~\bibnamefont{Anastasiou}},
  \bibinfo{author}{\bibfnamefont{E.~W.~N.} \bibnamefont{Glover}},
  \bibinfo{author}{\bibfnamefont{C.}~\bibnamefont{Oleari}}, \bibnamefont{and}
  \bibinfo{author}{\bibfnamefont{M.~E.} \bibnamefont{Tejeda-Yeomans}},
  \bibinfo{journal}{Phys. Lett.} \textbf{\bibinfo{volume}{B506}},
  \bibinfo{pages}{59} (\bibinfo{year}{2001}{\natexlab{b}}),
  \eprint{hep-ph/0012007}.

\bibitem[{\citenamefont{Glover and Tejeda-Yeomans}(2003)}]{Glover:2003cm}
\bibinfo{author}{\bibfnamefont{E.~W.~N.} \bibnamefont{Glover}}
  \bibnamefont{and} \bibinfo{author}{\bibfnamefont{M.~E.}
  \bibnamefont{Tejeda-Yeomans}}, \bibinfo{journal}{JHEP}
  \textbf{\bibinfo{volume}{06}}, \bibinfo{pages}{033} (\bibinfo{year}{2003}),
  \eprint{hep-ph/0304169}.

\bibitem[{\citenamefont{Bern et~al.}(2003)\citenamefont{Bern, De~Freitas, and
  Dixon}}]{Bern:2003ck}
\bibinfo{author}{\bibfnamefont{Z.}~\bibnamefont{Bern}},
  \bibinfo{author}{\bibfnamefont{A.}~\bibnamefont{De~Freitas}},
  \bibnamefont{and} \bibinfo{author}{\bibfnamefont{L.~J.} \bibnamefont{Dixon}},
  \bibinfo{journal}{JHEP} \textbf{\bibinfo{volume}{06}}, \bibinfo{pages}{028}
  (\bibinfo{year}{2003}), \bibinfo{note}{[Erratum: JHEP04,112(2014)]},
  \eprint{hep-ph/0304168}.

\bibitem[{\citenamefont{Gehrmann-De~Ridder
  et~al.}(2005{\natexlab{a}})\citenamefont{Gehrmann-De~Ridder, Gehrmann, and
  Glover}}]{GehrmannDeRidder:2005cm}
\bibinfo{author}{\bibfnamefont{A.}~\bibnamefont{Gehrmann-De~Ridder}},
  \bibinfo{author}{\bibfnamefont{T.}~\bibnamefont{Gehrmann}}, \bibnamefont{and}
  \bibinfo{author}{\bibfnamefont{E.~W.~N.} \bibnamefont{Glover}},
  \bibinfo{journal}{JHEP} \textbf{\bibinfo{volume}{09}}, \bibinfo{pages}{056}
  (\bibinfo{year}{2005}{\natexlab{a}}), \eprint{hep-ph/0505111}.

\bibitem[{\citenamefont{Gehrmann-De~Ridder
  et~al.}(2005{\natexlab{b}})\citenamefont{Gehrmann-De~Ridder, Gehrmann, and
  Glover}}]{GehrmannDeRidder:2005aw}
\bibinfo{author}{\bibfnamefont{A.}~\bibnamefont{Gehrmann-De~Ridder}},
  \bibinfo{author}{\bibfnamefont{T.}~\bibnamefont{Gehrmann}}, \bibnamefont{and}
  \bibinfo{author}{\bibfnamefont{E.~W.~N.} \bibnamefont{Glover}},
  \bibinfo{journal}{Phys. Lett.} \textbf{\bibinfo{volume}{B612}},
  \bibinfo{pages}{49} (\bibinfo{year}{2005}{\natexlab{b}}),
  \eprint{hep-ph/0502110}.

\bibitem[{\citenamefont{Gehrmann-De~Ridder
  et~al.}(2005{\natexlab{c}})\citenamefont{Gehrmann-De~Ridder, Gehrmann, and
  Glover}}]{GehrmannDeRidder:2005hi}
\bibinfo{author}{\bibfnamefont{A.}~\bibnamefont{Gehrmann-De~Ridder}},
  \bibinfo{author}{\bibfnamefont{T.}~\bibnamefont{Gehrmann}}, \bibnamefont{and}
  \bibinfo{author}{\bibfnamefont{E.~W.~N.} \bibnamefont{Glover}},
  \bibinfo{journal}{Phys. Lett.} \textbf{\bibinfo{volume}{B612}},
  \bibinfo{pages}{36} (\bibinfo{year}{2005}{\natexlab{c}}),
  \eprint{hep-ph/0501291}.

\bibitem[{\citenamefont{Daleo et~al.}(2007)\citenamefont{Daleo, Gehrmann, and
  Maitre}}]{Daleo:2006xa}
\bibinfo{author}{\bibfnamefont{A.}~\bibnamefont{Daleo}},
  \bibinfo{author}{\bibfnamefont{T.}~\bibnamefont{Gehrmann}}, \bibnamefont{and}
  \bibinfo{author}{\bibfnamefont{D.}~\bibnamefont{Maitre}},
  \bibinfo{journal}{JHEP} \textbf{\bibinfo{volume}{04}}, \bibinfo{pages}{016}
  (\bibinfo{year}{2007}), \eprint{hep-ph/0612257}.

\bibitem[{\citenamefont{Daleo et~al.}(2010)\citenamefont{Daleo,
  Gehrmann-De~Ridder, Gehrmann, and Luisoni}}]{Daleo:2009yj}
\bibinfo{author}{\bibfnamefont{A.}~\bibnamefont{Daleo}},
  \bibinfo{author}{\bibfnamefont{A.}~\bibnamefont{Gehrmann-De~Ridder}},
  \bibinfo{author}{\bibfnamefont{T.}~\bibnamefont{Gehrmann}}, \bibnamefont{and}
  \bibinfo{author}{\bibfnamefont{G.}~\bibnamefont{Luisoni}},
  \bibinfo{journal}{JHEP} \textbf{\bibinfo{volume}{01}}, \bibinfo{pages}{118}
  (\bibinfo{year}{2010}), \eprint{0912.0374}.

\bibitem[{\citenamefont{Gehrmann and Monni}(2011)}]{Gehrmann:2011wi}
\bibinfo{author}{\bibfnamefont{T.}~\bibnamefont{Gehrmann}} \bibnamefont{and}
  \bibinfo{author}{\bibfnamefont{P.~F.} \bibnamefont{Monni}},
  \bibinfo{journal}{JHEP} \textbf{\bibinfo{volume}{12}}, \bibinfo{pages}{049}
  (\bibinfo{year}{2011}), \eprint{1107.4037}.

\bibitem[{\citenamefont{Boughezal et~al.}(2011)\citenamefont{Boughezal,
  Gehrmann-De~Ridder, and Ritzmann}}]{Boughezal:2010mc}
\bibinfo{author}{\bibfnamefont{R.}~\bibnamefont{Boughezal}},
  \bibinfo{author}{\bibfnamefont{A.}~\bibnamefont{Gehrmann-De~Ridder}},
  \bibnamefont{and} \bibinfo{author}{\bibfnamefont{M.}~\bibnamefont{Ritzmann}},
  \bibinfo{journal}{JHEP} \textbf{\bibinfo{volume}{02}}, \bibinfo{pages}{098}
  (\bibinfo{year}{2011}), \eprint{1011.6631}.

\bibitem[{\citenamefont{Gehrmann-De~Ridder
  et~al.}(2012)\citenamefont{Gehrmann-De~Ridder, Gehrmann, and
  Ritzmann}}]{GehrmannDeRidder:2012ja}
\bibinfo{author}{\bibfnamefont{A.}~\bibnamefont{Gehrmann-De~Ridder}},
  \bibinfo{author}{\bibfnamefont{T.}~\bibnamefont{Gehrmann}}, \bibnamefont{and}
  \bibinfo{author}{\bibfnamefont{M.}~\bibnamefont{Ritzmann}},
  \bibinfo{journal}{JHEP} \textbf{\bibinfo{volume}{10}}, \bibinfo{pages}{047}
  (\bibinfo{year}{2012}), \eprint{1207.5779}.

\bibitem[{\citenamefont{Currie et~al.}(2013)\citenamefont{Currie, Glover, and
  Wells}}]{Currie:2013vh}
\bibinfo{author}{\bibfnamefont{J.}~\bibnamefont{Currie}},
  \bibinfo{author}{\bibfnamefont{E.~W.~N.} \bibnamefont{Glover}},
  \bibnamefont{and} \bibinfo{author}{\bibfnamefont{S.}~\bibnamefont{Wells}},
  \bibinfo{journal}{JHEP} \textbf{\bibinfo{volume}{04}}, \bibinfo{pages}{066}
  (\bibinfo{year}{2013}), \eprint{1301.4693}.

\bibitem[{\citenamefont{Currie et~al.}(2017{\natexlab{c}})\citenamefont{Currie,
  Glover, Gehrmann, Gehrmann-De~Ridder, Huss, and Pires}}]{Currie:2017ctp}
\bibinfo{author}{\bibfnamefont{J.}~\bibnamefont{Currie}},
  \bibinfo{author}{\bibfnamefont{E.~W.~N.} \bibnamefont{Glover}},
  \bibinfo{author}{\bibfnamefont{T.}~\bibnamefont{Gehrmann}},
  \bibinfo{author}{\bibfnamefont{A.}~\bibnamefont{Gehrmann-De~Ridder}},
  \bibinfo{author}{\bibfnamefont{A.}~\bibnamefont{Huss}}, \bibnamefont{and}
  \bibinfo{author}{\bibfnamefont{J.}~\bibnamefont{Pires}},
  \bibinfo{journal}{Acta Phys. Polon.} \textbf{\bibinfo{volume}{B48}},
  \bibinfo{pages}{955} (\bibinfo{year}{2017}{\natexlab{c}}),
  \eprint{1704.00923}.

\bibitem[{\citenamefont{Gleisberg et~al.}(2004)\citenamefont{Gleisberg,
  H{\"o}che, Krauss, Sch{\"a}licke, Schumann, and Winter}}]{Gleisberg:2003xi}
\bibinfo{author}{\bibfnamefont{T.}~\bibnamefont{Gleisberg}},
  \bibinfo{author}{\bibfnamefont{S.}~\bibnamefont{H{\"o}che}},
  \bibinfo{author}{\bibfnamefont{F.}~\bibnamefont{Krauss}},
  \bibinfo{author}{\bibfnamefont{A.}~\bibnamefont{Sch{\"a}licke}},
  \bibinfo{author}{\bibfnamefont{S.}~\bibnamefont{Schumann}}, \bibnamefont{and}
  \bibinfo{author}{\bibfnamefont{J.}~\bibnamefont{Winter}},
  \bibinfo{journal}{JHEP} \textbf{\bibinfo{volume}{02}}, \bibinfo{pages}{056}
  (\bibinfo{year}{2004}), \eprint{hep-ph/0311263}.

\bibitem[{\citenamefont{Gleisberg et~al.}(2009)\citenamefont{Gleisberg,
  H{\"o}che, Krauss, Sch\"{o}nherr, Schumann, Siegert, and
  Winter}}]{Gleisberg:2008ta}
\bibinfo{author}{\bibfnamefont{T.}~\bibnamefont{Gleisberg}},
  \bibinfo{author}{\bibfnamefont{S.}~\bibnamefont{H{\"o}che}},
  \bibinfo{author}{\bibfnamefont{F.}~\bibnamefont{Krauss}},
  \bibinfo{author}{\bibfnamefont{M.}~\bibnamefont{Sch\"{o}nherr}},
  \bibinfo{author}{\bibfnamefont{S.}~\bibnamefont{Schumann}},
  \bibinfo{author}{\bibfnamefont{F.}~\bibnamefont{Siegert}}, \bibnamefont{and}
  \bibinfo{author}{\bibfnamefont{J.}~\bibnamefont{Winter}},
  \bibinfo{journal}{JHEP} \textbf{\bibinfo{volume}{02}}, \bibinfo{pages}{007}
  (\bibinfo{year}{2009}), \eprint{0811.4622}.

\bibitem[{\citenamefont{H{\"o}che et~al.}(2012)\citenamefont{H{\"o}che, Krauss,
  Sch{\"o}nherr, and Siegert}}]{Hoeche:2011fd}
\bibinfo{author}{\bibfnamefont{S.}~\bibnamefont{H{\"o}che}},
  \bibinfo{author}{\bibfnamefont{F.}~\bibnamefont{Krauss}},
  \bibinfo{author}{\bibfnamefont{M.}~\bibnamefont{Sch{\"o}nherr}},
  \bibnamefont{and} \bibinfo{author}{\bibfnamefont{F.}~\bibnamefont{Siegert}},
  \bibinfo{journal}{JHEP} \textbf{\bibinfo{volume}{09}}, \bibinfo{pages}{049}
  (\bibinfo{year}{2012}), \eprint{1111.1220}.

\bibitem[{\citenamefont{H{\"o}che et~al.}(2013)\citenamefont{H{\"o}che, Krauss,
  Sch{\"o}nherr, and Siegert}}]{Hoeche:2012ft}
\bibinfo{author}{\bibfnamefont{S.}~\bibnamefont{H{\"o}che}},
  \bibinfo{author}{\bibfnamefont{F.}~\bibnamefont{Krauss}},
  \bibinfo{author}{\bibfnamefont{M.}~\bibnamefont{Sch{\"o}nherr}},
  \bibnamefont{and} \bibinfo{author}{\bibfnamefont{F.}~\bibnamefont{Siegert}},
  \bibinfo{journal}{Phys.Rev.Lett.} \textbf{\bibinfo{volume}{110}},
  \bibinfo{pages}{052001} (\bibinfo{year}{2013}), \eprint{1201.5882}.

\bibitem[{\citenamefont{Schumann and Krauss}(2008)}]{Schumann:2007mg}
\bibinfo{author}{\bibfnamefont{S.}~\bibnamefont{Schumann}} \bibnamefont{and}
  \bibinfo{author}{\bibfnamefont{F.}~\bibnamefont{Krauss}},
  \bibinfo{journal}{JHEP} \textbf{\bibinfo{volume}{03}}, \bibinfo{pages}{038}
  (\bibinfo{year}{2008}), \eprint{0709.1027}.

\bibitem[{\citenamefont{Catani and Seymour}(1997)}]{Catani:1996vz}
\bibinfo{author}{\bibfnamefont{S.}~\bibnamefont{Catani}} \bibnamefont{and}
  \bibinfo{author}{\bibfnamefont{M.~H.} \bibnamefont{Seymour}},
  \bibinfo{journal}{Nucl. Phys.} \textbf{\bibinfo{volume}{B485}},
  \bibinfo{pages}{291} (\bibinfo{year}{1997}), \eprint{hep-ph/9605323}.

\bibitem[{\citenamefont{Catani et~al.}(2002)\citenamefont{Catani, Dittmaier,
  Seymour, and Trocsanyi}}]{Catani:2002hc}
\bibinfo{author}{\bibfnamefont{S.}~\bibnamefont{Catani}},
  \bibinfo{author}{\bibfnamefont{S.}~\bibnamefont{Dittmaier}},
  \bibinfo{author}{\bibfnamefont{M.~H.} \bibnamefont{Seymour}},
  \bibnamefont{and}
  \bibinfo{author}{\bibfnamefont{Z.}~\bibnamefont{Trocsanyi}},
  \bibinfo{journal}{Nucl. Phys.} \textbf{\bibinfo{volume}{B627}},
  \bibinfo{pages}{189} (\bibinfo{year}{2002}), \eprint{hep-ph/0201036}.

\bibitem[{\citenamefont{Catani et~al.}(1991)\citenamefont{Catani, Webber, and
  Marchesini}}]{Catani:1990rr}
\bibinfo{author}{\bibfnamefont{S.}~\bibnamefont{Catani}},
  \bibinfo{author}{\bibfnamefont{B.~R.} \bibnamefont{Webber}},
  \bibnamefont{and}
  \bibinfo{author}{\bibfnamefont{G.}~\bibnamefont{Marchesini}},
  \bibinfo{journal}{Nucl. Phys.} \textbf{\bibinfo{volume}{B349}},
  \bibinfo{pages}{635} (\bibinfo{year}{1991}).

\bibitem[{\citenamefont{B{\"a}hr et~al.}(2008)}]{Bahr:2008pv}
\bibinfo{author}{\bibfnamefont{M.}~\bibnamefont{B{\"a}hr}}
  \bibnamefont{et~al.}, \bibinfo{journal}{Eur. Phys. J.}
  \textbf{\bibinfo{volume}{C58}}, \bibinfo{pages}{639} (\bibinfo{year}{2008}),
  \eprint{0803.0883}.

\bibitem[{\citenamefont{Bellm et~al.}(2016)}]{Bellm:2015jjp}
\bibinfo{author}{\bibfnamefont{J.}~\bibnamefont{Bellm}} \bibnamefont{et~al.},
  \bibinfo{journal}{Eur. Phys. J.} \textbf{\bibinfo{volume}{C76}},
  \bibinfo{pages}{196} (\bibinfo{year}{2016}), \eprint{1512.01178}.

\bibitem[{\citenamefont{Bellm et~al.}(2017)}]{Bellm:2017bvx}
\bibinfo{author}{\bibfnamefont{J.}~\bibnamefont{Bellm}} \bibnamefont{et~al.}
  (\bibinfo{year}{2017}), \eprint{1705.06919}.

\bibitem[{\citenamefont{Alwall et~al.}(2011)\citenamefont{Alwall, Herquet,
  Maltoni, Mattelaer, and Stelzer}}]{Alwall:2011uj}
\bibinfo{author}{\bibfnamefont{J.}~\bibnamefont{Alwall}},
  \bibinfo{author}{\bibfnamefont{M.}~\bibnamefont{Herquet}},
  \bibinfo{author}{\bibfnamefont{F.}~\bibnamefont{Maltoni}},
  \bibinfo{author}{\bibfnamefont{O.}~\bibnamefont{Mattelaer}},
  \bibnamefont{and} \bibinfo{author}{\bibfnamefont{T.}~\bibnamefont{Stelzer}},
  \bibinfo{journal}{JHEP} \textbf{\bibinfo{volume}{06}}, \bibinfo{pages}{128}
  (\bibinfo{year}{2011}), \eprint{1106.0522}.

\bibitem[{\citenamefont{Alioli et~al.}(2014)\citenamefont{Alioli, Badger,
  Bellm, Biedermann, Boudjema et~al.}}]{Alioli:2013nda}
\bibinfo{author}{\bibfnamefont{S.}~\bibnamefont{Alioli}},
  \bibinfo{author}{\bibfnamefont{S.}~\bibnamefont{Badger}},
  \bibinfo{author}{\bibfnamefont{J.}~\bibnamefont{Bellm}},
  \bibinfo{author}{\bibfnamefont{B.}~\bibnamefont{Biedermann}},
  \bibinfo{author}{\bibfnamefont{F.}~\bibnamefont{Boudjema}},
  \bibnamefont{et~al.}, \bibinfo{journal}{Comput.Phys.Commun.}
  \textbf{\bibinfo{volume}{185}}, \bibinfo{pages}{560} (\bibinfo{year}{2014}),
  \eprint{1308.3462}.

\bibitem[{\citenamefont{Buckley et~al.}(2015)\citenamefont{Buckley, Ferrando,
  Lloyd, Nordstr{\"o}m, Page, R{\"u}fenacht, Sch{\"o}nherr, and
  Watt}}]{Buckley:2014ana}
\bibinfo{author}{\bibfnamefont{A.}~\bibnamefont{Buckley}},
  \bibinfo{author}{\bibfnamefont{J.}~\bibnamefont{Ferrando}},
  \bibinfo{author}{\bibfnamefont{S.}~\bibnamefont{Lloyd}},
  \bibinfo{author}{\bibfnamefont{K.}~\bibnamefont{Nordstr{\"o}m}},
  \bibinfo{author}{\bibfnamefont{B.}~\bibnamefont{Page}},
  \bibinfo{author}{\bibfnamefont{M.}~\bibnamefont{R{\"u}fenacht}},
  \bibinfo{author}{\bibfnamefont{M.}~\bibnamefont{Sch{\"o}nherr}},
  \bibnamefont{and} \bibinfo{author}{\bibfnamefont{G.}~\bibnamefont{Watt}},
  \bibinfo{journal}{Eur. Phys. J.} \textbf{\bibinfo{volume}{C75}},
  \bibinfo{pages}{132} (\bibinfo{year}{2015}), \eprint{1412.7420}.

\bibitem[{\citenamefont{Gieseke et~al.}(2003)\citenamefont{Gieseke, Stephens,
  and Webber}}]{Gieseke:2003rz}
\bibinfo{author}{\bibfnamefont{S.}~\bibnamefont{Gieseke}},
  \bibinfo{author}{\bibfnamefont{P.}~\bibnamefont{Stephens}}, \bibnamefont{and}
  \bibinfo{author}{\bibfnamefont{B.}~\bibnamefont{Webber}},
  \bibinfo{journal}{JHEP} \textbf{\bibinfo{volume}{12}}, \bibinfo{pages}{045}
  (\bibinfo{year}{2003}), \eprint{hep-ph/0310083}.

\bibitem[{\citenamefont{Pl{\"a}tzer}(2013)}]{Platzer:2012bs}
\bibinfo{author}{\bibfnamefont{S.}~\bibnamefont{Pl{\"a}tzer}},
  \bibinfo{journal}{JHEP} \textbf{\bibinfo{volume}{08}}, \bibinfo{pages}{114}
  (\bibinfo{year}{2013}), \eprint{1211.5467}.

\bibitem[{\citenamefont{Bellm et~al.}(2018)\citenamefont{Bellm, Gieseke, and
  Pl{\"a}tzer}}]{Bellm:2017ktr}
\bibinfo{author}{\bibfnamefont{J.}~\bibnamefont{Bellm}},
  \bibinfo{author}{\bibfnamefont{S.}~\bibnamefont{Gieseke}}, \bibnamefont{and}
  \bibinfo{author}{\bibfnamefont{S.}~\bibnamefont{Pl{\"a}tzer}},
  \bibinfo{journal}{Eur. Phys. J.} \textbf{\bibinfo{volume}{C78}},
  \bibinfo{pages}{244} (\bibinfo{year}{2018}), \eprint{1705.06700}.

\bibitem[{\citenamefont{Pl{\"a}tzer and Gieseke}(2011)}]{Platzer:2009jq}
\bibinfo{author}{\bibfnamefont{S.}~\bibnamefont{Pl{\"a}tzer}} \bibnamefont{and}
  \bibinfo{author}{\bibfnamefont{S.}~\bibnamefont{Gieseke}},
  \bibinfo{journal}{JHEP} \textbf{\bibinfo{volume}{01}}, \bibinfo{pages}{024}
  (\bibinfo{year}{2011}), \eprint{0909.5593}.

\bibitem[{\citenamefont{Frixione et~al.}(2007)\citenamefont{Frixione, Nason,
  and Oleari}}]{Frixione:2007vw}
\bibinfo{author}{\bibfnamefont{S.}~\bibnamefont{Frixione}},
  \bibinfo{author}{\bibfnamefont{P.}~\bibnamefont{Nason}}, \bibnamefont{and}
  \bibinfo{author}{\bibfnamefont{C.}~\bibnamefont{Oleari}},
  \bibinfo{journal}{JHEP} \textbf{\bibinfo{volume}{11}}, \bibinfo{pages}{070}
  (\bibinfo{year}{2007}), \eprint{0709.2092}.

\bibitem[{\citenamefont{Alioli et~al.}(2010)\citenamefont{Alioli, Nason,
  Oleari, and Re}}]{Alioli:2010xd}
\bibinfo{author}{\bibfnamefont{S.}~\bibnamefont{Alioli}},
  \bibinfo{author}{\bibfnamefont{P.}~\bibnamefont{Nason}},
  \bibinfo{author}{\bibfnamefont{C.}~\bibnamefont{Oleari}}, \bibnamefont{and}
  \bibinfo{author}{\bibfnamefont{E.}~\bibnamefont{Re}}, \bibinfo{journal}{JHEP}
  \textbf{\bibinfo{volume}{06}}, \bibinfo{pages}{043} (\bibinfo{year}{2010}),
  \eprint{1002.2581}.

\bibitem[{\citenamefont{Alioli et~al.}(2011)\citenamefont{Alioli, Hamilton,
  Nason, Oleari, and Re}}]{Alioli:2010xa}
\bibinfo{author}{\bibfnamefont{S.}~\bibnamefont{Alioli}},
  \bibinfo{author}{\bibfnamefont{K.}~\bibnamefont{Hamilton}},
  \bibinfo{author}{\bibfnamefont{P.}~\bibnamefont{Nason}},
  \bibinfo{author}{\bibfnamefont{C.}~\bibnamefont{Oleari}}, \bibnamefont{and}
  \bibinfo{author}{\bibfnamefont{E.}~\bibnamefont{Re}}, \bibinfo{journal}{JHEP}
  \textbf{\bibinfo{volume}{04}}, \bibinfo{pages}{081} (\bibinfo{year}{2011}),
  \eprint{1012.3380}.

\bibitem[{\citenamefont{Sj{\"o}strand et~al.}(2014)\citenamefont{Sj{\"o}strand,
  Ask, Christiansen, Corke, Desai, Ilten, Mrenna, Prestel, Rasmussen, and
  Skands}}]{Sjostrand:2014zea}
\bibinfo{author}{\bibfnamefont{T.}~\bibnamefont{Sj{\"o}strand}},
  \bibinfo{author}{\bibfnamefont{S.}~\bibnamefont{Ask}},
  \bibinfo{author}{\bibfnamefont{J.~R.} \bibnamefont{Christiansen}},
  \bibinfo{author}{\bibfnamefont{R.}~\bibnamefont{Corke}},
  \bibinfo{author}{\bibfnamefont{N.}~\bibnamefont{Desai}},
  \bibinfo{author}{\bibfnamefont{P.}~\bibnamefont{Ilten}},
  \bibinfo{author}{\bibfnamefont{S.}~\bibnamefont{Mrenna}},
  \bibinfo{author}{\bibfnamefont{S.}~\bibnamefont{Prestel}},
  \bibinfo{author}{\bibfnamefont{C.~O.} \bibnamefont{Rasmussen}},
  \bibnamefont{and} \bibinfo{author}{\bibfnamefont{P.~Z.} \bibnamefont{Skands}}
  (\bibinfo{year}{2014}), \eprint{1410.3012}.

\bibitem[{\citenamefont{Magnea and Sterman}(1990)}]{Magnea:1990zb}
\bibinfo{author}{\bibfnamefont{L.}~\bibnamefont{Magnea}} \bibnamefont{and}
  \bibinfo{author}{\bibfnamefont{G.~F.} \bibnamefont{Sterman}},
  \bibinfo{journal}{Phys. Rev.} \textbf{\bibinfo{volume}{D42}},
  \bibinfo{pages}{4222} (\bibinfo{year}{1990}).

\bibitem[{\citenamefont{Ahrens et~al.}(2009)\citenamefont{Ahrens, Becher,
  Neubert, and Yang}}]{Ahrens:2008qu}
\bibinfo{author}{\bibfnamefont{V.}~\bibnamefont{Ahrens}},
  \bibinfo{author}{\bibfnamefont{T.}~\bibnamefont{Becher}},
  \bibinfo{author}{\bibfnamefont{M.}~\bibnamefont{Neubert}}, \bibnamefont{and}
  \bibinfo{author}{\bibfnamefont{L.~L.} \bibnamefont{Yang}},
  \bibinfo{journal}{Phys. Rev.} \textbf{\bibinfo{volume}{D79}},
  \bibinfo{pages}{033013} (\bibinfo{year}{2009}), \eprint{0808.3008}.

\bibitem[{\citenamefont{Buschmann et~al.}(2015)\citenamefont{Buschmann,
  Goncalves, Kuttimalai, Sch{\"o}nherr, Krauss, and Plehn}}]{Buschmann:2014sia}
\bibinfo{author}{\bibfnamefont{M.}~\bibnamefont{Buschmann}},
  \bibinfo{author}{\bibfnamefont{D.}~\bibnamefont{Goncalves}},
  \bibinfo{author}{\bibfnamefont{S.}~\bibnamefont{Kuttimalai}},
  \bibinfo{author}{\bibfnamefont{M.}~\bibnamefont{Sch{\"o}nherr}},
  \bibinfo{author}{\bibfnamefont{F.}~\bibnamefont{Krauss}}, \bibnamefont{and}
  \bibinfo{author}{\bibfnamefont{T.}~\bibnamefont{Plehn}},
  \bibinfo{journal}{JHEP} \textbf{\bibinfo{volume}{02}}, \bibinfo{pages}{038}
  (\bibinfo{year}{2015}), \eprint{1410.5806}.

\bibitem[{\citenamefont{Jones et~al.}(2018)\citenamefont{Jones, Kerner, and
  Luisoni}}]{Jones:2018hbb}
\bibinfo{author}{\bibfnamefont{S.~P.} \bibnamefont{Jones}},
  \bibinfo{author}{\bibfnamefont{M.}~\bibnamefont{Kerner}}, \bibnamefont{and}
  \bibinfo{author}{\bibfnamefont{G.}~\bibnamefont{Luisoni}},
  \bibinfo{journal}{Phys. Rev. Lett.} \textbf{\bibinfo{volume}{120}},
  \bibinfo{pages}{162001} (\bibinfo{year}{2018}), \eprint{1802.00349}.

\bibitem[{\citenamefont{Ellis et~al.}(1992)\citenamefont{Ellis, Kunszt, and
  Soper}}]{Ellis:1992qq}
\bibinfo{author}{\bibfnamefont{S.~D.} \bibnamefont{Ellis}},
  \bibinfo{author}{\bibfnamefont{Z.}~\bibnamefont{Kunszt}}, \bibnamefont{and}
  \bibinfo{author}{\bibfnamefont{D.~E.} \bibnamefont{Soper}},
  \bibinfo{journal}{Phys. Rev. Lett.} \textbf{\bibinfo{volume}{69}},
  \bibinfo{pages}{3615} (\bibinfo{year}{1992}), \eprint{hep-ph/9208249}.

\bibitem[{\citenamefont{Moch et~al.}(2018)\citenamefont{Moch, Eren, Lipka, Liu,
  and Ringer}}]{Moch:2018hgy}
\bibinfo{author}{\bibfnamefont{S.-O.} \bibnamefont{Moch}},
  \bibinfo{author}{\bibfnamefont{E.}~\bibnamefont{Eren}},
  \bibinfo{author}{\bibfnamefont{K.}~\bibnamefont{Lipka}},
  \bibinfo{author}{\bibfnamefont{X.}~\bibnamefont{Liu}}, \bibnamefont{and}
  \bibinfo{author}{\bibfnamefont{F.}~\bibnamefont{Ringer}},
  \bibinfo{journal}{PoS} \textbf{\bibinfo{volume}{LL2018}},
  \bibinfo{pages}{002} (\bibinfo{year}{2018}), \eprint{1808.04574}.

\bibitem[{\citenamefont{Winter et~al.}(2004)\citenamefont{Winter, Krauss, and
  Soff}}]{Winter:2003tt}
\bibinfo{author}{\bibfnamefont{J.-C.} \bibnamefont{Winter}},
  \bibinfo{author}{\bibfnamefont{F.}~\bibnamefont{Krauss}}, \bibnamefont{and}
  \bibinfo{author}{\bibfnamefont{G.}~\bibnamefont{Soff}},
  \bibinfo{journal}{Eur. Phys. J.} \textbf{\bibinfo{volume}{C36}},
  \bibinfo{pages}{381} (\bibinfo{year}{2004}), \eprint{hep-ph/0311085}.

\bibitem[{\citenamefont{Sj{\"o}strand et~al.}(2006)\citenamefont{Sj{\"o}strand,
  Mrenna, and Skands}}]{Sjostrand:2006za}
\bibinfo{author}{\bibfnamefont{T.}~\bibnamefont{Sj{\"o}strand}},
  \bibinfo{author}{\bibfnamefont{S.}~\bibnamefont{Mrenna}}, \bibnamefont{and}
  \bibinfo{author}{\bibfnamefont{P.}~\bibnamefont{Skands}},
  \bibinfo{journal}{JHEP} \textbf{\bibinfo{volume}{05}}, \bibinfo{pages}{026}
  (\bibinfo{year}{2006}), \eprint{hep-ph/0603175}.

\end{thebibliography}

\end{document}